\documentclass[12pt]{article} 
\pdfoutput=1
\usepackage{hyperref}
\usepackage{url}
\usepackage{setspace}
\usepackage{amssymb,amscd,braket}
\usepackage{amsfonts}
\usepackage{amsmath}
\usepackage{color}
\usepackage{graphicx}
\usepackage{simplewick}
\usepackage{cancel}
\usepackage{tikz-feynman}
\usepackage{mathrsfs}

\usepackage{cite}

\newcommand{\ra}{\rightarrow}

\newcommand{\lra}{\leftrightarrow}

\newcommand{\<}{\langle}
\renewcommand{\>}{\rangle}
\newcommand{\be}{\begin{equation}}
\newcommand{\ee}{\end{equation}}
\newcommand{\ba}{\begin{aligned}}
\newcommand{\ea}{\end{aligned}}
\newcommand{\bea}{\begin{eqnarray}}
\newcommand{\eea}{\end{eqnarray}}
\newcommand{\benn}{\begin{equation*}}
\newcommand{\eenn}{\end{equation*}}
\newcommand{\bi}{\begin{itemize}}  
\newcommand{\ei}{\end{itemize}}
\newcommand{\bpm}{\begin{pmatrix}}
\newcommand{\epm}{\end{pmatrix}}

\newcommand{\p}{\partial}

\newcommand{\eps}{\varepsilon}

\newcommand{\De}{\Delta}
\newcommand{\de}{\delta}
\newcommand{\G}{\Gamma}

\newcommand{\Ecal}{{\mathcal E}}

\newcommand{\Kcal}{{\mathcal K}}

\newcommand{\Ncal}{{\mathcal N}}
\newcommand{\Ocal}{{\mathcal O}}
\newcommand{\Pcal}{{\mathcal P}}

\newcommand{\nn}{\nonumber}
\newcommand{\xp}{x^+}
\newcommand{\xm}{x^-}

\newcommand{\fr}{\frac}
\newcommand{\tfr}{\tfrac}
\newcommand{\half}{\frac{1}{2}}

\newcommand{\zb}{\bar{z}}

\newcommand{\comm}[2]{[#1,#2]}

\newcommand{\aslash}[1]{\,\,{\raise.15ex\hbox{/}\mkern-12mu #1}}
\newcommand{\bslash}[1]{\,\,{\raise.15ex\hbox{/}\mkern-9mu #1}}

\newcommand{\x}[2]{x_{#1,#2}^-}
\newcommand{\f}[2]{f_{#1,#2}}
\newcommand{\g}[2]{g_{#1,#2}}

\numberwithin{equation}{section}

\newcommand\rref[1]{(\ref{#1})}

\newcommand\lrpar{\raise .8ex\hbox{$^\leftrightarrow$} \hspace{-9pt}
\partial}
\newcommand\lpar{\raise .8ex\hbox{$^\leftarrow$} \hspace{-9pt}
\partial}
\newcommand\rpar{\raise .8ex\hbox{$^\rightarrow$} \hspace{-9pt}
\partial}
\newcommand\lrd{\raise .8ex\hbox{$^\leftrightarrow$} \hspace{-9pt}
\nabla}

\newcommand{\gsim}{\lower.7ex\hbox{$\;\stackrel{\textstyle>}{\sim}\;$}}
\newcommand{\lsim}{\lower.7ex\hbox{$\;\stackrel{\textstyle<}{\sim}\;$}}

\renewcommand{\title}[1]{\vbox{\center\LARGE{#1}}\vspace{5mm}}
\renewcommand{\author}[1]{\vbox{\center#1}\vspace{5mm}}
\newcommand{\address}[1]{\vbox{\center\footnotesize\em#1}}
\newcommand{\email}[1]{\vbox{\center\footnotesize\tt#1}\vspace{5mm}}

\begin{document}

\begin{titlepage}

\hfill \\
\begin{flushright}
\hfill{\tt CERN-TH-2020-200}
\end{flushright}

\begin{center}

\hfill \\

\title{On the Stress Tensor Light-ray Operator Algebra}

\author{Alexandre Belin$^{a}$, Diego M. Hofman$^{b}$, Gr\'egoire Mathys$^{b}$, Matthew T.\ Walters$^{c}$
}

\address{
${}^a$CERN, Theory Division,\\
CH-1211 Geneva, Switzerland\\
\vspace{0.5cm}
${}^b$Institute for Theoretical Physics, University of Amsterdam,\\ 
1090 GL Amsterdam, The Netherlands \\
\vspace{0.5cm}
${}^c$ Institute of Physics, \'Ecole Polytechnique F\'ed\'erale de Lausanne (EPFL),\\
CH-1015 Lausanne, Switzerland
}

\email{\href{mailto:a.belin@cern.ch}{a.belin@cern.ch},\, \href{mailto:d.m.hofman@uva.nl}{d.m.hofman@uva.nl},\, \href{mailto:g.o.mathys@uva.nl}{g.o.mathys@uva.nl},\, \href{mailto:matthew.walters@epfl.ch}{matthew.walters@epfl.ch} }

\end{center}

\abstract{ We study correlation functions involving generalized ANEC operators of the form $\int dx^- \left(x^-\right)^{n+2} T_{--}(\vec{x})$ in four dimensions. We compute two, three, and four-point functions involving external scalar states in both free and holographic Conformal Field Theories. From this information, we extract the algebra of these light-ray operators. We find a global subalgebra spanned by $n=\{-2, -1, 0, 1, 2\}$ which annihilate the conformally invariant vacuum and transform among themselves under the action of the collinear conformal group that preserves the light-ray. Operators outside this range give rise to an infinite central term, in agreement with previous suggestions in the literature. In free theories, even some of the operators inside the global subalgebra fail to commute when placed at spacelike separation on the same null-plane. This lack of commutativity is not integrable, presenting an obstruction to the construction of a well defined light-ray algebra at coincident $\vec{x}$ coordinates. For holographic CFTs the behavior worsens and operators with $n \neq -2$ fail to commute at spacelike separation. We reproduce this result in the bulk of AdS where we present new exact shockwave solutions dual to the insertions of these (exponentiated) operators on the boundary. 

}

\vfill

\end{titlepage}

\eject

\tableofcontents
\addtocontents{toc}{\protect\setcounter{tocdepth}{2}}

\section{Introduction and summary of results}

Two-dimensional conformal field theories occupy a special place in the landscape of all Conformal Field Theories (CFTs). In two dimensions, conformal invariance of a field theory implies the existence of an infinite-dimensional symmetry -- the Virasoro symmetry \cite{Belavin:1984vu}. The presence of this symmetry has far-reaching consequences, going from the existence of CFTs with a finite number of (Virasoro) primary operators (rational CFTs) to the actual solvability of the conformal bootstrap in such cases \cite{Polyakov:1974gs}.

Virasoro symmetry also plays an important role for quantum gravity in AdS$_3$ through the AdS/CFT correspondence. Gravitational dynamics are much simpler in three-dimensions due to their topological nature, in fact so simple that they are completely universal. The CFT analog of this statement is that the stress-tensor sector of a 2d CFT should be completely universal, which is indeed the case as enforced by Virasoro symmetry. These observations have enabled a powerful machinery to derive gravitational dynamics in holographic two-dimensional CFTs by assuming the dominance of the Virasoro identity block \cite{Hartman:2013mia,Fitzpatrick:2014vua,Asplund:2014coa,Roberts:2014ifa,Fitzpatrick:2015zha,Anous:2016kss,Fitzpatrick:2016ive,Anand:2017dav,Anous:2019yku}.\footnote{Note that not all observables are necessarily reproduced by the identity block even in holographic CFTs, see for example \cite{Belin:2017nze}.}

Given the successes of Virasoro symmetry in $d=2$, it is natural to ask whether such a symmetry can exist in higher-dimensional CFTs. A intuitive way to think about the Virasoro symmetry is that in two dimensions, only the stress-tensor $T$ and its composites can appear in the $T\times T$ OPE, with coefficients uniquely determined by the central charge. This immediately presents serious challenges for higher-dimensional CFTs since all (neutral)  operators of the theory
can in principle appear in the stress-tensor OPE, making any form of universality seem hopeless. Potential ways around this obstruction have been suggested. Let us consider two such proposals.\footnote{A third proposal that we will not discuss in the context of this paper relates to the algebra of chiral operators placed on a 2d plane in four and six dimensional Superconformal Theories \cite{Beem:2013sza,Beem:2014kka}.}

The first is to consider a class of non-local CFT observables known as light-ray operators \cite{Casini:2017roe,Kravchuk:2018htv,Cordova:2018ygx,Belin:2019mnx,Kologlu:2019bco,Kologlu:2019mfz,Chang:2020qpj}. The operators we will be particularly interested in are built from null-integrals of the stress-tensor as follows
\be \label{lightrayintro}
\Ecal_f(x^+,\vec{x}^\perp) \equiv \int_{-\infty}^\infty dx^- f(x^-) T_{--}(x) \,.
\ee
For $f(\xm)=1$, this operator satisfies the averaged null energy condition (ANEC) \cite{Faulkner:2016mzt,Hartman:2016lgu} meaning that the operator is positive in any quantum field theory (QFT). This fact has far reaching consequences and it encodes important constraints on QFTs consistent with causality. In particular, it imposes bounds on the range of central charges of unitary CFTs. This was first suggested in \cite{Hofman:2008ar} and finally proven in \cite{Hofman:2016awc} (see \cite{Cordova:2017zej,Meltzer:2018tnm,Chowdhury:2017vel} for other bounds). 

A family of operators of the form \rref{lightrayintro} all living at the same $\vec{x}^\perp$ are embedded in a two-dimensional plane. This offers a promising framework to look for a Virasoro algebra. In fact, by considering properties of modular Hamiltonians on deformed half-spaces, \cite{Casini:2017roe} proposed that the operators
\be \label{generalintro}
L_n(x^+,\vec{x}^\perp) \equiv \Ecal_{x^{n+2}}(x^+,\vec{x}^\perp)  \,,
\ee
satisfy the Virasoro algebra (in $d=4$)
\be
\left[L_m(\vec{x}^\perp),L_n(\vec{y}^\perp)\right] =-i \delta^{(2)}(\vec{x}^\perp-\vec{y}^\perp)(m-n)L_{m+n+1}(\vec{y}^\perp)\,.\label{eq:Virasoro4dintro}
\ee
This algebra is really a Witt algebra rather than Virasoro since one does not obtain a finite central charge (see \cite{Huang:2020ycs} for a proposal to do so by balancing UV and IR divergences). A version of this algebra was proposed to hold  for arbitrary CFTs in $d>2$. A part of the ``global" version of this algebra was also argued for in \cite{Cordova:2018ygx} (see also \cite{Donnay:2020fof}), with explicit checks in free field theory. A careful consideration of this proposal for a Virasoro algebra in $d=4$ will be presented in this work.

The second possible way to find a Virasoro algebra is to consider very special CFTs: holographic large $N$ CFTs in $d$ dimensions which capture the dynamics of Einstein gravity in AdS$_{d+1}$. While a generic CFT will have all kinds of operators appearing in the stress-tensor OPE, holographic CFTs have the special property that
\be
T \times T \sim T + \text{composites} + \mathcal{O}(1/N^2)+\mathcal{O}(1/\Delta_{\text{gap}}) \,,
\ee
where $\Delta_{\text{gap}}$ controls the corrections to Einstein gravity in the bulk. This suggests that while the stress-tensor sector is highly non-universal in a generic CFT, it becomes universal at large $N$ and large $\Delta_{\text{gap}}$. Some evidence has been gathered in this direction \cite{Camanho:2014apa,Afkhami-Jeddi:2016ntf,Kulaxizi:2017ixa,Li:2017lmh,Costa:2017twz,Meltzer:2017rtf,Fitzpatrick:2019zqz,Huang:2019fog,Kulaxizi:2019tkd,Fitzpatrick:2019efk,Karlsson:2019dbd,Karlsson:2020ghx,Fitzpatrick:2020yjb}. One may hope that this universality is controlled by a Virasoro symmetry, emergent at large $N$ and large $\Delta_{\text{gap}}$, which can recast gravitational dynamics of Einstein gravity in terms of a symmetry. 

It is with this overarching goal in mind that we will study the algebra of light-ray operators \rref{lightrayintro}. In the CFT context we will study mostly free theories and will comment on how to use the conformal block decomposition to extrapolate some of these results to holographic CFTs. We then turn to computations in AdS gravity where we explicitly obtain shockwave solutions that allow us to explore the algebra of these operators directly. As we will see, the algebra \rref{eq:Virasoro4dintro} as advocated for in \cite{Casini:2017roe}, does not seem to hold, neither in free field theory nor in holographic CFTs where one would expect the most universality\footnote{Possible obstructions to these type of constructions were already put forward in \cite{Kologlu:2019bco,Kologlu:2019mfz} by pointing out issues with the convergence properties of \rref{lightrayintro}. We will comment on this as we encounter these issues in our computations.}.

\subsection{Summary of results}

In this paper, we present various results for the expectation values and commutators of operators \rref{lightrayintro} in certain states. We provide calculations in free field theories and holographic theories in $d=4$, for which we discuss both the gravitational and CFT sides of the computations. While we give numerous explicit computations throughout the paper, we would like to highlight the following three results.

\subsubsection*{Collinear transformations and a family of five light-ray operators}

There is a subset of the conformal group known as the collinear subgroup \cite{Braun:2003rp}. The group action maps five light-ray operators into one another. These operators are the $L_n$ operators of \rref{generalintro} for $-2\leq n \leq 2$, the simplest of which is the ANEC operator ($L_{-2}$). These operators combine into a five-dimensional representation of the collinear algebra and it thus natural to discuss them together. It is interesting to note that from the point of view of the Virasoro algebra \rref{eq:Virasoro4dintro}, one may want to call the operators $-2\leq n \leq 0$ the ``global" part, but we will see that the action of these operators on the vacuum make it more natural to refer to this whole family of five operators as global. In general $d$ dimensional CFTs this global algebra has dimension $d+1$, consistent with the well known three-dimensional global subalgebra of the Virasoro generators in CFTs in $d=2$.

\subsubsection*{A breakdown of the algebra}

We will explicitly see that the operators \rref{lightrayintro} do not in general commute when inserted on the same null-plane even at finite spacelike separation.  For example, in free field theory we have\footnote{See section \ref{conv} for conventions.}
\be
\braket{\phi(x_1)[L_1(x_2),L_0(x_3)]\phi(x_4)}=-\frac{\left(x_1^--\frac{| \vec{x}_{12}^\perp|^2}{x_{12}^+}\right)-\left(x_4^- +\frac{|\vec{x}_{24}^\perp|^2}{x_{24}^+}\right)}{3\pi^2x_{12}^+x_{24}^+|\vec{x}_{23}^\perp|^2}\, .\label{eq:L1L0commintro}
\ee
For a holographic CFT,  the problem actually worsens and lower modes of the $L_n$'s fail to commute. A simple way to visualize the result in this case is to quote the expression for operators inserted on the celestial sphere at different angles as is familiar in collider experiments \cite{Hofman:2008ar}. For example, taking spherically symmetric scalar states of definite timelike momentum $p^0$, we find
\be
\fr{\<\Ocal(p^0)|\comm{L_{-1}}{L_{-2}}|\Ocal(p^0)\>}{\<\Ocal(p^0)|\Ocal(p^0)\>} = -i \fr{p^0}{16\pi^2} \big( 1 + 3 \cos \theta_{12} \big) \,,
\ee
where $\theta_{12}$ is the angle between the two operators on the celestial sphere.

The fact that the operators do not commute at spacelike separation makes it extremely challenging to define an algebra. The result \rref{eq:L1L0commintro} is not integrable in the $\vec{x}^\perp$ direction rendering the short distance singularity ambiguous. We will expand on this issue throughout the paper and in the discussion section.

Similar observations were made in \cite{Kologlu:2019bco}, where it was shown that four-point functions involving two light-ray operators $L_n$ and $L_m$ are only unambiguously defined (i.e.~that the integrals of the Wightman functions are absolutely convergent) provided $n+m$ satisfies a bound that depends on the Regge intercept of the CFT. We will discuss how our results connect to this statement.

\subsubsection*{Generalized shockwave geometries in AdS}

It is long known that the gravity dual of ANEC operator insertions are shockwaves \cite{Hofman:2008ar}. In this paper, we present new exact solutions to Einstein's equations which are generalized shockwaves in AdS, with a source given by one of the global $L_{n}$'s at the boundary. We were able to find exact solutions for all operator insertions but $L_0$, for reasons that we detail in the main text (see section \ref{gensho} for explicit metrics corresponding to $L_{-1}$ and $L_2$). By scattering waves through these shocks, we can compute correlators in the bulk and compare to the computation in a holographic CFT. We find perfect agreement with the CFT answer, and find that these shocks do not commute.
\vspace{12pt}

This paper is organized as follows: In section 2, we present our conventions and define carefully the operators we want to investigate. In section 3, we study the action of the collinear subgroup, which is the subgroup of conformal transformations that maps the light-ray onto itself, and identify a family of five light-ray operators that map into one another under the group action. In section 3, we evaluate two- and three-point functions involving light-ray operators and compute the would-be central charge of the algebra we are investigating. In section 4, we compute the four-point functions as well as the commutator involving two global light-ray operators in free field theory, and investigate the algebra of the five global operators. In section 6, we explain the finite transverse separation contribution in the commutator of two light-ray operators by studying a subset of the OPE of two light-ray operators, with the specific example of $[L_1,L_1]$ in mind. In section 7, we perform a conformal block decomposition relevant for holographic CFTs. In section 8, we describe the gravitational shockwaves dual to the generalized ANEC operators. We conclude in section 9. Many details are provided in the appendices.

\section{Generalized ANEC operators}
\label{sec:Conventions}

In this section, we introduce the conventions that we use throughout this work, as well as precisely define the light-ray operators that we consider.

\subsection{Conventions}\label{conv}

We will be working in $d=3+1$ spacetime dimensions, with the coordinates
\be
x^\pm \equiv x^0 \pm x^3, \quad \vec{x}^\perp \equiv (x^1,x^2),
\ee
and the associated metric
\be
ds^2 = -dx^+ dx^- + d\vec{x}^{\perp2}.
\ee
Coordinates with lowered indices are
\be
x_\pm \equiv \half(x_0 \pm x_3) = -\half x^\mp, \quad \vec{x}_\perp \equiv (x_1,x_2) = \vec{x}^\perp,
\ee
and the invariant distance is
\be
x^2 = -x^+ x^- + |\vec{x}^\perp|^2 = -4x_+ x_- + |\vec{x}_\perp|^2.
\ee

Throughout this work, we will be studying Wightman functions of the general form
\be \label{eq:np Wightman}
\<\Ocal_1(x_1) \cdots \Ocal_n (x_n)\>\, .
\ee
We can ensure this particular fixed ordering in Lorentzian signature (with $\Ocal_i$ to the left of $\Ocal_{i+1}$) by using the following $i\epsilon$ prescription \cite{Hartman:2015lfa}
\be
x_i^\pm \ra x_i^\pm - i\epsilon_i,
\ee
with $\epsilon_1 > \cdots > \epsilon_n$. In practice, one can do this by using the following prescription: when evaluating an $n-$point Wightman function of the form \eqref{eq:np Wightman}, choose
\be 
x_1^{\pm} \rightarrow x_1^\pm - n\, i\epsilon\, ,\qquad x_2^\pm \rightarrow x_2^\pm -(n-1)\, i \epsilon\, ,\qquad \dots\, ,\qquad x_n^\pm \rightarrow x_n^\pm - i \epsilon\, . \label{iepsprescription}
\ee
When evaluating integrals over $x^-$, we will often encounter poles from the OPE singularity where the distance between two operators goes to zero, i.e.~$x_{ij}^2=0$. The locations of these poles take the general form
\be
x_j^- = x_i^- - \fr{|\vec{x}_{ij}^\perp|^2}{x_{ij}^+}.
\ee
It will be convenient to represent this combination of coordinates with the shorthand notation
\be
\x{i}{j} \equiv x_i^- - \fr{|\vec{x}_{ij}^\perp|^2}{x_{ij}^+}.
\ee
Note that the first index indicates the location of the pole (in $x^-$), while the second index indicates which variable was integrated over. The coordinate $x_{i,j}^-$ is thus specifically the location of a pole in the $x_j^-$ plane. 
%

\subsection{Definition of generalized ANEC operators \label{genANECop}}

We study a set of light-ray operators that have been considered previously in~\cite{Casini:2017roe,Cordova:2018ygx,Kologlu:2019bco,Huang:2020ycs}. They are generalizations of the ANEC operator where we integrate $T_{--}(x)$ along the null direction $x^-$, weighted by an arbitrary function $f(x^-)$,
\be
\Ecal_f(x^+,\vec{x}^\perp) \equiv \int_{-\infty}^\infty dx^- f(x^-) T_{--}(x)\,. \label{eq:epsilonf}
\ee
In particular, for $f(x^-)=1$ we recover the ANEC operator.  One must be careful when inserting these operators in Wightman functions, since the resulting integral over $\xm$ may not converge. This of course depends on the behaviour of the function $f$ near infinity. We will detail below the precise function class from which we draw $f$.

As explained in the introduction, we will consider the functions $f(x^-) = (x^-)^{n+2}$, in analogy with Virasoro generators in $d=2$. We will denote the associated operators by $L_n$, thus defined as 
\be
L_n(x^+,\vec{x}^\perp) \equiv \Ecal_{x^{n+2}}(x^+,\vec{x}^\perp) = \int_{-\infty}^\infty dx^- (x^-)^{n+2} T_{--}(x^+,x^-,\vec{x}^\perp) \,. \label{eq:DefLn}
\ee
Our convention \eqref{eq:DefLn} is slightly different than that used in \cite{Casini:2017roe,Kologlu:2019bco,Huang:2020ycs} (the label $n$ is shifted by 1). It will become clear in section~\ref{sec:algebra} why we find our convention more convenient. Note that throughout this work, we will use a slight abuse of notation and write $L_n(x) \equiv L_n (x^+, \vec{x}^\perp)$. 

Let us now return to the function class from which we would like to draw the functions $f(\xm)$. For the integral to converge inside arbitrary correlation functions, it is manifest that the function $f$ should have nice boundedness properties near infinity. In particular, choosing $f(\xm)=(\xm)^{n+2}$ will be ill-behaved for sufficiently high $n$. Rather than directly working with bounded functions, we will define the operators $L_n$ through a limiting procedure. We define
\bea
L_n(x^+,\vec{x}^\perp)&=&\lim_{\delta\to 0^+} \int_{-\infty}^\infty dx^- e^{i \delta \xm} (x^-)^{n+2} T_{--}(x^+,x^-,\vec{x}^\perp)  \notag \\
&=& \oint  dx^- (x^-)^{n+2} T_{--}(x^+,x^-,\vec{x}^\perp)  \label{lightraycontour}\,,
\eea
where we close the integration contour in the upper half-plane before taking the $\delta\to0^+$ limit. 
This contour integral is now well-behaved inside arbitrary correlation functions. 
To avoid cluttering the equations, we will omit the limiting procedure and not explicitly write out the contour integral in the rest of the paper, but it should be understood that we implement this procedure on all operators. This procedure is harmless when considering convergent real integrals (like the ones we will study) and amounts to a particular regularization when they don't. A physical way of thinking about this is to restrict the matrix elements of the operators involved to states with support confined to a localized enough region in $x^-$.

That being said, in this paper we will mostly focus on $f(\xm)=(\xm)^{n+2}$ for $-2\leq n \leq 2$. We will show that in all correlation functions we consider, the integrand for this set of functions is bounded at infinity, with no additional poles added by $f(x^-)$, and we can close the contour without any need for a regularization procedure. This observation will be useful in practice when evaluating integrals in the following sections.

For $n < -2$, the function $f(x^-) = (x^-)^{n+2}$ introduces a pole at $x^-=0$. In defining the light-ray operators $L_n$, we must choose a prescription for this pole along the initial line of integration. We will choose to move this pole into the upper half-plane, such that it is enclosed by our final integration contour,
\be 
L_{-n}(x^+,\vec{x}^\perp) = \int_{-\infty}^\infty dx^- \frac{1}{(x^--i\epsilon)^{n-2}}T_{--}(x^+,x^-,\vec{x}^\perp)\, , \qquad n>2\, .
\label{eq:NegativeLnDef}
\ee
This choice of prescription ensures that these operators act as creation operators on the vacuum, in analogy with two-dimensional CFTs.

\section{Conformal transformations of light-ray operators\label{sec:algebra}}

Before studying the structure of correlation functions involving the light-ray operators $L_n$, it will be useful to first understand their behavior under conformal transformations. In particular, we shall focus on the so-called ``collinear'' subgroup of conformal transformations, which preserve the null line $x^+ = \vec{x}^\perp = 0$. Under these transformations, the operators $L_n$ in general dimensions behave similarly to their $d=2$ inspiration, with a set of ``global'' operators forming a finite-dimensional representation of the collinear subgroup, and the remaining $L_n$ grouped into two infinite towers of ``Virasoro'' operators.

\subsection{Collinear subgroup}

In $d$ dimensions, the conformal group $SO(d,2)$ is built from translations $P_\mu$, Lorentzian boosts and rotations $M_{\mu\nu}$, dilatations $D$, and the special conformal transformations $K_\mu$. We will follow the conventions of~\cite{Braun:2003rp} for the commutation relations of these conformal generators.

We are particularly interested in the set of conformal transformations which map the light-ray along the $x^-$ direction,
\be
x^\mu = \alpha n^\mu, \quad n^\mu \equiv (n^+,n^-,\vec{n}^\perp) = (0,1,\vec{0}),
\ee
to itself. These transformations are generated by the four generators $D$, $P_-$, $M_{+-}$, and $K_+$, which form the \emph{collinear subalgebra} of the full conformal algebra. If we arrange these generators into the useful form
\be
J_{-1} \equiv iP_-, \quad J_0 \equiv \fr{i}{2}(D-2M_{+-}), \quad J_1 \equiv -i K_+,
\ee
we see that they satisfy the familiar algebra of $SL(2,\mathbb{R})$,
\be
\comm{J_0}{J_{\pm1}} = \mp J_{\pm1}, \quad \comm{J_1}{J_{-1}} = 2J_0,
\ee
similar to the ``global'' conformal algebra in $d=2$. We also have the remaining combination
\be
\bar{J}_0 \equiv \fr{i}{2}(D+2M_{+-}),
\ee
which commutes with all $J_i$ and measures the collinear ``twist'' $\bar{h} \equiv \half(\De - m)$, where $m$ is the spin component in the $x^\pm$ plane.

Under a general collinear transformation, the coordinate $x^-$ transforms as
\be
x^- \ra \fr{ax^- + b}{cx^- + d},
\label{eq:TransformationSL2}
\ee
with the constraint $ad-bc=1$. The remaining coordinates transform as
\be
x^+ \ra x^+ - \fr{c |\vec{x}^\perp|^2}{cx^- + d}, \qquad \vec{x}^\perp \ra \fr{\vec{x}^\perp}{cx^- + d}.
\ee
We therefore clearly see that these transformations preserve the null line $x^+ = \vec{x}^\perp = 0$.

The action of the collinear generators on a general primary operator $\Ocal(x^-)$ located on this null line is
\begin{align}
\comm{J_{-1}}{\Ocal(x^-)} &= \p_-\Ocal(x^-),\nn \\
\comm{J_0}{\Ocal(x^-)} &= \Big( h + x^- \p_- \Big)\Ocal(x^-), \\
\comm{J_1}{\Ocal(x^-)} &= \Big( 2h x^- + (x^-)^2 \p_- \Big)\Ocal(x^-),\nn
\end{align}
where $h \equiv \half(\De+m)$. We are specifically interested in the operator $T_{--}$, which at arbitrary $x$ transforms as
\begin{align}
\comm{J_{-1}}{T_{--}(x)} &= \p_-T_{--}(x), \nn\\
\comm{J_0}{T_{--}(x)} &= \Big( h_T + x^- \p_- + \half \vec{x}^\perp \cdot \vec{\p}_\perp \Big) T_{--}(x), \label{eq:CollTransTmm}\\
\comm{J_1}{T_{--}(x)} &= \Big(2h_T x^- + (x^-)^2 \p_- + x^- \vec{x}^\perp \cdot \vec{\p}_\perp) + |\vec{x}^\perp|^2 \p_+ \Big) T_{--}(x) + 2\vec{x}^\perp \cdot \vec{T}_{-\perp}(x),\nn
\end{align}
with $h_T = 3$ (and $\bar{h}_T = 1$) for $d=4$. Away from $\vec{x}^\perp = 0$, $T_{--}$ therefore mixes with other components of the stress tensor under collinear transformations.

\subsection{Transformations of generalized ANEC operators}

Let's now consider the behavior of the light-ray operators $L_n$ under general collinear transformations, which can be derived from that of $T_{--}$ in eq.~\eqref{eq:CollTransTmm}. These transformations are simplest for $\vec{x}^\perp = 0$, in which case we find\footnote{In deriving these expressions, we have used the fact that all support for $T_{--}(x)$ vanishes at $x^- \ra \infty$. This follows from the limiting procedure \rref{lightraycontour}.}
\be
\ba
\comm{J_{-1}}{L_n(x^+)} &= -(n+2) L_{n-1}(x^+), \\
\comm{J_0}{L_n(x^+)} &= -n L_n(x^+), \\
\comm{J_1}{L_n(x^+)} &= -(n-2) L_{n+1}(x^+).
\ea
\ee
The operator $L_n$ thus has collinear weight $-n$. Unsurprisingly, we see that $J_{\pm 1}$ act as raising and lowering operators, moving us between the different $L_n$.

\begin{figure}[t!]
\begin{center}
\includegraphics[width=0.6\textwidth]{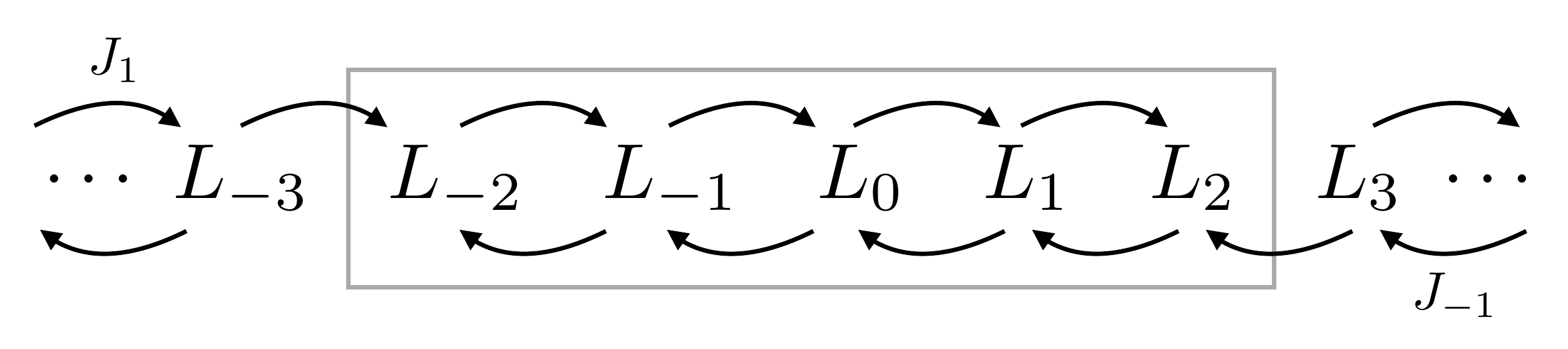}
\caption{Schematic representation of the action of the collinear generators $J_{\pm 1}$ on the light-ray operators $L_n$. The ``global'' operators with $n=\{-2,-1,0,1,2\}$ form a five-dimensional representation of $SL(2,\mathbb{R})$, with the remaining operators in two infinite towers.}
\label{fig:LightRayTower} 
\end{center}
\end{figure}

However, if we specifically look at the ANEC operator $L_{-2}$, we see that it is \emph{annihilated} by $J_{-1}$,
\be
\comm{J_{-1}}{L_{-2}(x^+)} = 0,
\ee
which simply follows from the fact that $L_{-2}$ is translation-invariant along $x^-$. If we now repeatedly act with $J_1$ on $L_{-2}$, we move through the higher $L_n$ until we reach $L_2$, which is annihilated by $J_1$,
\be
\comm{J_1}{L_2(x^+)} = 0.
\ee
The central five operators
\benn
\{L_{-2}, \, L_{-1}, \, L_0, \, L_1, \, L_2\},
\eenn
thus form a finite-dimensional representation of $SL(2,\mathbb{R})$ when acting at $\vec{x}^\perp = 0$. We shall refer to these five operators as ``global'' light-ray operators, in analogy with two dimensions. None of the remaining generalized ANEC operators $L_{\pm n}$ with $n \geq 3$ are annihilated by the $J_i$, and instead form two infinite towers with respect to the collinear subgroup, as shown schematically in figure~\ref{fig:LightRayTower}.

We can also consider the behavior of these light-ray operators under the finite $SL(2,\mathbb{R})$ transformation~\eqref{eq:TransformationSL2}. At $\vec{x}^\perp=0$, the stress tensor simply transforms as
\be
T_{--}(x^+,x^-) \ra (cx^- + d)^{2h_T} T_{--}(x^+,x^-),
\ee
which we can use to derive the transformation of the general light-ray operator
\be
L_n(x^+) \ra \int dx^- \fr{(ax^-+b)^{n+2}}{(cx^-+d)^{n-2}} T_{--}(x^+,x^-).
\ee
Expanding this expression as a series in $x^-$, we see that for $|n| > 2$ the operator $L_n$ mixes with an infinite number of other light-ray operators. However, the five ``global'' operators only mix with each other, as expected. The transformation of the five global operators under a general collinear transformation are   
\begin{align}
L_{-2} &\ra d^4 L_{-2} + 4cd^3 L_{-1} + 6c^2d^2 L_0 + 4 c^3d L_1 + c^4 L_2\, ,\nn\\
L_{-1} &\ra bd^3 L_{-2} + d^2(ad + 3bc)L_{-1} + 3cd(ad + bc)L_0+ c^2(3ad + bc)L_{1} + ac^3 L_2\, ,\nonumber\\
L_0 &\ra b^2 d^2 L_{-2} + 2bd(bc +ad)L_{-1}+ (a^2d^2 + 4abcd + b^2c^2)L_0 + 2ac(ad + bc) L_1 + a^2 c^2 L_2 \, ,\nonumber \\
L_1&\ra b^3 d L_{-2}+ b^2(3ad + bc)L_{-1} + 3ab(ad + bc)L_0 + a^2(ad + 3bc)L_1 + a^3c L_2 \, ,\nonumber\\
L_2 &\ra b^4L_{-2} + 4ab^3 L_{-1} + 6a^2b^2 L_0 + 4a^3 b L_{1} + a^4 L_2\label{eq:GeneralAnecTrafo}\, ,
\end{align}
where all operators are functions of $x^+$ with $\vec{x}^\perp=0$. 

Things become more complicated if we move the light-ray operators to general $\vec{x}^\perp$, because the collinear tranformations were specifically chosen to preserve light-rays along $\vec{x}^\perp = 0$. At finite $\vec{x}^\perp$, the direction of the null line changes under general collinear transformations, such that $T_{--}$ mixes with other components of the stress tensor, as we saw in eq.~\eqref{eq:CollTransTmm}. At arbitrary $x$, we therefore obtain the general transformations
\begin{align}
\comm{J_{-1}}{L_n(x)} &= -(n+2) L_{n-1}(x), \nn\\
\comm{J_0}{L_n(x)} &= -\Big(n-\half \vec{x}^\perp \cdot \vec{\p}_\perp \Big) L_n(x), \label{eq:CollTransLn} \\
\comm{J_1}{L_n(x)} &= -\Big(n - 2 - \vec{x}^\perp \cdot \vec{\p}_\perp \Big) L_{n+1}(x) + |\vec{x}^\perp|^2 \p_+ L_n(x) + 2\int dx^- (x^-)^{n+2} \vec{x}^\perp \cdot \vec{T}_{-\perp}(x).\nn
\end{align}
Finally, we can consider the action of the generator $\bar{J}_0$, obtaining
\be 
\comm{\bar{J}_0}{L_n(x)} = \Big(\bar{h}_T + x^+ \p_+ + \half \vec{x}^\perp \cdot \vec{\p}_\perp \Big) L_n(x).
\ee
Unsurprisingly, every light-ray operator has the same collinear twist $\bar{h}_T=1$, since the raising and lowering operators commute with $\bar{J}_0$.

\section{Two- and three-point functions and the tale of integration}
\label{sec:2ptfct}

In this section, we would like to gain some intuition for the behaviour of these light-ray operators. In addition, we want to explain with the simplest examples how to perform the light-ray integral that converts $(x^-)^{n+2}T_{--}$ into a light-ray operator $L_n$. The easiest playgrounds to probe this question are the two- and three-point functions:
\benn
\<T_{--}(x_1) T_{--}(x_2)\>, \quad\qquad  \<\Ocal(x_1) T_{--}(x_2) \Ocal(x_3)\>,
\eenn
whose structure is fixed by conformal symmetry. In this fairly simple setup, we can understand how light-ray operators act on the vacuum and on scalar operators, as well as gather information about the candidate central charge that a possible algebra could have.

\subsection{Two-point functions of light-ray operators }

The stress-tensor two-point function in $d$ dimensions takes the following form~\cite{Osborn:1993cr}
\be
\<T_{\mu\nu}(x) T_{\rho\sigma}(0)\> = \fr{c_T}{x^{2d}} \left[ \half \Big( I_{\mu\rho}(x) I_{\nu\sigma}(x) + I_{\mu\sigma}(x) I_{\nu\rho}(x) \Big) - \fr{1}{d} \eta_{\mu\nu} \eta_{\rho\sigma} \right],
\ee
where we have defined
\be
I_{\mu\nu}(x) \equiv \eta_{\mu\nu} - 2\fr{x_\mu x_\nu}{x^2}.
\ee
To compute correlators involving light-ray operators, we are specifically interested in the component $T_{--}$ in $d=4$, which has the simple correlator
\be
\<T_{--}(x) T_{--}(0)\> = 4 c_T\fr{(x_-)^4}{x^{12}} = \fr{c_T}{4}\fr{(x^+)^4}{x^{12}}.
\ee

\subsubsection*{Two-point function involving global operators}

Using this two-point function, we can integrate over one of the positions, weighted by a function $f(x^-)$ which we first take to be holomorphic up to possible poles at infinity, which are outside the contour of integration due to the regularization procedure~\eqref{lightraycontour}. To ensure that the operators are properly ordered, we use the $i\epsilon$ prescription given in \rref{iepsprescription}, namely
\be 
x_{12}^\pm \rightarrow x_{12}^\pm -i\epsilon\, .
\ee 
We now want to compute 
\be
\ba
\<\Ecal_f(x_1) T_{--}(x_2)\> &\equiv \int_{-\infty}^\infty dx_1^- \, f(x_1^-) \<T_{--}(x_1) T_{--}(x_2)\> \\
&= \fr{c_T}{4} \int_{-\infty}^\infty dx_1^- \, f(x_1^-) \fr{(x_{12}^+)^4}{[-(x_{12}^+-i\epsilon)(x_{12}^- - i\epsilon) + |\vec{x}_{12}^\perp|^2]^6} \,.
\ea
\ee
The integral over $x_i^-$ can be evaluated using Cauchy's theorem (recall that the integral has been transformed into a contour integral by the limiting procedure described in section~\ref{sec:Conventions}). There is a sixth-order pole for $x_1^-$, located at
\be
x_1^- = x_2^- + \fr{|\vec{x}_{12}^\perp|^2}{x_{12}^+} + i\epsilon \equiv \x{2}{1} + i\epsilon \,,
\ee
from which we obtain 
\be
\<\Ecal_f(x_1) T_{--}(x_2)\> = (2\pi i) \fr{c_T}{20} \fr{f^{(5)}(\x{2}{1})}{(x_{12}^+)^2} \,. \label{eq:EfTmm}
\ee
If we instead integrate over the position $x_2^-$ of the second operator, there is no pole in the contour and we obtain
\begin{align}
\<T_{--}(x_1) \Ecal_f(x_2)\> = 0\,. \label{eq:TmmEcal}
\end{align}
The appearance of a fifth derivative in \eqref{eq:EfTmm} is rather important. From this, we immediately see that the five ``global'' operators that form a finite-dimensional representation of the collinear subalgebra annihilate the vacuum when acting both to the left and to the right
\be
\<L_n(x_1) T_{--}(x_2)\> = \<T_{--}(x_1) L_n(x_2)\> = 0\, ,\qquad (|n| \leq 2).
\ee
This observation follows directly from the fact that the functions
\be 
f(x^-) \in \lbrace (x^-)^0,\, (x^-)^1,\, (x^-)^2,\, (x^-)^3,\, (x^-)^4\rbrace \,,
\ee
have a vanishing fifth derivative. For these five functions, there is no pole at infinity in the two-point function, such that we could have closed the integration contour in either direction, without the need for our regularization procedure.

\subsubsection*{Two-point function involving non-global $L_{n}$}

We now turn to the remaining $L_n$ operators with $|n| > 2$. Let's start with the operators with higher powers of $x^-$, i.e.~$n \geq 3$. Since the only poles in this choice of $f(x^-)$ are at infinity, the procedure is exactly the same as for the global operators, and we find
\begin{align}
\braket{T_{--}(x_1)L_n(x_2)} &= 0\, , \label{eq:LmTmm0} \\
\braket{L_n(x_1)T_{--}(x_2)}&=  (n+2)(n+1)n(n-1)(n-2)\frac{i\pi c_T (\x{2}{1})^{n-3}}{240(x_{12}^+)^2}\, .\label{eq:LmTmm}
\end{align}
These $L_n$ therefore act like annihilation operators on the vacuum. Note that because the integrands contained poles at infinity, unlike the case for the global operators, it was necessary to implement our regularization procedure in evaluating these expressions.

For the operators $L_{-n}$ with $n\geq 3$, the story is slightly more involved. The integral that we need to perform is given by 
\be
\<T_{--}(x_1)L_{-n}(x_2)\> = \fr{c_T}{4} \int_{-\infty}^\infty dx_2^- \,  \fr{(x_{12}^+)^4}{(x_2^--i\epsilon)^{n-2}[-(x_{12}^+-i\epsilon)(x_{12}^- - i\epsilon) + |\vec{x}_{12}^\perp|^2]^6}\, .\label{eq:TmmLmn}
\ee
The new feature is that the integrand in \eqref{eq:TmmLmn} now has two poles, one of which is introduced by the function $f$ itself (see eq.~\eqref{eq:NegativeLnDef}). They are located at 
\be 
x_2^-\rightarrow i\epsilon\, , \qquad x_2^- \rightarrow \x{1}{2}-i\epsilon\, .
\ee 
One is in the upper half-plane while the other is in the lower half-plane, so only a single pole contributes to the contour integral. We obtain
\be
\<T_{--}(x_1)L_{-n}(x_2)\> = (n+2)(n+1)n(n-1)(n-2)\frac{i\pi c_T }{240(x_{12}^+)^2(\x{1}{2})^{n+3}}\, ,
\ee
which implies that $L_{-n}$ does not annihilate the vacuum when acting to the right. 

For the opposite ordering, both poles are now in the upper half plane and the integral thus vanishes
\be
\<L_{-n}(x_1)T_{--}(x_2)\> = 0.
\ee
The operators $L_{-n}$ for $n\geq 3$ therefore act as creation operators on the vacuum.

\subsection{What about the central charge?} 

We now consider the two-point function between two light-ray operators. From the discussion above, it is clear that the only non-vanishing two-point function is 
\be 
\braket{L_m(x_1)L_{n}(x_2)},
\ee
with $m> 2$ and $n<-2$. For notational simplicity, we can relabel $n\rightarrow -n$ such that both $m >2$ and $n>2$. The integral to perform is then
\be
\<L_m(x_1) L_{-n}(x_2)\> = \fr{c_T}{4} \int_{-\infty}^\infty dx_1^-dx_2^- \, \frac{(x_1^--2i\epsilon)^{m+2}}{(x_2^--i\epsilon)^{n-2} }\fr{(x_{12}^+-i\epsilon)^4}{[-(x_{12}^+-i\epsilon)(x_{12}^- - i\epsilon) + |\vec{x}_{12}^\perp|^2]^6}.
\ee
The easiest approach is to evaluate the $x_1^-$ integral first to obtain~\eqref{eq:LmTmm} and then perform the $x_2^-$ integral. We find
\begin{align}
\<L_m(x_1) L_{-n}(x_2)\> 
 &=(2\pi i)^2 \frac{c_T \Gamma(n-m)}{480\Gamma(n-2)\Gamma(-m-2)}\frac{(-1)^{n}|\vec{x}_{12}^\perp|^{2(m-n)}}{(x_{12}^+-i\epsilon)^{m-n+2} }\, ,
\end{align}
which is  symmetric under the simultaneous exchange $m\leftrightarrow -n$ and $x_1\leftrightarrow x_2$, and vanishes when $m \leq 2$ or $n\leq2$. Notice that this implies that these operators are not an orthogonal basis for the algebra.

This correlator is also the commutator, because the reversed ordering vanishes due to~\eqref{eq:LmTmm0}. We can thus look at the case where $m=n$ to probe a possible central charge in the algebra,
\begin{align}
\braket{[L_m(x_1),L_{-m}(x_2)]} 
&= -\frac{(m+2)(m+1)m(m-1)(m-2)\pi^2 }{120(x_{12}^+-i\epsilon)^2}c_T  \, .\label{eq:centralcharge}
\end{align} 

To probe the algebra, one would specifically want to consider the case $x_1^+ = x_2^+$. However, \eqref{eq:centralcharge} diverges once we further take the $\epsilon \rightarrow 0 $ limit. At face value, this would suggest that the central charge of a putative algebra is infinity, which matches the observations made in \cite{Wall:2011hj,Casini:2017roe}. An important remark is that this divergent central term is not strictly speaking the one that one would guess by the form of the proposed Witt algebra discussed in \cite{Casini:2017roe}. That term is forbidden by the collinear conformal group. Note however that a term of the form~\eqref{eq:centralcharge} has appeared before in \cite{Huang:2020ycs}.

One may try to regularize the operators to obtain a finite answer, but we will not attempt do so here (see \cite{Huang:2020ycs} for a discussion of such an approach). We come back to this issue briefly in the discussion, see section \ref{sec:conclu}.

\subsection{Three-point functions}\label{sec:3ptF}

Three-point functions are also fixed by conformal symmetry (up to OPE coefficients), so we can compute three-point functions involving light-ray operators without specifying a CFT (unlike four-point functions, which will be the focus of the subsequent sections).

In four dimensions, the three-point function between the stress-energy tensor $T_{--}$ and a general scalar operator $\Ocal$ takes the form \cite{Osborn:1993cr}
\be
\<\Ocal(x_1) T_{--}(x_2) \Ocal(x_3)\> = - \fr{\De}{6\pi^2} \fr{1}{x_{13}^{2\De-2}} \left( \fr{(x_{12}^+)^2}{x_{12}^6 x_{23}^2} + 2 \fr{x_{12}^+ x_{23}^+}{x_{12}^4 x_{23}^4} + \fr{(x_{23}^+)^2}{x_{12}^2 x_{23}^6} \right)\,,
\label{eq:OOTmm}
\ee
where we have normalized the operator $\Ocal$ such that
\be \<\Ocal(x) \Ocal(0)\> = \fr{1}{x^{2\De}}\, . \label{twoptnorm}
\ee
We can now integrate~\eqref{eq:OOTmm} to obtain three-point functions involving light-ray operators.

\subsubsection{$\<\Ocal \Ecal_f \Ocal\>$}

Let's first consider a general function $f(x^-)$ to obtain the generalized ANEC correlator
\be
\<\Ocal(x_1) \Ecal_f(x_2) \Ocal(x_3)\> \equiv \int_{-\infty}^\infty dx_2^- f(x_2^-) \<\Ocal(x_1) T_{--}(x_2) \Ocal(x_3)\>,
\ee
where we assume for now that the inclusion of the function $f$ does not introduce new singularities in the integrand. There are therefore two poles in the $x_2^-$ plane:
\be
x_2^- = \x{1}{2} - i\epsilon, \qquad x_2^- = \x{3}{2} + i\epsilon.
\ee

Using the integrals from appendix~\ref{ap:UsefulContourIntegrals3pt}, we can then evaluate this expression, obtaining either the residue of the pole in the lower half-plane,
\be
\<\Ocal(x_1) \Ecal_f(x_2) \Ocal(x_3)\> = - \fr{i\De}{6\pi} \fr{1}{x_{12}^+ x_{23}^+ x_{13}^{2\De-2}} \left[ \fr{f''(\x{1}{2})}{(\x{1}{2} - \x{3}{2})} - 6 \fr{f'(\x{1}{2})}{(\x{1}{2} - \x{3}{2})^2} + 12 \fr{f(\x{1}{2})}{(\x{1}{2} - \x{3}{2})^3} \right].
\label{eq:OEfO_left}
\ee
or the residue of the pole in the upper half-plane,
\be
\<\Ocal(x_1) \Ecal_f(x_2) \Ocal(x_3)\> = - \fr{i\De}{6\pi} \fr{1}{x_{12}^+ x_{23}^+ x_{13}^{2\De-2}} \left[ \fr{f''(\x{3}{2})}{(\x{1}{2} - \x{3}{2})} + 6 \fr{f'(\x{3}{2})}{(\x{1}{2} - \x{3}{2})^2} + 12 \fr{f(\x{3}{2})}{(\x{1}{2} - \x{3}{2})^3} \right].
\label{eq:OEfO_right}
\ee
Note that these two expressions are equivalent \emph{only} for functions $f$ which introduce no additional singularities. As we will now discuss, for our set of functions $f(\xm)=(\xm)^{n+2}$ this specifically corresponds to the case $-2\leq n \leq 2$, which are the five global operators.

\subsubsection{$\<\Ocal L_n \Ocal\>$}

Let's now focus on the operators $L_n$ with $n \geq 3$. These functions introduce a pole at infinity, in which case we must follow our regularization procedure~\eqref{lightraycontour} and close in the upper half-plane. We can then use eq.~\rref{eq:OEfO_right} for the case $f(x^-) = (x^-)^{n+2}$, obtaining
\begin{align}
\<\Ocal(x_1) L_{n}(x_2) \Ocal(x_3)\> &= - \fr{i\De}{6\pi} \fr{(\x{3}{2})^{n}}{x_{12}^+ x_{23}^+ (\x{1}{2} - \x{3}{2})^3 x_{13}^{2\De-2} } \Big[ (n+1)(n+2) (\x{1}{2})^2 \nonumber\\
&\phantom{=}- 2(n+2)(n-2)\x{1}{2} \x{3}{2} + (n-1)(n-2) (\x{3}{2})^2 \Big].
\label{eq:OLnO_right}
\end{align}

Next, we can consider $L_{-n}$ with $n \geq 3$. With our $i\epsilon$ prescription, these functions introduce a new pole at $x_2^- = i\epsilon$. We therefore must include the contribution of this pole when closing in the upper half-plane. Equivalently, we can evaluate the contour in the lower half-plane, as this function introduces no pole at infinity. Either way, we obtain the resulting expression
\begin{align}
\<\Ocal(x_1) L_{-n}(x_2) \Ocal(x_3)\>  &= - \fr{i\De}{6\pi} \fr{(\x{1}{2})^{-n}}{x_{12}^+ x_{23}^+ (\x{1}{2} - \x{3}{2})^3 x_{13}^{2\De-2} } \Big[ (n+1)(n+2)(\x{1}{2})^2\nonumber\\
&\phantom{=} - 2(n+2)(n-2)\x{1}{2}\x{3}{2} + (n-1)(n-2)(\x{3}{2})^2 \Big].
\label{eq:OLnO_left}
\end{align}

Finally, we have the global operators $L_n$ with $|n| \leq 2$. These operators do not introduce either a pole at zero or a pole at infinity for these three-point functions, in which case we can safely close the contour in either direction. Indeed, one can explicitly check that \eqref{eq:OLnO_right} and \eqref{eq:OLnO_left} agree for $L_n$ with
\be
n=\lbrace -2,\,-1,\, 0,\, 1,\, 2\, \rbrace.
\ee
For reference, let us explicitly write out the resulting expressions for these five special operators in scalar three-point functions:
\be\label{eq:ThreePtFct}
\ba
\<\Ocal(x_1) L_{-2}(x_2) \Ocal(x_3)\> &= - \fr{2i\De}{\pi} \fr{1}{x_{12}^+ x_{23}^+ (\x{1}{2} - \x{3}{2})^3 x_{13}^{2\De-2}}, \\
\<\Ocal(x_1) L_{-1}(x_2) \Ocal(x_3)\> &= - \fr{i\De}{\pi} \fr{\x{1}{2} + \x{3}{2}}{x_{12}^+ x_{23}^+ (\x{1}{2} - \x{3}{2})^3 x_{13}^{2\De-2} }, \\
\<\Ocal(x_1) L_{0}(x_2) \Ocal(x_3)\> &= - \fr{i\De}{3\pi} \fr{(\x{1}{2})^2 + 4 \x{1}{2} \x{3}{2} + (\x{3}{2})^2}{x_{12}^+ x_{23}^+ (\x{1}{2} - \x{3}{2})^3 x_{13}^{2\De-2} }, \\
\<\Ocal(x_1) L_1(x_2) \Ocal(x_3)\> &= - \fr{i\De}{\pi} \fr{\x{1}{2} \x{3}{2}(\x{1}{2}+\x{3}{2})}{x_{12}^+ x_{23}^+ (\x{1}{2} - \x{3}{2})^3 x_{13}^{2\De-2} }, \\
\<\Ocal(x_1) L_2(x_2) \Ocal(x_3)\> &= - \fr{2i\De}{\pi} \fr{(\x{1}{2})^2 (\x{3}{2})^2}{x_{12}^+ x_{23}^+ (\x{1}{2} - \x{3}{2})^3 x_{13}^{2\De-2} }.
\ea
\ee
Because these operators form a finite-dimensional representation of the collinear subgroup, their correlation functions are related by the action of the generators $J_i$ in~\eqref{eq:CollTransLn}.

%
%

For the rest of this work, we will leave the operators with $|n| > 2$ aside and only focus on the five global operators. From now on, when we write a function $f$ we will therefore implicitly mean that the function is $f(\xm)=(\xm)^{n+2}$ for $-2 \leq n \leq 2$.

\section{Four-point functions in free field theory}

So far, we have only considered properties of light-ray operators which are fixed by conformal symmetry. In the coming sections, we will now turn to explicit computations of four-point functions involving two generalized ANEC operators. Since such correlators are highly theory-dependent, we must specify the CFT in which we want to compute them. We will start in free field theory, before moving to holographic CFTs in section \ref{sec:confblock}. 

In $d=4$, a free field $\phi$ has dimension $\Delta =1$, whose two-point function we will normalize as in \rref{twoptnorm}.\footnote{While this normalization is natural from a CFT perspective, it differs from the usual convention for a free scalar by a factor of $\fr{1}{4\pi^2}$.} The associated Wightman functions can all be constructed via Wick contraction, and involve derivatives of two-point functions of the field $\phi$. It will therefore be useful to consider the building block two-point function
\be
\<\p_-^m\phi(x) \p_-^n\phi(y)\> = \fr{(-1)^n(m+n)!(x^+ - y^+)^{m+n}}{(x-y)^{2(m+n+1)}}.
\ee
In free-field theory, the stress-energy tensor is given by
\be
T_{\mu\nu}(x) = \fr{1}{6\pi^2} \p_\mu\phi(x) \p_\nu\phi (x)- \fr{1}{24\pi^2} \eta_{\mu\nu} \p_\sigma\phi(x) \p^\sigma \phi(x) - \fr{1}{12\pi^2} \phi(x) \p_\mu \p_\nu \phi(x).
\ee
We will mainly be interested in the component $T_{--}$,
\be
T_{--}(x)= \fr{1}{6\pi^2}\left[ (\p_-\phi(x))^2 - \fr{1}{2} \phi(x) \p_-^2 \phi(x)\right],\label{eq:Tmmfreefield}
\ee
which has been normalized according to the three-point function~\eqref{eq:OOTmm}. We are interested in computing the following correlation functions 
\begin{align} \label{fg4pt}
\braket{\phi(x_1)\mathcal{E}_f(x_2)\mathcal{E}_g(x_3)\phi(x_4)} &= \int dx_2^- dx_3^- \, f(x_2^-)\, g(x_3^-)\, \<\phi(x_1)T_{--}(x_2)
T_{--}(x_3)\phi(x_4)\> \,,
\end{align}
for
\be 
f(x^-),g(x^-) \in \lbrace (x^-)^0, (x^-)^1, (x^-)^2, (x^-)^3, (x^-)^4\rbrace\, ,\label{eq:allowedf}
\ee
corresponding to the five global operators $\lbrace L_{-2}, L_{-1}, L_0, L_1, L_2\rbrace$. Note that for any correlation function in this class, the integrand never has poles at zero or infinity. We can therefore safely close each of the integration contours in either the lower or upper half-plane, without worrying about the regularization procedure \rref{lightraycontour}. Moreover, the order in which the integrals are performed also does not affect the resulting expressions.

\begin{figure}[t!]
\begin{tikzpicture}[x=0.75pt,y=0.75pt,yscale=-1,xscale=1]

\draw    (90.46,189.27) -- (60.4,241.1) ;
\draw [shift={(60.4,241.1)}, rotate = 120.11] [color={rgb, 255:red, 0; green, 0; blue, 0 }  ][fill={rgb, 255:red, 0; green, 0; blue, 0 }  ][line width=0.75]      (0, 0) circle [x radius= 3.35, y radius= 3.35]   ;
\draw [shift={(90.46,189.27)}, rotate = 120.11] [color={rgb, 255:red, 0; green, 0; blue, 0 }  ][fill={rgb, 255:red, 0; green, 0; blue, 0 }  ][line width=0.75]      (0, 0) circle [x radius= 3.35, y radius= 3.35]   ;
\draw    (90.46,189.27) -- (169.46,189.27) ;
\draw    (169.46,189.27) -- (199.63,240.79) ;
\draw [shift={(199.63,240.79)}, rotate = 59.64] [color={rgb, 255:red, 0; green, 0; blue, 0 }  ][fill={rgb, 255:red, 0; green, 0; blue, 0 }  ][line width=0.75]      (0, 0) circle [x radius= 3.35, y radius= 3.35]   ;
\draw [shift={(169.46,189.27)}, rotate = 59.64] [color={rgb, 255:red, 0; green, 0; blue, 0 }  ][fill={rgb, 255:red, 0; green, 0; blue, 0 }  ][line width=0.75]      (0, 0) circle [x radius= 3.35, y radius= 3.35]   ;
\draw    (369.46,189.27) -- (260.4,241.1) ;
\draw [shift={(260.4,241.1)}, rotate = 154.58] [color={rgb, 255:red, 0; green, 0; blue, 0 }  ][fill={rgb, 255:red, 0; green, 0; blue, 0 }  ][line width=0.75]      (0, 0) circle [x radius= 3.35, y radius= 3.35]   ;
\draw [shift={(369.46,189.27)}, rotate = 154.58] [color={rgb, 255:red, 0; green, 0; blue, 0 }  ][fill={rgb, 255:red, 0; green, 0; blue, 0 }  ][line width=0.75]      (0, 0) circle [x radius= 3.35, y radius= 3.35]   ;
\draw    (290.46,189.27) -- (369.46,189.27) ;
\draw    (290.46,189.27) -- (400.4,240.1) ;
\draw [shift={(400.4,240.1)}, rotate = 24.81] [color={rgb, 255:red, 0; green, 0; blue, 0 }  ][fill={rgb, 255:red, 0; green, 0; blue, 0 }  ][line width=0.75]      (0, 0) circle [x radius= 3.35, y radius= 3.35]   ;
\draw [shift={(290.46,189.27)}, rotate = 24.81] [color={rgb, 255:red, 0; green, 0; blue, 0 }  ][fill={rgb, 255:red, 0; green, 0; blue, 0 }  ][line width=0.75]      (0, 0) circle [x radius= 3.35, y radius= 3.35]   ;
\draw    (599.4,240.1) -- (459.4,241.1) ;
\draw [shift={(459.4,241.1)}, rotate = 179.59] [color={rgb, 255:red, 0; green, 0; blue, 0 }  ][fill={rgb, 255:red, 0; green, 0; blue, 0 }  ][line width=0.75]      (0, 0) circle [x radius= 3.35, y radius= 3.35]   ;
\draw [shift={(599.4,240.1)}, rotate = 179.59] [color={rgb, 255:red, 0; green, 0; blue, 0 }  ][fill={rgb, 255:red, 0; green, 0; blue, 0 }  ][line width=0.75]      (0, 0) circle [x radius= 3.35, y radius= 3.35]   ;
\draw   (489.46,189.6) .. controls (489.46,186.01) and (507.14,183.1) .. (528.96,183.1) .. controls (550.77,183.1) and (568.46,186.01) .. (568.46,189.6) .. controls (568.46,193.19) and (550.77,196.1) .. (528.96,196.1) .. controls (507.14,196.1) and (489.46,193.19) .. (489.46,189.6) -- cycle ;
\draw  [fill={rgb, 255:red, 0; green, 0; blue, 0 }  ,fill opacity=1 ] (492.48,189.6) .. controls (492.48,187.93) and (491.12,186.58) .. (489.46,186.58) .. controls (487.79,186.58) and (486.44,187.93) .. (486.44,189.6) .. controls (486.44,191.27) and (487.79,192.62) .. (489.46,192.62) .. controls (491.12,192.62) and (492.48,191.27) .. (492.48,189.6) -- cycle ;
\draw  [fill={rgb, 255:red, 0; green, 0; blue, 0 }  ,fill opacity=1 ] (571.48,189.6) .. controls (571.48,187.93) and (570.12,186.58) .. (568.46,186.58) .. controls (566.79,186.58) and (565.44,187.93) .. (565.44,189.6) .. controls (565.44,191.27) and (566.79,192.62) .. (568.46,192.62) .. controls (570.12,192.62) and (571.48,191.27) .. (571.48,189.6) -- cycle ;

\draw (40,250.5) node [anchor=north west][inner sep=0.75pt]    {$\phi ( x_{1})$};
\draw (240,250.5) node [anchor=north west][inner sep=0.75pt]    {$\phi ( x_{1})$};
\draw (180,250.5) node [anchor=north west][inner sep=0.75pt]    {$\phi ( x_{4})$};
\draw (380,250.5) node [anchor=north west][inner sep=0.75pt]    {$\phi ( x_{4})$};
\draw (579,250.5) node [anchor=north west][inner sep=0.75pt]    {$\phi ( x_{4})$};
\draw (439,250.5) node [anchor=north west][inner sep=0.75pt]    {$\phi ( x_{1})$};
\draw (70,160.5) node [anchor=north west][inner sep=0.75pt]    {$T( x_{2})$};
\draw (271,160.5) node [anchor=north west][inner sep=0.75pt]    {$T( x_{2})$};
\draw (470,160.5) node [anchor=north west][inner sep=0.75pt]    {$T( x_{2})$};
\draw (150,160.5) node [anchor=north west][inner sep=0.75pt]    {$T( x_{3})$};
\draw (351,160.5) node [anchor=north west][inner sep=0.75pt]    {$T( x_{3})$};
\draw (550,160.5) node [anchor=north west][inner sep=0.75pt]    {$T( x_{3})$};
\draw (225,204.1) node [anchor=north west][inner sep=0.75pt]    {$+$};
\draw (427,204.1) node [anchor=north west][inner sep=0.75pt]    {$+$};

\end{tikzpicture}
\caption{The three different topologies of Wick contraction that appear in $\braket{\phi(x_1)T_{--}(x_2)T_{--}(x_3)\phi(x_4)}$, where we wrote $T(x_2)\equiv T_{--}(x_2)$. The lines indicates how the fields are contracted since $T_{--}(x)$ is bilinear in $\phi(x)$. Only the first topology contributes to the integrated correlators for the contours at hand.}
\label{figFeynman}
\end{figure}
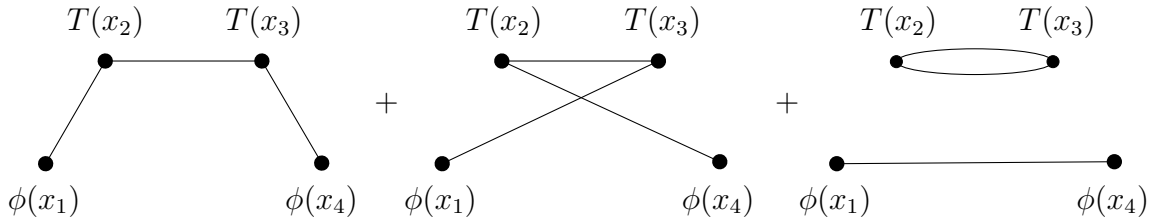

To compute \rref{fg4pt}, we first need the correlator $\braket{\phi T_{--}T_{--}\phi}$. Using \eqref{eq:Tmmfreefield}, it is a straightforward Wick contraction exercise to compute the full correlator, and the full expression is shown in appendix \ref{ap:UsefulContourIntegrals4pt}. However, it turns out that not all Wick contractions are relevant for computing the correlation functions of light-ray operators, since several of them will vanish once the integrals are evaluated. The simplest way to see this is to close both contours outwards, which we can represent as 
\begin{align}
\braket{\phi(x_1)\mathcal{E}_f(x_2)\mathcal{E}_g(x_3)\phi(x_4)} &= \int dx_2^- dx_3^- f(x_2^-)g(x_3^-)\contraction{\<\phi(}{x_1}{)T_{--}}{(x_2)}
\contraction{\<\phi(x_1)T_{--}(x_2)T_{--}(}{x_3}{)\phi}{(x_4)}
\<\phi(x_1)T_{--}(x_2)T_{--}(x_3)\phi(x_4)\>\nn,
\end{align}
where this notation means that we integrate $x_2^-$ by picking up the singularity when $x_2\rightarrow x_1$, and integrate $x_3^-$ by picking up the singularity when $x_3 \rightarrow x_4$. Because of this, it is clear that the only terms that will contribute to the final result are terms in $\braket{\phi T_{--}T_{--}\phi}$ that have a denominator of the form $x_{12}^a x_{34}^b$ for positive $a$ and $b$. All other terms will vanish once we integrate. The different contractions of this four-point function lead to three possible topologies, shown schematically in figure~\ref{figFeynman}. Based on the preceding argument, it is clear that only the first topology contributes upon integration. 

To compute a four-point function with light-ray operators, we can therefore concentrate solely on the subset of the full correlator given by the Wick contractions appearing in the first topology. This is explicitly given by
\be
\ba
&\<\phi(x_1) T_{--}(x_2) T_{--}(x_3) \phi(x_4)\> \\
& \qquad \supset \fr{1}{x_{12}^2 x_{23}^{10} x_{34}^2} \bigg( \fr{1}{36\pi^4} (x_{23}^+)^4 + \fr{1}{9\pi^4} x_{23}^4 (x_{23}^+)^2  \left( \fr{x_{12}^+}{x_{12}^2} + \fr{x_{23}^+}{x_{23}^2} \right) \left( \fr{x_{23}^+}{x_{23}^2} + \fr{x_{34}^+}{x_{34}^2} \right) \\
& \qquad \qquad \qquad \qquad \quad + \, \fr{1}{36\pi^4} x_{23}^8 \left( \fr{x_{12}^+}{x_{12}^2} + \fr{x_{23}^+}{x_{23}^2} \right)^2 \left( \fr{x_{23}^+}{x_{23}^2} + \fr{x_{34}^+}{x_{34}^2} \right)^2 \bigg).
\ea
\label{eq:PhiTTPhi}
\ee
We are now ready to perform the integrals and compute the four-point function $\braket{\phi \Ecal_f \Ecal_g\phi}$.

\subsection{The four-point function $\<\phi \Ecal_f \Ecal_g \phi\>$ and the commutator}
In this section, we compute both the four-point correlator with two distinct light-ray operators $\mathcal{E}_f$ and $\mathcal{E}_g$, as well as the resulting commutator. The relevant part of the local four-point function is given by \rref{eq:PhiTTPhi} and we now need to integrate it, picking up poles as $x_2\to x_1$ and $x_3\to x_4$.

The very first term of \eqref{eq:PhiTTPhi} has a single pole in both $x_2^-$ and $x_3^-$. The next term has both a single pole and a double pole for both coordinates, while the last term has poles up to third order. These integrals are displayed in full glory in appendix \ref{ap:UsefulContourIntegrals4pt} (cf eqs.~\eqref{eq:4ptint1} to \eqref{eq:4ptint3}). Combining the results of all the integrals together, we then obtain the full light-ray correlator,
\begin{align}
\label{eq:phiEEphi}
&\<\phi(x_1) \Ecal_f(x_2) \Ecal_g(x_3) \phi(x_4)\> \\
& \qquad = (2\pi i)^2 \Bigg( \fr{6}{\pi^4} \fr{f(\x{1}{2}) g(\x{4}{3}) \, (x_{23}^+-i\epsilon)^4}{(x_{12}^+) (x_{34}^+) [-(x_{23}^+-i\epsilon)(\x{1}{2} - \x{4}{3}) + |\vec{x}_{23}^\perp|^2]^5} \nn\\
& \qquad \qquad \qquad + \, \fr{3}{2\pi^4} \fr{\Big( f'(\x{1}{2}) g(\x{4}{3}) - f(\x{1}{2}) g'(\x{4}{3}) \Big) (x_{23}^+-i\epsilon)^3}{(x_{12}^+) (x_{34}^+) [-(x_{23}^+-i\epsilon)(\x{1}{2} - \x{4}{3}) + |\vec{x}_{23}^\perp|^2]^4}\nn \\
& \qquad \qquad \qquad - \, \fr{1}{2\pi^4} \fr{f'(\x{1}{2}) g'(\x{4}{3}) \, (x_{23}^+-i\epsilon)^2}{(x_{12}^+) (x_{34}^+) [-(x_{23}^+-i\epsilon)(\x{1}{2} - \x{4}{3}) + |\vec{x}_{23}^\perp|^2]^3}\nn \\
& \qquad \qquad \qquad + \, \fr{1}{12\pi^4} \fr{\Big( f''(\x{1}{2}) g(\x{4}{3}) + f(\x{1}{2}) g''(\x{4}{3}) \Big) (x_{23}^+-i\epsilon)^2}{(x_{12}^+) (x_{34}^+) [-(x_{23}^+-i\epsilon)(\x{1}{2} - \x{4}{3}) + |\vec{x}_{23}^\perp|^2]^3}\nn \\
& \qquad \qquad \qquad - \, \fr{1}{24\pi^4} \fr{\Big( f''(\x{1}{2}) g'(\x{4}{3}) - f'(\x{1}{2}) g''(\x{4}{3}) \Big) (x_{23}^+-i\epsilon)}{(x_{12}^+) (x_{34}^+) [-(x_{23}^+-i\epsilon)(\x{1}{2} - \x{4}{3}) + |\vec{x}_{23}^\perp|^2]^2}\nn \\
& \qquad \qquad \qquad + \, \fr{1}{144\pi^4} \fr{f''(\x{1}{2}) g''(\x{4}{3})}{(x_{12}^+) (x_{34}^+) [-(x_{23}^+-i\epsilon)(\x{1}{2} - \x{4}{3}) + |\vec{x}_{23}^\perp|^2]} \Bigg).\nn
\end{align}
\enlargethispage{\baselineskip}In expression \eqref{eq:phiEEphi} we have anticipated which factors of $i\epsilon$ will be necessary to compute the commutator and have suppressed all factors of $i\epsilon$ that are irrelevant.\footnote{To reinstate all the $i\epsilon$ where needed, one just needs to perform the following replacements in \eqref{eq:phiEEphi}: $\x{1}{2}\rightarrow \x{1}{2}-i\epsilon$, $\x{4}{3}\rightarrow \x{4}{3}+ i\epsilon$, $x_{12}^+\rightarrow x_{12}^+ -i\epsilon$, and $x_{34}^+\rightarrow x_{34}^+-i\epsilon$.}
To compute the correlator with the opposite ordering for the light-ray operators, we simply take \rref{eq:phiEEphi} and replace $f\leftrightarrow g$ and $x_2 \leftrightarrow x_3$. We can then take the difference of these two expressions to obtain the commutator.

Because we are interested in the case where both light-ray operators are on the same null-plane, we need to be particularly careful with the $i\epsilon$ prescription between $x_2$ and $x_3$ when we subtract correlators of the form \eqref{eq:phiEEphi}. To simplify the analysis, we will study the various terms independently, starting with the term that has no derivatives acting on the functions $f(x^-)$ and $g(x^-)$. We denote this specific term as 
\begin{align}
\left.\<\phi \comm{\Ecal_f(x_2)}{\Ecal_g(x_3)} \phi\>\right|_{fg} 
&\sim \fr{f(\x{1}{2}) g(\x{4}{3} ) \, (x_{23}^+-i\epsilon)^4}{(x_{12}^+) (x_{34}^+) [-(x_{23}^+-i\epsilon)(\x{1}{2} - \x{4}{3}) + |\vec{x}_{23}^\perp|^2]^5} \notag \\ 
&-\fr{f(\x{4}{2}) g(\x{1}{3} ) \, (x_{23}^++i\epsilon)^4}{(x_{13}^+) (x_{24}^+) [(x_{23}^++i\epsilon)(\x{1}{3} - \x{4}{2}) + |\vec{x}_{23}^\perp|^2]^5} 
\end{align}
If we did not include the correct $i\epsilon$ prescription, then the numerator would naively vanish for $x_{23}^+=0$ and one would conclude that this contribution is zero. Instead, with the correct $i\epsilon$ prescription, we obtain the following expression in the limit where $x_2^+=x_3^+$,
\be
\ba
 \left.\<\phi \comm{\Ecal_f(x_2)}{\Ecal_g(x_3)} \phi\>\right|_{fg}
& \ra \fr{(i\epsilon)^4}{x_{12}^+ x_{24}^+} \left( \fr{f(\x{1}{2}) g(\x{4}{3})}{[|\vec{x}_{23}^\perp|^2 + i\epsilon(\x{1}{2} - \x{4}{3})]^5} - \fr{f(\x{4}{2}) g(\x{1}{3})}{[|\vec{x}_{23}^\perp|^2 + i\epsilon(\x{1}{3} - \x{4}{2})]^5} \right).\label{eq:fgcom1}
\ea
\ee
Looking at this expression, we see that it vanishes in the limit $\epsilon \ra 0$ provided $|\vec{x}_{23}^\perp|^2 \neq 0$. However, if $|\vec{x}_{23}^\perp|^2 = 0$, then this expression actually diverges as $\epsilon \ra 0$. It is hence clear that this expression should be thought of as a distribution proportional to a delta function in the transverse separation between the two light-ray operators: $\delta^{(2)}(\vec{x}_{23}^\perp)$. To extract this delta function contribution, we need the following relation
\be 
\lim_{\epsilon\rightarrow 0} \frac{(i\epsilon)^{a-1}}{\left[|\vec{x}^\perp|^2+ i \epsilon y\right]^a } = \frac{\pi}{(a-1)y^{a-1}}\delta^{(2)}(\vec{x}^\perp)\, , 
\label{eq:deltafctrel1}
\ee
which is valid provided $a\geq 2$. The derivation of this result is presented in appendix \ref{ap:deltafunction}.\footnote{See appendix \ref{app:deltaperp} for a different way of extracting the delta function, by integrating over the transverse coordinates.}  
Using \eqref{eq:deltafctrel1}, the term \eqref{eq:fgcom1} hence becomes
\begin{align}
 \left.\<\phi \comm{\Ecal_f(x_2)}{\Ecal_g(x_3)} \phi\>\right|_{fg}
&\sim \fr{\pi \de^{(2)}(\vec{x}_{23}^\perp)}{4x_{12}^+ x_{24}^+ (\x{1}{2} - \x{4}{2})^4}  \Big(f(\x{1}{2}) g(\x{4}{2}) - f(\x{4}{2}) g(\x{1}{2})\Big).
\end{align}
Based on this analysis, we can now compute the general commutator contribution, which is valid for $j\leq 3$ and $0\leq k \leq j$.
\begin{align}
&\fr{(x_{23}^+)^{4-j} f^{(k)}(\x{1}{2}) g^{(j-k)}(\x{4}{3})}{x_{12}^+ x_{34}^+ [-x_{23}^+(\x{1}{2} - \x{4}{3}) + |\vec{x}_{23}^\perp|^2]^{5-j}} - \fr{(-x_{23}^+)^{4-j} f^{(j-k)}(\x{4}{2}) g^{(k)}(\x{1}{3})}{x_{13}^+ x_{24}^+ [x_{23}^+(\x{1}{3} - \x{4}{2}) + |\vec{x}_{23}^\perp|^2]^{5-j}} \\
& \quad = \fr{ (-1)^{j}\pi\de^{(2)}(\vec{x}_{23}^\perp)}{(4-j)x_{12}^+ x_{24}^+(\x{1}{2} - \x{4}{2})^{4-j}}  \Big(f^{(k)}(\x{1}{2}) g^{(j-k)}(\x{4}{2}) - f^{(j-k)}(\x{4}{2}) g^{(k)}(\x{1}{2}) \Big) \, .\nn
\end{align} 
Note that the commutator only has support when $|\vec{x}_{23}^\perp|=0$, except for when $j=4$, which has a finite contribution when $|\vec{x}_{23}^\perp|\neq0$, and is much more subtle, as we will discuss below. 

We can now present the full commutator, which is given by 
\begin{align}
&\<\phi(x_1) \comm{\Ecal_f(x_2)}{\Ecal_g(x_3)} \phi(x_4)\>\label{eq:PhiComm} \\
& \quad = \fr{1}{\pi x_{12}^+ x_{24}^+} \de^{(2)}(\vec{x}_{23}^\perp) \Bigg[ -6\fr{\f{1}{2}\,\g{4}{2} - \f{4}{2}\,\g{1}{2}}{(\x{1}{2} - \x{4}{2})^4} + \, 2  \fr{\left(\f{1}{2}'\,\g{4}{2}- \f{4}{2}\,\g{1}{2}'\right)-\left(\f{1}{2}\,\g{4}{2}' -\f{4}{2}'\,\g{1}{2} \right)}{(\x{1}{2} - \x{4}{2})^3}\nn\\
& \qquad \quad   - \, \fr{1}{6} \fr{\left(\f{1}{2}'' \,\g{4}{2} - \f{4}{2}  \g{1}{2}''\,\right)-6\left(\f{1}{2}'\, \g{4}{2}' - \f{4}{2}'\, \g{1}{2}'\right)+\left(\f{1}{2}\, \g{4}{2}'' -  \f{4}{2}'' \,\g{1}{2}\right)}{(\x{1}{2} - \x{4}{2})^2} \nn \\
& \qquad \quad  - \, \fr{1}{6} \fr{\left(\f{1}{2}''\, \g{4}{2}' - \f{4}{2}'\, \g{1}{2}''\right)-\left(\f{1}{2}'\, \g{4}{2}'' - \f{4}{2}'' \,\g{1}{2}'\right)}{\x{1}{2} - \x{4}{2}}  \Bigg] \nn \\
& \qquad \quad  - \, \fr{1}{36\pi^2x_{12}^+x_{24}^+}\left[ \fr{\f{1}{2}''\,\g{4}{3}''}{ [-(x_{23}^+-i\epsilon)(\x{1}{2} - \x{4}{3}) + |\vec{x}_{23}^\perp|^2]}-\fr{\f{4}{2}'' \,\g{1}{3}''}{[(x_{23}^++i\epsilon)(\x{1}{3}- \x{4}{2}) + |\vec{x}_{23}^\perp|^2]}\right]\nn \,,
\end{align}
where we have defined the shorthand notation,
\be 
\f{i}{j}\equiv f(\x{i}{j})\, ,\qquad \g{i}{j} \equiv g(\x{i}{j})\,.
\ee
In \eqref{eq:PhiComm}, the two terms displayed on the last line which are proportional to $f'' g''$, do not admit a representation as a distribution proportional to a delta function. They correspond to \eqref{eq:deltafctrel1} with $a=1$, whose numerator is finite. In appendix \ref{ap:deltafunction}, we demonstrate that this contribution is not integrable and explain why we cannot write a meaningful transverse function $\delta^{(2)}(\vec{x}_{23}^\perp)$. Nevertheless, these two terms admit a well-defined limit when $\epsilon \rightarrow 0$, provided $|\vec{x}_{23}^\perp|^2 \neq 0$.

Before analyzing this finite-separation contribution, let us first focus our attention on computations where the $f''g''$ term does not contribute. This is the case for all computations where $\Ecal_f$ or $\Ecal_g$ are one of the three operators $\lbrace L_{-2},\, L_{-1},\, L_0\rbrace$. This subset of operators is special in the sense that they are the local versions of some of the charge operators of the conformal algebra, and are thus constrained by the associated Ward identities \cite{Cordova:2018ygx}.


\subsection{The algebra of light-ray operators}
\label{sec:FreeAlgebra}

We have now obtained the general commutator \eqref{eq:PhiComm} valid for the operators:
\be
\Ecal_f, \Ecal_g \in \{L_{-2},\,  L_{-1},\,  L_0,\,  L_1,\, L_2\}.
\ee
In addition, we have seen that there is a subtlety involving a finite piece at spacelike separation when $f''g''$ does not vanish, namely for certain commutators involving $L_1$ and $L_2$. Let us start by discussing the algebra in the absence of this finite piece.

\subsubsection{Algebra in the absence of the finite separation term}

We can use \eqref{eq:PhiComm} to directly evaluate commutators. The simplest scenario is the commutator of two ANEC operators
\be
\<\phi(x_1) \comm{L_{-2}(x_2)}{L_{-2}(x_3)} \phi(x_4)\> = 0\, ,
\ee
which in fact vanishes not only in free field theory but in arbitrary CFTs \cite{Kologlu:2019bco} leading to CFT sum rules with important consequences for large $N$ theories \cite{Belin:2019mnx,Kologlu:2019bco}. 

Next, consider the commutator $[L_{-2},L_{-1}]$. Using \rref{eq:PhiComm}, we find
\be
\<\phi(x_1) \comm{L_{-1}(x_2)}{L_{-2}(x_3)} \phi(x_4)\> = -\fr{2}{\pi} \fr{1}{x_{12}^+ x_{24}^+ (\x{1}{2} - \x{4}{2})^3} \de^{(2)}(\vec{x}_{23}^\perp) \,,
\ee
which has a nonzero contact term. Note that here we specifically assumed that $x_2^+ = x_3^+$, but kept the transverse components $\vec{x}_1^\perp$, $\vec{x}_2^\perp$, and $\vec{x}_4^\perp$ arbitrary. Up to a prefactor, the function multiplying the delta function in $\vec{x}_{23}^\perp$ turns out to be the one-point function on the ANEC operator, namely
\be
\<\phi(x_1) \comm{L_{-1}(x_2)}{L_{-2}(x_3)} \phi(x_4)\> = -i \de^{(2)}(\vec{x}_{23}^\perp) \<\phi(x_1) L_{-2}(x_2) \phi(x_4)\> \,.
\label{eq:Lm1Lm2Comm}
\ee
This easily generalizes to many other commutators as we summarize below (see appendix \ref{ap:commutators} for details):\begin{align} \label{toomanycom}
[L_{-2}(x),L_{-2}(0)]&=0              &  [L_{-1}(x),L_{-1}0)]&=0 \notag \\     
[L_{-1}(x),L_{-2}(0)]&=-i\delta^{(2)}(\vec{x}^\perp)L_{-2}(0)         &[L_{0}(x),L_{-1}(0)]&=-i\delta^{(2)}(\vec{x}^\perp)L_{0}(0) \notag \\    
[L_0(x),L_{-2}(0)]&=-2i\delta^{(2)}(\vec{x}^\perp)L_{-1}(0)    & [L_{1}(x),L_{-1}(0)]&=-2i\delta^{(2)}(\vec{x}^\perp)L_{1}(0) \notag     \\       
[L_1(x),L_{-2}(0)]&=-3i\delta^{(2)}(\vec{x}^\perp)L_{0}(0)   & [L_{2}(x),L_{-1}(0)]&=-3i\delta^{(2)}(\vec{x}^\perp)L_{2}(0)  \notag   \\
[L_2(x),L_{-2}(0)]&=-4i\delta^{(2)}(\vec{x}^\perp)L_{1}(0)   & [L_{0}(x),L_{0}(0)]&=0 \,.
\end{align}
These relations are not an accident, we are reproducing part of the Witt algebra mentioned in the introduction, namely
\be 
\left[L_m(\vec{x}^\perp),L_n(\vec{y}^\perp)\right] =-i \delta^{(2)}(\vec{x}^\perp-\vec{y}^\perp)(m-n)L_{m+n+1}(\vec{y}^\perp)\, ,\label{eq:Virasoro4d}
\ee
This algebra was advocated in \cite{Casini:2017roe}, and some commutators were checked in \cite{Cordova:2018ygx} for free field theories for correlators evaluated on the same null-plane (i.e.~$x^+=y^+$). We have now reproduced those results, and generated many more terms in the algebra.\footnote{Note that in \eqref{eq:Virasoro4d}, it naively seems that the quantum numbers don't match on both sides of the equation, but in fact they do because the transverse delta function $\delta^{(2)}(\vec{x}^\perp-\vec{y}^\perp)$ carries the appropriate compensating dimension. }

So what becomes of the commutators not included in \rref{toomanycom}? These are precisely the ones where the finite piece coming from $f''g''$ does not vanish. As we will now see, they seem to present an obstruction for the algebra.

\subsubsection{Commutators involving the finite separation term}


Let's finally discuss the commutators that involve a contribution at $|\vec{x}_{23}^\perp|\neq0$. At finite separation we can use the general result derived in \eqref{eq:PhiComm}, 
\begin{align}
&\braket{\phi(x_1)[L_1(x_2),L_0(x_3)]\phi(x_4)}=-\frac{\x{1}{2}-\x{4}{2}}{3\pi^2x_{12}^+x_{24}^+|\vec{x}_{23}^\perp|^2}\, .\label{eq:L1L0com}
\end{align}
This expression produces an ambiguity at $\vec{x}_{23}^\perp=0$. The problem is that the finite separation commutator diverges in that limit as $|x^\perp|^{-2}$, and this singularity cannot be integrated against arbitrary test-functions. It is therefore not possible to extract a delta function term at coincident points like for previous examples. This spells doom for the proposed operator algebra \rref{eq:Virasoro4d}. We expand on this point in appendix \ref{ap:deltafunction}. 

The breakdown of the algebra was already hinted at in \cite{Kologlu:2019bco}, where it was argued that Wightman functions which do not converge absolutely under the lightlike integrals when $x_{23}^+ = 0$ can spoil commutation at spacelike separation. However, the integral of the double commutator might still converge, because the double commutator is better behaved in singular limits than the Wightman function, in which case the spacelike commutator is well-defined but nonzero. This is precisely the scenario we have for eq.~\eqref{eq:L1L0com}, where the integral of the double commutator is absolutely convergent at $x_{23}^+=0$ while the integral of the Wightman function is not.


For completeness, we also give the other commutators, which have similar structure. Restricting to $|\vec{x}_{23}^\perp|\neq0$, where the finite separation contribution is well-behaved, we have
\begin{align}
\braket{\phi(x_1)[L_2(x_2),L_0(x_3)]\phi(x_4)}&=-\frac{2}{3\pi^2}\frac{(\x{1}{2}-\x{4}{2})(\x{1}{2}+\x{4}{2})}{x_{12}^+x_{24}^+|\vec{x}_{23}^\perp|^2}\, ,\\
\braket{\phi(x_1)[L_2(x_2),L_1(x_3)]\phi(x_4)}&=\frac{2}{\pi^2}\frac{(\x{1}{3}(\x{4}{2})^2-(\x{1}{2})^2\x{4}{3})}{\pi^2x_{12}^+x_{24}^+|\vec{x}_{23}^\perp|^2}\, .
\end{align}
Once again, we obtain non-integrable finite separation contributions.

The final two commutators are the diagonal ones, for which the delta-function contribution vanishes. The finite-separation piece is thus the only non-zero contribution,
\begin{align}
\<\phi(x_1) \comm{L_1(x_2)}{L_1(x_3)} \phi(x_4)\> &= - \, \fr{1}{\pi^2} \fr{\x{1}{2} \x{4}{3} - \x{4}{2} \x{1}{3}}{x_{12}^+ x_{24}^+ |\vec{x}_{23}^\perp|^2},\label{eq:L1L1comm} \\
\<\phi(x_1) \comm{L_2(x_2)}{L_2(x_3)} \phi(x_4)\> &= - \, \fr{4}{\pi^2} \fr{(\x{1}{2})^2 (\x{4}{3})^2 - (\x{4}{2})^2 (\x{1}{3})^2}{x_{12}^+ x_{24}^+ |\vec{x}_{23}^\perp|^2}\label{eq:L2L2comm}.
\end{align}
The divergences are integrable in this case and there is no $|\vec{x}_{23}^\perp| = 0$ contribution.

The three off-diagonal commutators pose a serious challenge to any possible algebra of light-ray operators. Moreover, we would like to emphasize that these results did not depend on the choice of the external state: all commutators discussed in this section have finite contributions in other scalar states. For example, we can easily see that
\benn 
\braket{\phi^n(x_1)[L_1(x_2),L_1(x_3)]\phi^n(x_4)} = n^2 \braket{\phi^{n-1}(x_1)\phi^{n-1}(x_4)}\braket{\phi(x_1)[L_1(x_2),L_1(x_3)]\phi(x_4)}\,,
\eenn
which obviously also contains a finite separation term. In the next section, we demonstrate how to reproduce this contribution using the OPE in free field theory.  

\subsubsection{$T_{--} \times \phi$ OPE in free field theory}
\label{TphiOPEsec}

The choice of contours from the previous section is such that the relevant OPE channel to study our correlators is $T_{--}(x) \times \phi(y)$. In this section, we explore what becomes of this OPE under the null integral over $x^-$ in free field theory. 

Using the definition of $T_{--}(x)$ in free field theory \eqref{eq:Tmmfreefield}, we can look at the OPE of $T_{--}(x)$ with $\phi(y)$, focusing only on the terms that are proportional to the operator $\phi$ itself. We find
\begin{align} 
T_{--}(x)\phi(y) &\sim \left[\frac{1}{6\pi^2}(\partial_-\phi)^2(x) -\frac{1}{12\pi^2}\phi\partial_-^2 \phi(x) \right]\phi(y) \nn \\
&= \frac{1}{12\pi^2}\left(\frac{-2(x^+-y^+)^2}{(x-y)^6}+\frac{4(x^+-y^+)}{(x-y)^4}\partial_- - \frac{1}{(x-y)^2}\partial_-^2\right)\phi(x)\, \label{eq:TmmPhiOPE2} .
\end{align}
Note that this OPE is not written in the canonical way, because the resulting operator on the right-hand side is evaluated at position $x$ instead of position $y$. If we would like an OPE evaluated at $\phi(y)$, then we still need to expand $\phi(x)$ around the position $y$. This implies that $\phi(x)$ in \eqref{eq:TmmPhiOPE2} is already the resummed version of that Taylor expansion. Because of this, one should keep in mind that we are really keeping infinitely many descendants.

We can use the OPE \eqref{eq:TmmPhiOPE2} twice in $\braket{\phi(x_1)T_{--}(x_2)T_{--}(x_3)\phi(x_4}$, as $x_2\rightarrow x_1$ and as $x_3\rightarrow x_4$ to obtain a differential operator acting on the two-point function $\braket{\phi(x_2)\phi(x_3)}$. This implies 
\begin{align}
\braket{\phi(x_1)T_{--}(x_2)T_{--}(x_3)\phi(x_4)} 
&= \frac{1}{36\pi^4}\left(\frac{(x_{21}^+)^2(x_{34}^+)^2}{x_{21}^6x_{34}^6}+\dots \right)
\braket{\phi(x_2)\phi(x_3)}\, ,\label{eq:DoubleOPETT}
\end{align}
where we have written only the most singular term coming from the double OPE, as this is the only term that contributes to the finite separation term. We can then multiply \eqref{eq:DoubleOPETT} by $(x_2^-)^3(x_3^-)^3$ and perform the two integrals as $x_2\rightarrow x_1$ and $x_3\rightarrow x_4$, we arrive at $\braket{\phi(x_1)L_1(x_2)L_1(x_3)\phi(x_4)}$. Once we send $\epsilon \rightarrow 0$, the contribution from \eqref{eq:DoubleOPETT} is 
\be 
\<\phi(x_1)L_1(x_2)L_1(x_3) \phi(x_4)\> = - \, \fr{1}{\pi^2} \fr{\x{1}{2} \x{4}{3}}{x_{12}^+ x_{24}^+ |\vec{x}_{23}^\perp|^2},
\ee
which is half the commutator from \eqref{eq:L1L1comm}. The other half is obtained the same way, but for the other ordering by performing the OPE as $x_3\rightarrow x_1$ and $x_2\rightarrow x_4$. 

In addition, we can ask how to reproduce these results in the other OPE channel, namely by fusing $T_{--}(x)$ with $T_{--}(y)$. This is what we will discuss in the next section. 

\section{The OPE of light-ray operators}

In the previous section, we saw that in free field theory commutators involving the light-ray operators $L_1$ and $L_2$ do not vanish at finite transverse separation. We would now like to understand whether this nonvanishing commutator can be understood as arising from the integrals of \emph{local} operators $\Ocal$ in the $T \times T$ OPE,
\be
\comm{L_n(x)}{L_n(y)} = \sum_\Ocal \fr{1}{|\vec{x}^\perp - \vec{y}^\perp|^{\alpha_\Ocal}} \int dy^- f_\Ocal(y^-) \Ocal(y),
\label{eq:LnOPE}
\ee
where, for simplicity, we will focus on the diagonal case where the two generalized ANEC operators are the same.

The OPE of two light-ray operators has been studied several times in the literature~\cite{Hofman:2008ar,Kravchuk:2018htv,Kologlu:2019mfz,Chang:2020qpj}, but here we will largely not make direct contact with these general results, and instead choose to focus on the particular example of free field theory. It would be very interesting in future work to understand the structure of generalized ANEC commutators more broadly, using the technology developed in those works.

To begin, we need to first rephrase the calculation of the commutator $\<\phi\comm{L_n}{L_n}\phi\>$ from the previous section in order to make the connection with the $T \times T$ OPE more manifest. There, we originally computed the commutator by evaluating
\begin{align}
\braket{\phi(x_1)[L_n(x_2),L_n(x_3)]\phi(x_4)} &= \int dx_2^- dx_3^- (x_2^-)^{n+2}(x_3^-)^{n+2}\contraction{\<\phi(}{x_1}{)T_{--}}{(x_2)}
\contraction{\<\phi(x_1)T_{--}(x_2)T_{--}(}{x_3}{)\phi}{(x_4)}
\<\phi(x_1)T_{--}(x_2)T_{--}(x_3)\phi(x_4)\>\nonumber\\
&-\int dx_2^- dx_3^- (x_2^-)^{n+2}(x_3^-)^{n+2}\bcontraction{\<\phi(}{x_1}{)T_{--}}{(x_3)}
\bcontraction{\<\phi(x_1)T_{--}(x_3)T_{--}(}{x_2}{)\phi}{(x_4)}
\<\phi(x_1)T_{--}(x_3)T_{--}(x_2)\phi(x_4)\>\, ,\nonumber
\end{align}
where the contraction symbols indicate which poles we considered while performing the integrals over $x_2^-$ and $x_3^-$ (for example, in the first line we integrate $x_2^-$ by picking up the OPE singularity at $x_{12}^2\rightarrow 0$). In order to obtain an expression of the form~\eqref{eq:LnOPE}, we need to evaluate only one of these two integrals (for concreteness, $x_2^-$), such that we can write the commutator in terms of the single remaining light-ray integral (over $x_3^-$).

While the particular contraction shown above was useful in practice, it clearly makes use of the OPE in the $T \times \phi$ channel, obscuring any connection with the $T \times T$ OPE. Instead, we can choose to close the contour for $x_2^-$ in the upper half-plane for both orderings, represented by the contractions
\begin{align}
\braket{\phi(x_1)[L_n(x_2),L_n(x_3)]\phi(x_4)} &= \int dx_2^- dx_3^- (x_2^-)^{n+2}(x_3^-)^{n+2}\contraction{\<\phi(x_1)T_{--}}{(x_2)}{T_{--}}{(x_3)}
\contraction[2ex]{\<\phi(x_2)T_{--}}{(x_2)}{T_{--}(x_3)\phi}{x_4)}
\<\phi(x_1)T_{--}(x_2)T_{--}(x_3)\phi(x_4)\>\nonumber\\
&-\int dx_2^- dx_3^- (x_2^-)^{n+2}(x_3^-)^{n+2}\bcontraction{\<\phi(x_1)T_{--}(x_3)T_{--}}{(x_2)}{\phi}{(x_4)}
\<\phi(x_1)T_{--}(x_3)T_{--}(x_2)\phi(x_4)\>\, ,\nonumber\\\label{eq:LnLncomorder2}
\end{align}
In the first term, we now need to pick up both OPE singularities in $x_{23}^2$ and $x_{24}^2$, while in the second term, we only need to pick up the singularity in $x_{24}^2$ (as before). One can check explicitly that in this correlator the two contributions from the $x_{24}^2$ OPE singularity cancel when evaluating the commutator and we are left with\footnote{As an operator statement, one generally cannot deform the integration contour for $x_2^-$ to represent the commutator $\comm{\Ocal_2}{\Ocal_3}$ as an integral around $x_3^-$. Conceptually, this is due to the $i\epsilon$ prescription for $x_{23}^+$, where the two orderings on the LHS of~\eqref{eq:LnLncomorder2} are functions of $x_{23}^+ \pm i\epsilon$, while the RHS only depends on $x_{23}^+ - i\epsilon$. However, for this particular correlation function involving $\phi$, where we are focused solely on the finite separation piece when both light-ray operators are on the same null plane, one can check explicitly that there is no dependence on this subtlety, and \eqref{eq:LnLncomorder3} holds.}
\begin{align}
\braket{\phi(x_1)[L_n(x_2),L_n(x_3)]\phi(x_4)}&= \int dx_3^- (x_3^-)^{n+2}\left[\int dx_2^- (x_2^-)^{n+2}\contraction{\<\phi(x_1)T_{--}}{(x_2)}{T_{--}}{(x_3)}
\<\phi(x_1)T_{--}(x_2)T_{--}(x_3)\phi(x_4)\>\right]\, .\label{eq:LnLncomorder3}
\end{align}
With this, we can compute the $x_2^-$ integral, picking up only the singularity at $x_{23}^2\ra0$. If we now expand the integrand with the $T \times T$ OPE, we thus obtain an expression of the form~\eqref{eq:LnOPE}. We now need to understand this OPE in free field theory, which we do below for the case of $\comm{L_1}{L_1}$.

\subsection{Example: $\comm{L_1}{L_1}$ \label{sec:L1L1OPE}}

To illustrate how to obtain the finite separation term from the OPE we will focus on the concrete example of $[L_1(x_2),L_1(x_3)]$ (we also explain how to adapt this to other commutators). The goal is to understand the leading divergence in this commutator as $|\vec{x}_{23}^\perp|^2\to0$ by identifying which operators in the $T \times T$ OPE are responsible for it.

One might have hoped that this finite separation contribution arises from the integral of a single local operator. However, we will see that this is not the case and we need to include an infinite number of terms in the sum~\eqref{eq:LnOPE}. This infinite sum can also be rewritten as the light-ray integral of a single non-local operator, which we construct explicitly in appendix~\ref{ap:L1L1OPEnonlocal}.  

The commutator we wish to compute via the OPE is given in \eqref{eq:L1L1comm}, but we reproduce it here for convenience
\begin{align}
\<\phi(x_1) \comm{L_1(x_2)}{L_1(x_3)} \phi(x_4)\> &=
 -\frac{1}{\pi^2}\frac{\left(x_1^--\frac{| \vec{x}_{12}^\perp|^2}{x_{13}^+}\right)\left(x_4^-+ \frac{|\vec{x}_{34}^\perp|^2}{x_{34}^+}\right)-\left(x_4^- +\frac{|\vec{x}_{24}^\perp|^2}{x_{34}^+}\right)\left(x_1^- -\frac{|\vec{x}_{13}^\perp|^2}{x_{13}^+}\right)}{x_{13}^+ x_{34}^+ |\vec{x}_{23}^\perp|^2}\nonumber\, .
\end{align}
We will not aim to derive the complete contribution from the stress-tensor OPE, but will concentrate solely on the terms that are responsible for the leading singularity in the transverse separation $|\vec{x}_{23}^\perp|$,
\be
\braket{\phi(x_1)[L_1(x_2),L_1(x_3)]\phi(x_4)} \Big|_{\vec{x}_2^\perp\rightarrow \vec{x}_3^\perp}
\approx \frac{2(x_{23}^\perp)^I}{\pi^2 |\vec{x}_{23}^\perp|^2(x_{13}^+x_{34}^+)^2}\left[-(x_{13}^\perp)^I x_{34}^+ \x{4}{3} + (x_{34}^\perp)^I x_{13}^+ \x{1}{3}\right], \label{eq:L1L1commexp1}
\ee
where $I={1,2}$ runs over the perpendicular indices. Note that the naive leading singularity $\sim 1/|x_{23}^\perp|^2$ vanishes due to antisymmetry. It is worthwhile to mention that the rest of the expansion truncates, with only one subleading correction which is regular as $|\vec{x}_{23}^\perp| \ra 0$.

We now want to understand the leading term \eqref{eq:L1L1commexp1} as arising from the integral of local operators in the $T \times T$ OPE. The only operators which can contribute must satisfy the following set of constraints. First, given the overall factor of $(x_{23}^\perp)^I$, it is clear that the operators must all be vectors in the transverse direction. In addition, the operators cannot contain $\partial_+$ derivatives, as these would be contracted with factors of $x_{23}^+$, which is zero. Finally, the operator must contain only two insertions of $\phi$, otherwise the resulting three-point function would vanish. With this, we can construct the most general set of operators that have the required properties and are consistent with scale and boost invariance, giving us the general sum
\be
[L_1(x_2),L_1(x_3)]\Big|_{\text{l.s}} = \frac{(x_{23}^\perp)^I}{|\vec{x}_{23}^\perp|^2} \sum_{m=0}^\infty a_m \int dx_3^- (x_3^-)^{m+3} \, \phi\overset{\leftrightarrow}{\partial}_I\overset{\leftrightarrow\,\,\,}{\partial^m_-} \phi(x_3),\label{eq:L1L1comopf}
\ee
where the subscript $\textrm{l.s.}$ stands for leading singularity in the $\phi$ correlator. Concretely, the operators written schematically in the RHS correspond to (descendants of) the conserved higher-spin currents,
\be
\phi\overset{\leftrightarrow}{\partial}_I\overset{\leftrightarrow\,\,\,}{\partial^m_-} \phi(x) \equiv \begin{cases} \p_I J_{-\cdots-}(x) & m \textrm{ even}, \\ J_{I-\cdots-}(x) & m \textrm{ odd},\end{cases}
\ee
whose explicit form is given by~\cite{Mikhailov:2002bp,Penedones:2010ue}
\be 
J_{i_1\dots i_s}(x)= \sum_{k=0}^{s}\frac{(-1)^k}{\G^2(k+1)\G^2(s-k+1)}\partial_{i_1}\dots \partial_{i_k}\phi(x)\,\partial_{i_{k+1}}\dots \partial_{i_s}\phi(x)-\text{traces},\label{eq:Ji1is}
\ee
where the indices are symmetrized. For example, the term with $m=0$ corresponds to $\partial_I\phi^2$, while the next term with $m=1$ is the stress tensor component $T_{I-}$.

\begin{figure}[t!]
\begin{center}
\includegraphics[width=0.7\textwidth]{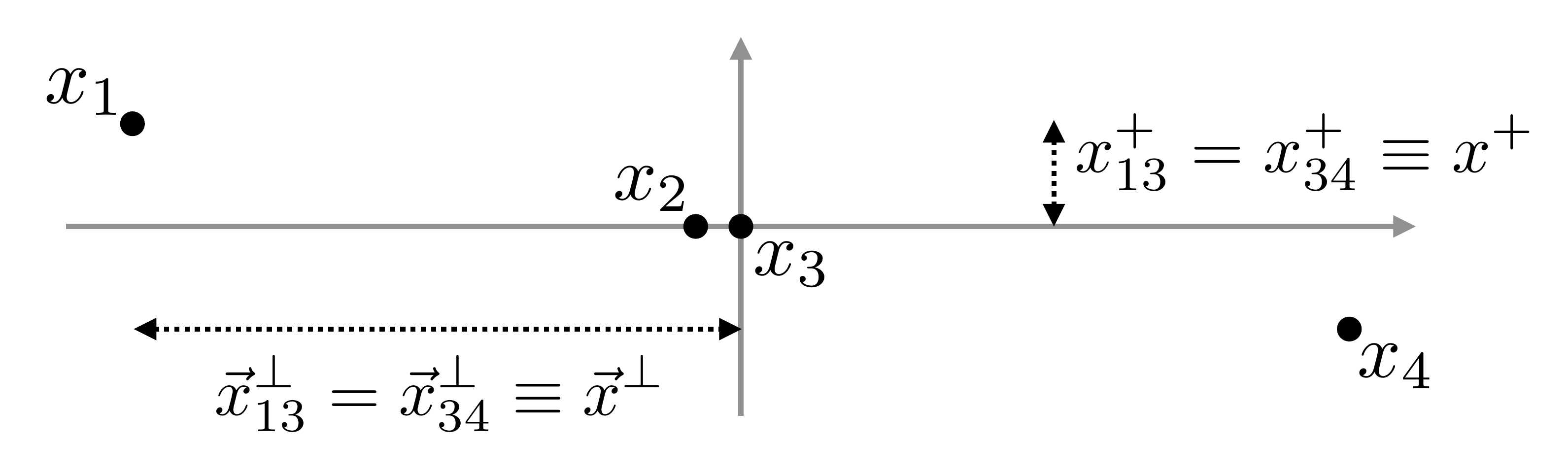}
\caption{Setup for the operators in the $x^+$ and $\vec{x}^\perp$ directions in our simplifying limit.}
  \label{fig:boat1}
\end{center}
\end{figure}

The overall coefficients $a_m$ can in principle be determined from the OPE coefficients in $T \times T$. For our purposes here, however, we merely wish to understand the structural form of this expansion.
To do so, we will consider a simplifying kinematic limit and compute the resulting contributions in the sum. Specifically, we consider the following arrangement:
\be
x_{13}^+ = x_{34}^+ \equiv x^+, \quad \vec{x}_{13}^\perp = \vec{x}_{34}^\perp \equiv \vec{x}^\perp, \quad |\vec{x}^\perp|^2 \gg |\vec{x}_{23}^\perp|^2.
\ee
We thus align all four operators on a line in the two transverse directions, with $x_1$ and $x_4$ at large equal distance on either side of the light-ray operators. This simplifying arrangement is displayed in figure~\ref{fig:boat1}.
In this setup, the leading singularity~\eqref{eq:L1L1commexp1} takes the simple form
\be
\braket{\phi(x_1)[L_1(x_2),L_1(x_3)]\phi(x_4)}\Big|_{\vec{x}_2^\perp\rightarrow \vec{x}_3^\perp}
\approx \frac{-4(x_{23}^\perp)^I(x^\perp)^I|\vec{x}^\perp|^2 }{\pi^2 (x^+)^4 |\vec{x}_{23}^\perp|^2}.\label{eq:L1L1x2closex3limit}
\ee
In appendix \ref{ap:L1L1OPE}, we explicitly compute the terms in the sum \eqref{eq:L1L1comopf} evaluated inside the correlation function, which we denote as
\be
\mathcal{G}^m(x_i) \equiv \frac{(x_{23}^\perp)^I}{|x_{23}^\perp|^2} \int dx_3^- (x_3^-)^{m+3} \<\phi(x_1) \, \phi\overset{\leftrightarrow}{\partial}_I\overset{\leftrightarrow\,\,\,}{\partial^m_-} \phi(x_3) \, \phi(x_4)\>.
\ee
We then show that in the limit described above these expressions all have the same functional form, up to an overall coefficient,
\be 
\mathcal{G}^m(x_i) \approx \frac{b_m (x_{23}^\perp)^I (x^\perp)^I |\vec{x}^\perp|^2}{\pi^2 (x^+)^4|\vec{x}_{23}^\perp|^2},
\ee
where $b_m$ are numerical factors that can be found in the appendix \ref{ap:L1L1OPE}. We can therefore clearly see that all operators in the OPE~\eqref{eq:L1L1comopf} contribute to the leading term, and we are ultimately just resumming the infinite series of coefficients.

We have therefore seen that the nonvanishing commutator $\<\phi\comm{L_1}{L_1}\phi\>$ can be understood directly as an infinite sum of light-ray operators built from the $T \times T$ OPE. Let us briefly comment on two aspects of this calculation.

First, if we want to understand the finite separation contributions of other commutators, then we just need to modify the sum of operators in~\eqref{eq:L1L1comopf} to account for the dimension and spin of the specific commutator we are considering. For example, in the case of $\comm{L_2}{L_2}$, we just need to replace $(x_3^-)^3\rightarrow (x_3^-)^5$ in~\eqref{eq:L1L1comopf} (with new coefficients $\tilde{a}_m$). 

Second, the integrand of $\braket{\phi\comm{L_1}{L_1}\phi}$ does not contain a pole at infinity, as one can explicitly confirm. As we discuss in more detail in appendix \ref{ap:L1L1OPE}, the first three terms of the sum $m=\lbrace 0,1,2\rbrace$ individually each contain a pole at infinity in their respective integrands. Nevertheless, in the limit used above, this contribution vanishes and the sum is well-behaved. Because the full expression that this sum is replacing has no pole at infinity, we expect the general infinite sum to also not have any such poles. This concludes our discussion of the generalized ANEC commutators in free field theory.

\section{Conformal block decomposition at large $N$}
\label{sec:confblock}

So far, we have only studied commutators of light-ray operators in the example of free field theory. In this section, we generalize this analysis by considering the commutator $\comm{L_m}{L_n}$ (at finite transverse separation) in a four-point function with a general scalar operator $\Ocal$ of dimension $\De$,
\be
\<\Ocal(x_1) \comm{L_m(x_2)}{L_n(x_3)} \Ocal(x_4)\> \qquad (\vec{x}_2^\perp \neq \vec{x}_3^\perp; \, |m|,|n| \leq 2).
\ee
Specifically, we study the contribution to this commutator from the conformal block associated with $\Ocal$ itself,
\be
\Ocal \times T \ra \Ocal \ra T \times \Ocal.
\ee
This particular contribution is interesting for two reasons. First, it is universal, as the OPE coefficient is fixed by conformal Ward identities. Second, this particular block is the leading contribution to the correlation function of light-ray operators in CFTs with large $N$ and a large gap, as we explain below by slightly generalizing an argument from~\cite{Kravchuk:2018htv,Kologlu:2019bco}. We can therefore compare the resulting commutator from $\Ocal$ exchange to gravitational computations in AdS, which we do in section~\ref{sec:Shockwaves}.

In practice, the easiest way to compute the contribution of individual conformal blocks is to map the null plane $x^+=0$ to the celestial sphere. After briefly reviewing this map, we demonstrate that $\Ocal$ exchange gives rise to a \emph{nonzero} commutator at finite separation, even for the simplest case of $\comm{L_{-1}}{L_{-2}}$, which suggests the existence of a non-trivial sum rule, where subleading conformal blocks must cancel this finite-separation commutator in physical CFTs.

\subsection{Contributions in holographic CFTs}

One of our main motivations in studying the commutators of generalized ANEC operators is to gain insight on possible universal structure in gravitational theories in AdS. We are therefore particularly interested in CFTs with a weakly-coupled bulk dual. Such theories generally must have the structure of generalized free field theory, with corrections suppressed by a large parameter $N$,
\be
\<\Ocal(x_1) T(x_2) T(x_3) \Ocal(x_4)\> = \<\Ocal(x_1) \Ocal(x_4)\> \<T(x_2) T(x_3)\> + O(1/N^2),
\ee
as well as a large gap $\De_{\textrm{gap}}$ in the scaling dimensions of the lowest single-trace operators with spin $j > 2$.

\begin{figure}[t!]
\begin{center}
\includegraphics[width=0.7\textwidth]{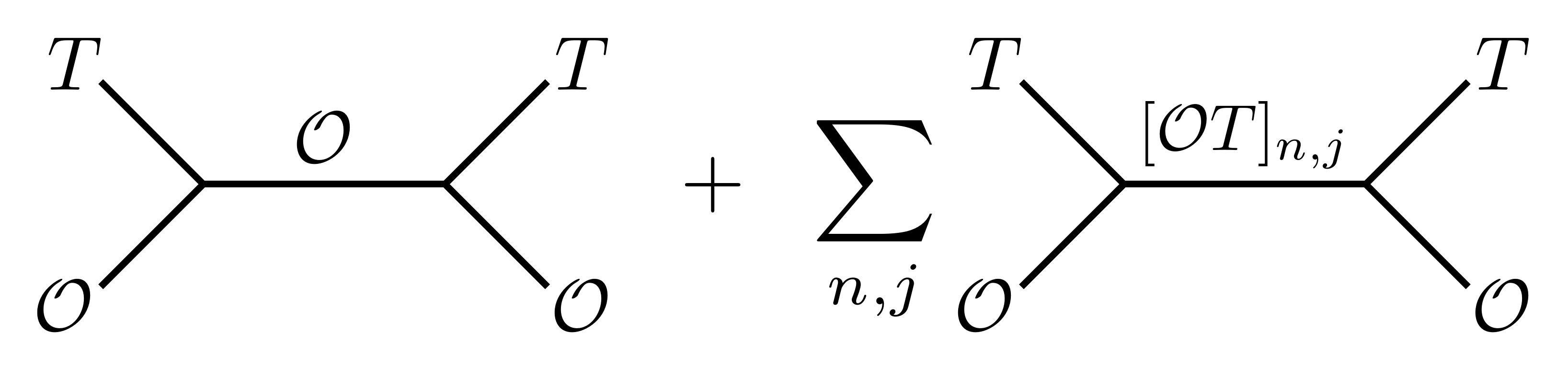}
\caption{Leading contributions to the four-point function $\<\Ocal TT \Ocal\>$ in CFTs with large $N$ and large gap. When integrating to obtain correlators of the global light-ray operators $L_m$, the contribution from double-trace operators (right) vanishes, leaving only the $\Ocal$ exchange.}
\label{fig:Blocks} 
\end{center}
\end{figure}

In such theories, the leading correction to the four-point function $\<\Ocal T T \Ocal\>$ can only come from $\Ocal$ itself, as well as the leading $1/N$ corrections to the infinite family of double-trace operators $[\Ocal T]_{n,j}$ corresponding to two-particle states in AdS. This is shown schematically in figure~\ref{fig:Blocks}.

Note that we are specifically interested in correlation functions of the global light-ray operators $L_m$ with $|m| \leq 2$, which annihilate the vacuum in either direction. In this case, we can rewrite their correlation function as the double-commutator
\be
\ba
&\<\Ocal(x_1) L_m(x_2) L_n(x_3) \Ocal(x_4)\> = \<\comm{\Ocal(x_1)}{L_m(x_2)} \comm{L_n(x_3)}{\Ocal(x_4)}\> \\
& \qquad \qquad = \int dx_2^-\, dx_3^- (x_2^-)^{m+2} (x_3^-)^{n+2} \<\comm{\Ocal(x_1)}{T_{--}(x_2)} \comm{T_{--}(x_3)}{\Ocal(x_4)}\> \\
& \qquad \qquad = \int dx_2^-\, dx_3^- (x_2^-)^{m+2} (x_3^-)^{n+2} \textrm{dDisc}\Big[ \<\Ocal(x_1) T_{--}(x_2) T_{--}(x_3) \Ocal(x_4)\> \Big].
\ea
\ee
Four-point functions involving the global $L_m$ therefore correspond to weighted integrals of the double-discontinuity of $\<\Ocal TT \Ocal\>$. However, the leading correction due to double-trace operators has no double-discontinuity, which means that their contribution is suppressed by additional powers of $1/N$~\cite{Caron-Huot:2017vep}. At leading order in $1/N$ and $1/\De_{\textrm{gap}}$, the only contribution to the commutators $\comm{L_m}{L_n}$ therefore comes from the exchange of $\Ocal$.

\subsection{Event shapes on the celestial sphere}

Up to this point, we have focused on light-ray operators located on a null plane at finite $x^+$. However, we can also consider the limit where these operators are taken to future null infinity (i.e.~$x^+ \ra \infty$),\footnote{The rescaling by $x^+$ ensures that the resulting correlation functions are finite, while the factor of $2$ simply provides a useful normalization for the resulting expressions.}
\be
L_m(\infty) \equiv \lim_{x^+ \ra \infty} \left( \fr{x^+}{2} \right)^2 L_m(x).
\ee
More generally, we can use rotations to take light-ray operators to future null infinity in any direction, parametrized by the null vector
\be
n^\mu = (1,\vec{n}), \quad n^2 = 0,
\ee
where $\vec{n}$ is a unit-normalized vector indicating a point on the celestial sphere. In the following discussion, we will label these light-ray operators at general positions on the celestial sphere as $L_m(n)$, suppressing the $\infty$ in the argument for notational simplicity.

\begin{figure}[t!]
\begin{center}
\includegraphics[width=0.85\textwidth]{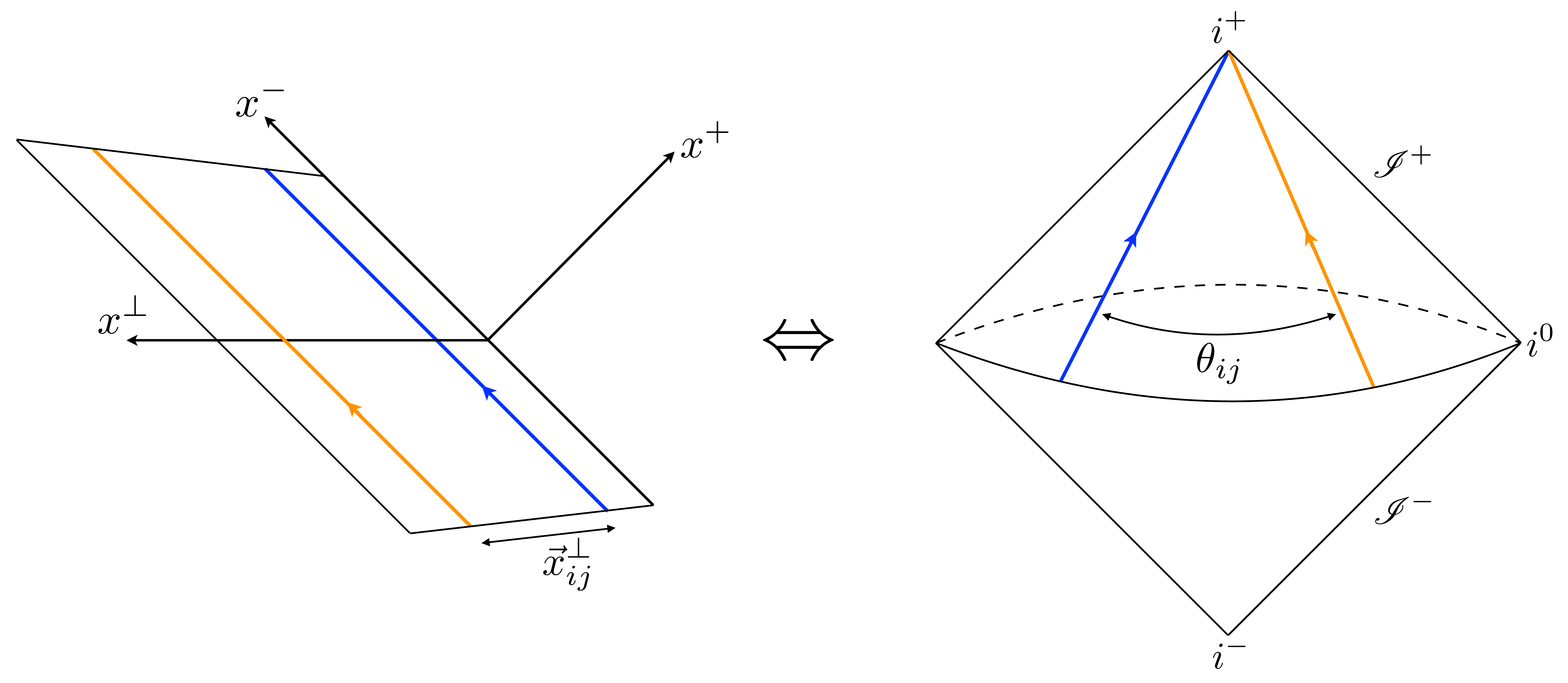}
\caption{Two light-ray operators (indicated by the blue and orange lines) on the null plane $x^+ = 0$ (left) can be mapped via the conformal transformation~\eqref{eq:MapToInfty} to future null infinity $\mathscr{I}^+$ (right). The transverse separation $\vec{x}_{ij}^\perp$ between the two operators on the plane leads to an angular separation $\theta_{ij}$ on the celestial sphere.}
\label{fig:CelestialSphere} 
\end{center}
\end{figure}

We can also map the null plane $x^+ = 0$ to future null infinity $\mathscr{I}^+$ via the conformal transformation
\be
x^+ \ra -\fr{1}{x^+}, \qquad x^- \ra x^- - \fr{|\vec{x}^\perp|^2}{x^+}, \qquad \vec{x}^\perp \ra \fr{\vec{x}^\perp}{x^+}.
\label{eq:MapToInfty}
\ee
In this case, light-ray operators at distinct transverse coordinates $\vec{x}_i^\perp$ on the null plane map to light-ray operators inserted in different directions $\vec{n}_i$ on the celestial sphere, as shown in figure~\ref{fig:CelestialSphere}. The relative angle $\vec{n}_i \cdot \vec{n}_j = \cos\theta_{ij}$ between them is given by
\be
\fr{1-\cos \theta_{ij}}{2} = \fr{|\vec{x}^\perp_{ij}|^2}{(1+|\vec{x}_i^\perp|^2)(1+|\vec{x}_j^\perp|^2)}.
\ee
We can therefore map correlation functions involving light-ray operators on the same null plane to so-called event shapes~\cite{Basham:1977iq,Basham:1978bw,Basham:1978zq,Belitsky:2013xxa} involving light-ray operators at null infinity,
\be
\<\Ocal|L_{m_1}(x_1) \cdots L_{m_k}(x_k)|\Ocal\> \quad \Leftrightarrow \quad \<\Ocal|L_{m_1}(n_1) \cdots L_{m_k}(n_k)|\Ocal\>.
\ee
Physically, these event shapes can be thought of as the correlation between detectors located at different points on the celestial sphere.

The important point is that we can study the commutators of light-ray operators at finite separation by computing event shapes for light-ray operators separated by a finite angle. As we will see, the calculation of the resulting commutator will be much simpler for event shapes than on the null plane.

We can already see this simplification in three-point functions involving the insertion of a single light-ray operator. For example, if we consider the three-point function with the ANEC operator $L_{-2}$ from eq.~\eqref{eq:ThreePtFct}, at $x^+ \ra \infty$ we obtain
\be
\lim_{x_2^+ \ra 0} \left(\fr{x_2^+}{2}\right)^2 \<\Ocal(x_1) L_{-2}(x_2) \Ocal(x_3)\> = \fr{i\De}{2\pi} \fr{1}{(x_{13}^-)^3 x_{13}^{2\De-2}}.
\ee
Generalizing this expression to an arbitrary direction on the celestial sphere, we have
\be
\<\Ocal(x_1) L_{-2}(n) \Ocal(x_3)\> = \fr{i\De}{2\pi} \fr{1}{(-n\cdot x_{13})^3 x_{13}^{2\De-2}},
\ee
which reduces to the above expression for $n^+ = 2$, $n^- = \vec{n}^\perp = 0$.

Note that this expression depends only on the relative distance between the two insertions of $\Ocal$, which means that if we Fourier transform to momentum space we find that momentum is conserved~\cite{Bautista:2019qxj},
\be
\ba
\<\Ocal(p)|L_{-2}(n)|\Ocal(p')\> &= \int d^4x_1 d^4x_3 \, e^{-i(p\cdot x_1 - p'\cdot x_3)} \<\Ocal(x_1) L_{-2}(n) \Ocal(x_3)\> \\
&= (2\pi)^4 \de^4(p-p') \fr{\pi^2 (-p^2)^\De}{2^{2\De-4} (-n \cdot p)^3\G(\De-1)\G(\De)}.
\ea
\label{eq:EventShape1ANEC}
\ee
As is well-known, if we divide this expression by the norm of the external state,
\be
\<\Ocal(p)|\Ocal(p')\> = (2\pi)^4 \de^4(p-p') \fr{\pi^3(-p^2)^{\De-2}}{2^{2\De-6} \G(\De-1)\G(\De)},
\ee
we measure the energy deposited in a given direction on the celestial sphere, which for a spherically-symmetric state with $\vec{p}=0$ is simply
\be
\ba
\<L_{-2}(n)\> \equiv \fr{\<\Ocal(p^0)|L_{-2}(n)|\Ocal(p^0)\>}{\<\Ocal(p^0)|\Ocal(p^0)\>} = \fr{p^0}{4\pi}.
\ea
\ee

We can also consider three-point functions involving the other global light-ray operators at null infinity, such as $L_{-1}$,
\be
\<\Ocal(x_1) L_{-1}(n) \Ocal(x_3)\> = \fr{i\Delta}{4\pi} \fr{(-n \cdot x_1) + (-n \cdot x_3)}{(-n \cdot x_{13})^3 x_{13}^{2\De-2}}.
\ee
As we can see, this expression is almost the same as for $L_{-2}$, except for the positions in the numerator. We can use this fact to simplify the Fourier transform to momentum space,
\be
\ba
&\<\Ocal(p)|L_{-1}(n)|\Ocal(p')\> = \fr{i\Delta}{4\pi}\int d^4x_1 d^4x_3 \, e^{-i(p\cdot x_1 - p'\cdot x_3)} \fr{(-n \cdot x_1) + (-n \cdot x_3)}{(-n \cdot x_{13})^3 x_{13}^{2\De-2}} \\
& \qquad = -\fr{i}{2}\Big( (n\cdot \p_p) - (n\cdot \p_{p'}) \Big) \fr{i\Delta}{2\pi}  \int d^4x_1 d^4x_2 \, e^{-i(p\cdot x_1 - p'\cdot x_3)} \fr{1}{(-n \cdot x_{13})^3 x_{13}^{2\De-2}} \\
& \qquad = -\fr{i}{2} \Big( (n\cdot \p_p) - (n\cdot \p_{p'}) \Big) \<\Ocal(p)|L_{-2}(n)|\Ocal(p')\>.
\ea
\ee
The three-point function with $L_{-1}$ is therefore given by a differential operator acting on the three-point function with $L_{-2}$,
\begin{align}\label{L1holo}
\<\Ocal(p)|L_{-1}(n)|\Ocal(p')\> &= -(2\pi)^4 \de^4(p-p') \fr{i\pi^2 \De (-p^2)^{\De-1}}{2^{2\De-4} (-n \cdot p)^2\G(\De-1)\G(\De)} \\
& \qquad - \, \Big( (n \cdot \p_p) (2\pi)^4 \de^4(p-p') \Big) \fr{i\pi^2 (-p^2)^\De}{2^{2\De-4} (-n\cdot p)^3 \G(\De-1)\G(\De)}\, .\nn
\end{align}
Notice that the second term gives a singular contribution to the expectation value in a given momentum eigenstate. This is a common feature related to the non-normalizability of definite momentum states \cite{Hofman:2008ar}. It can easily be solved by spreading the wave function over a small neighbourhood around $p$. By doing this one can integrate the second term by parts. The upshot is that the expectation value becomes independent of the direction $n$, just like that of $L_{-2}$,

\be\label{normdelta}
\ba
\<L_{-1}(n)\> \equiv \fr{\<\Ocal(p^0)|L_{-1}(n)|\Ocal(p^0)\>}{\<\Ocal(p^0)|\Ocal(p^0)\>} =  \fr{i\De}{4\pi}\, .
\ea
\ee

A similar calculation can be performed for the remaining global light-ray generators $L_m$ with $m=0,1,2$. We can still connect them to the three point function involving $L_{-2}$ albeit by acting with higher-order differential operators,
\be
\ba
\<\Ocal(p)|L_0(n)|\Ocal(p')\> &= -\fr{1}{6} \Big( (n \cdot \p_p)^2 - 4 (n\cdot \p_p)(n\cdot \p_{p'}) + (n \cdot \p_{p'})^2 \Big) \<\Ocal(p)|L_{-2}(n)|\Ocal(p')\>, \\
\<\Ocal(p)|L_1(n)|\Ocal(p')\> &= -\fr{i}{2} (n \cdot \p_p) (n \cdot \p_{p'}) \Big( (n \cdot \p_p) - (n \cdot \p_{p'}) \Big) \<\Ocal(p)|L_{-2}(n)|\Ocal(p')\>, \\
\<\Ocal(p)|L_2(n)|\Ocal(p')\> &= (n \cdot \p_p)^2 (n \cdot \p_{p'})^2 \<\Ocal(p)|L_{-2}(n)|\Ocal(p')\>.
\ea
\ee

\subsection{Commutators at finite separation}

Let's now consider event shapes involving two light-ray operators. In general, we can compute this correlation function by inserting a complete set of intermediate states, which in momentum space takes the simple form
\be
\ba
&\<\Ocal(p)|L_{m_1}(n_1) L_{m_2}(n_2)|\Ocal(p')\> \\
& \qquad = \sum_{\Ocal'} \int \fr{d^4q}{(2\pi)^4} \fr{\<\Ocal(p)|L_{m_1}(n_1)|\Ocal'(q)\>\<\Ocal'(q)|L_{m_2}(n_2)|\Ocal(p')\>}{\<\Ocal'(q)|\Ocal'(q)\>},
\ea
\ee
where the sum is over primary operators $\Ocal'$ in the $\Ocal \times T$ OPE. Here we will specifically focus on the case where the only contribution is from $\Ocal$ itself. In this case, we can easily obtain the resulting four-point function from the three-point functions computed in the previous section.

To start, let's consider the case where both light-ray operators are the ANEC operator $L_{-2}$. In this case, the delta function in eq.~\eqref{eq:EventShape1ANEC} fixes the intermediate momentum $q$, leaving us with the straightforward result
\be
\<\Ocal(p)|L_{-2}(n_1) L_{-2}(n_2)|\Ocal(p')\> = (2\pi)^4 \de^4(p-p') \fr{\pi (-p^2)^{\De+2}}{2^{2\De-2} (-n_1 \cdot p)^3 (-n_2\cdot p)^3 \G(\De-1)\G(\De)}.
\label{eq:EventShape2ANEC}
\ee
This expression is symmetric under the exchange $n_1 \lra n_2$, which means that the ANEC operators commute for any relative position on the celestial sphere~\cite{Hofman:2008ar,Kologlu:2019bco},
\be
\<\Ocal(p)|\comm{L_{-2}(n_1)}{L_{-2}(n_2)}|\Ocal(p')\>\Big|_\Ocal = 0,
\ee
where the subscript indicates that this is specifically the contribution of the $\Ocal$ conformal block to the commutator.

Next, we can study the commutator between $L_{-1}$ and $L_{-2}$. Using the differential operator derived in the previous section, we can write the resulting commutator as
\be
\ba
&\<\Ocal(p)|\comm{L_{-1}(n_1)}{L_{-2}(n_2)}|\Ocal(p')\> \\
& \qquad = \int \fr{d^4q}{(2\pi)^4} \fr{\left( -\fr{i}{2} (n_1 \cdot \p_p - n_1 \cdot \p_{q})\<\Ocal(p)|L_{-2}(n_1)|\Ocal(q)\>\right)\<\Ocal(q)|L_{-2}(n_2)|\Ocal(p')\>}{\<\Ocal(q)|\Ocal(q)\>} \\
& \qquad \quad - \int \fr{d^4q}{(2\pi)^4} \fr{\<\Ocal(p)|L_{-2}(n_2)|\Ocal(q)\>\left( -\fr{i}{2} (n_1 \cdot \p_q - n_1 \cdot \p_{p'})\<\Ocal(q)|L_{-2}(n_1)|\Ocal(p')\>\right)}{\<\Ocal(q)|\Ocal(q)\>}.
\ea
\ee
Using integration by parts, we can massage this expression into the form
\be
\ba
&\<\Ocal(p)|\comm{L_{-1}(n_1)}{L_{-2}(n_2)}|\Ocal(p')\> \\
& \qquad = -\fr{i}{2} \Big( (n_1 \cdot \p_p) + (n_1 \cdot \p_{p'}) \Big) \<\Ocal(p)|L_{-2}(n_1) L_{-2}(n_2)|\Ocal(p')\> \\
& \qquad \quad - \, \fr{i}{2} \int \fr{d^4q}{(2\pi)^4} \<\Ocal(p)|L_{-2}(n_1)|\Ocal(q)\>\<\Ocal(q)|L_{-2}(n_2)|\Ocal(p')\> \left( (n_1 \cdot \p_q) \fr{1}{\<\Ocal(q)|\Ocal(q)\>}\right) \\
& \qquad \quad - \, \fr{i}{2} \int \fr{d^4q}{(2\pi)^4} \fr{\<\Ocal(p)|L_{-2}(n_1)|\Ocal(q)\>\left( (n_1 \cdot \p_{q})\<\Ocal(q)|L_{-2}(n_2)|\Ocal(p')\>\right)}{\<\Ocal(q)|\Ocal(q)\>} \\
& \qquad \quad + \, \fr{i}{2} \int \fr{d^4q}{(2\pi)^4} \fr{\<\Ocal(p)|L_{-2}(n_2)|\Ocal(q)\>\left( (n_1 \cdot \p_{q})\<\Ocal(q)|L_{-2}(n_1)|\Ocal(p')\>\right)}{\<\Ocal(q)|\Ocal(q)\>}.
\ea
\ee
While this expression is admittedly somewhat complicated, it makes a few useful facts manifest. First, all derivatives of the overall momentum-conserving delta function cancel, which is not true in general correlation functions involving $L_{-1}$. Next, the last two terms cancel if the two null directions are the same, $n_1^\mu = n_2^\mu$, which means their contribution must depend solely on the angular dependence of the ANEC correlator~\eqref{eq:EventShape2ANEC} and is proportional to $n_1 \cdot n_2$. Finally, any proportionality to the scaling dimension $\De$ of the external operator comes from the first two terms, and this dependence clearly cancels.

Evaluating this expression, we then obtain the resulting commutator
\be
\ba
&\<\Ocal(p)|\comm{L_{-1}(n_1)}{L_{-2}(n_2)}|\Ocal(p')\> \\
& \quad = (2\pi)^4 \de^4(p-p') \fr{-i\pi (-p^2)^{\De+1}}{2^{2\De-4} (-n_1 \cdot p)^2 (-n_2\cdot p)^3 \G(\De-1)\G(\De)} \left( 1 - \fr{3}{4} \fr{(-n_1 \cdot n_2)(-p^2)}{(-n_1 \cdot p)(-n_2 \cdot p)} \right).
\ea
\ee
In a spherically symmetric state with $\vec{p}=0$, this expression reduces to the expectation value
\be\label{resultconfblo}
\<\comm{L_{-1}(n_1)}{L_{-2}(n_2)}\> = -\fr{ip^0}{4\pi^2} \left( 1 - \fr{3}{4} (-n_1 \cdot n_2)\right) = -\fr{ip^0}{16\pi^2} \Big( 1 + 3 \cos \theta_{12} \Big).
\ee
Using the map from null infinity to the null plane in eq.~\eqref{eq:MapToInfty}, we therefore find
\be
\<\Ocal(x_1)\comm{L_{-1}(x_2)}{L_{-2}(x_3)}\Ocal(x_4)\> \Big|_\Ocal \neq 0 \qquad (\vec{x}_2^\perp \neq \vec{x}_3^\perp).
\ee
We confirm this directly in position space in appendix \ref{app:NullPlaneBlocks} (see \rref{finitecompos}).

To sum up, $L_{-1}$ and $L_{-2}$ do not commute perturbatively at leading order in $1/N$ and $1/\De_{\textrm{gap}}$. In section~\ref{sec:Shockwaves} we reproduce this result, as well as the exact form of the non-vanishing commutator, explicitly from the AdS side.

\begin{figure}[t!]
\begin{center}
\includegraphics[width=0.6\textwidth]{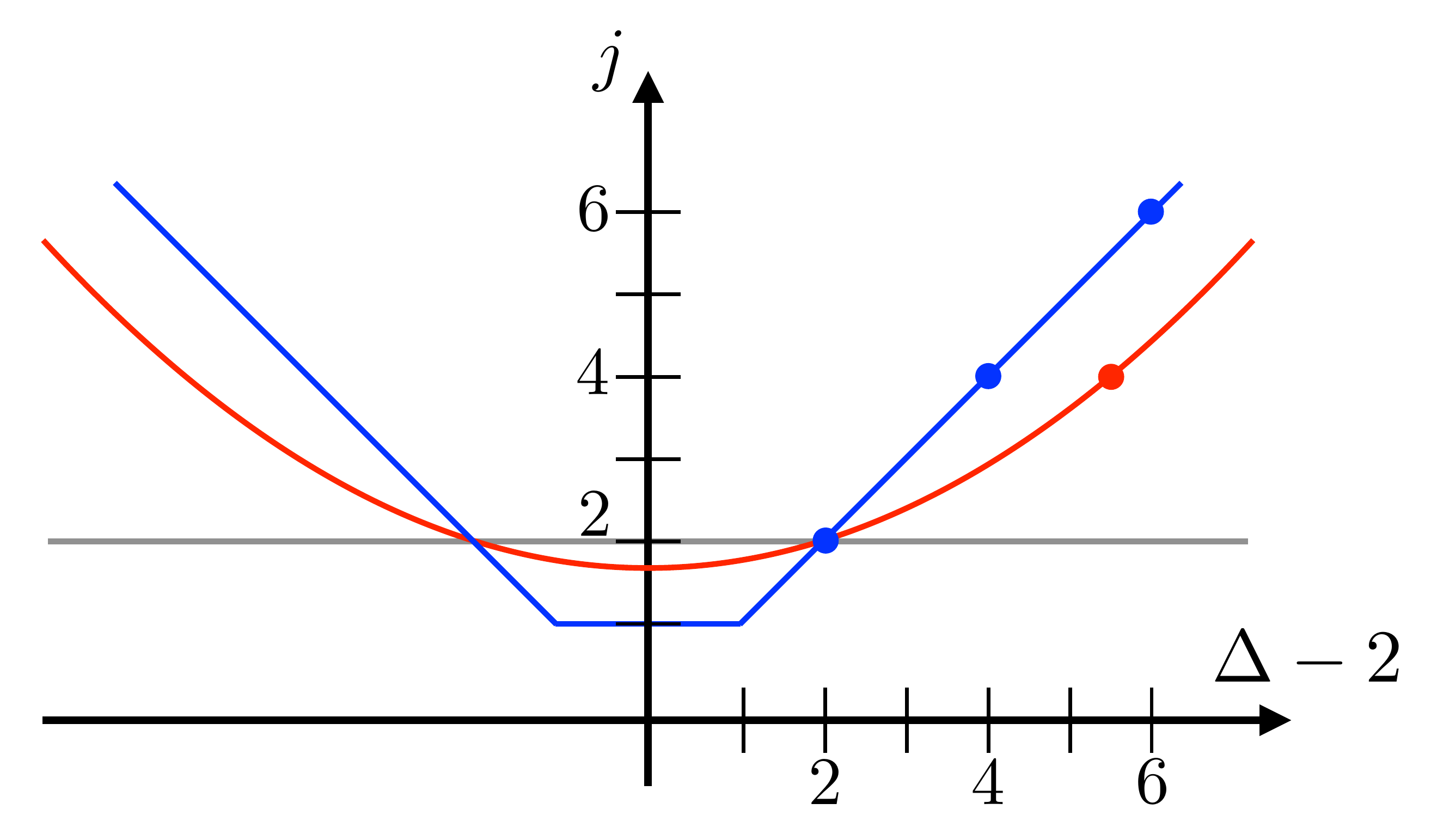}
\caption{Schematic representation of the leading Regge trajectory at large $N$ for the example of $\Ncal=4$ super Yang-Mills at zero (blue), strong (red), and infinite (gray) coupling. In free theory, the Regge trajectory corresponds to the tower of higher-spin conserved currents (blue), with intercept $j_0 = 1$. At infinite coupling, the higher-spin currents are lifted to $\De_{\textrm{gap}} \ra \infty$, resulting in a flat Regge trajectory (gray) with intercept $j_0 = 2$.}
\label{fig:ReggeIntercept} 
\end{center}
\end{figure}

On the one hand, this result is not surprising. In~\cite{Kologlu:2019bco}, it was argued that the commutator $\comm{L_{m_1}}{L_{m_2}}$ is controlled by the intercept $j_0$ of the leading Regge trajectory in the $T \times T$ OPE, with the expectation
\be
m_1 + m_2 < -(j_0+1) \quad \Rightarrow \quad \comm{L_{m_1}(x_1)}{L_{m_2}(x_2)} = 0 \quad (\vec{x}_1^\perp \neq \vec{x}_2^\perp).
\label{eq:InterceptBound}
\ee
In the limit $\De_{\textrm{gap}} \ra \infty$, the Regge intercept $j_0 = 2$, as shown schematically in figure~\ref{fig:ReggeIntercept}. The commutator $\comm{L_{-1}}{L_{-2}}$ therefore lies outside the bounds of guaranteed commutation. The authors of \cite{Kologlu:2019bco} do not commit on the fate of such operators since the integral of the Wightman function needs to be regularized, but they mention the possibility that this leads to non-commutativity at spacelike separation, as we have now confirmed.

However, it was also shown in~\cite{Kologlu:2019bco} that nonperturbatively any CFT should satisfy $j_0 \leq 1$. This indicates that in a physical theory with finite $N$ and $\De_{\textrm{gap}}$, the nonperturbative contributions from additional intermediate states must perfectly cancel the $\Ocal$ contribution, such that the commutator vanishes for any finite separation. This parallels the discussion in section~\ref{sec:FreeAlgebra}: if we had considered $\phi^2$ as an external state, the contribution from the infinite tower of higher-spin conserved currents in the $\phi^2 \times T$ OPE would have exactly cancelled the contribution from $\phi^2$ itself,
\be\label{sumrule1}
\<\phi^2|\comm{L_{-1}(x_2)}{L_{-2}(x_3)}|\phi^2\> \Big|_{\phi^2} + \sum_{j=2}^\infty \<\phi^2|\comm{L_{-1}(x_2)}{L_{-2}(x_3)}|\phi^2\> \Big|_{\Ocal_j} = 0, \quad (\vec{x}_2^\perp \neq \vec{x}_3^\perp).
\ee

It would be interesting to study this sum rule in more detail in future work, especially in the context of holography, as it requires nonperturbative effects in a UV complete theory of gravity to contribute an $O(1)$ correction to an inherently IR observable.

One can now repeat this procedure for other commutators of global operators, such as $\comm{L_{-1}}{L_{-1}}$. While the exact form of the final expression is not important, the crucial point is that those commutators are also nonzero at finite separation. The naive expectation from~\eqref{eq:InterceptBound}, along with the bound $j_0 \leq 1$ from~\cite{Kologlu:2019bco}, is that in general physical CFTs only $\comm{L_{-2}}{L_{-2}}$ and $\comm{L_{-1}}{L_{-2}}$ vanish nonperturbatively. However, we have seen that in the case of free field theory all global commutators with $m_1 + m_2 \leq 0$ vanish at finite separation. It would be useful to determine whether this is specifically a property of free field theory, or if such correlators are well-behaved in a larger class of CFTs.

\section{Generalized shockwaves in AdS}
\label{sec:Shockwaves}

In this section, we discuss the gravitational counterpart of the generalized ANEC operators in holographic CFTs. It is not surprising that linearized solutions, representing the insertion of generalized ANEC operators in the boundary, can be obtained both by solving the Einstein equations directly or by performing the bulk dual of the conformal transformations discussed for the CFT formulation. Maybe more surprisingly, we present exact solutions in the bulk which contain the information of higher $n$-point correlators in an analogous fashion to known results for the ANEC operator \cite{Hofman:2008ar,Hofman:2009ug}. This is a consequence of the collinear algebra explored in section \ref{sec:algebra}. Even more unexpectedly, there exists a gauge where the linearized solutions can be made exact. We explain why this is the case and why it fails for the shock dual to the $L_0$ operator.

Using these results we discuss correlators and commutators in the gravitational theory. We see that the naive commutation relations found in free theories fail to materialize in this setup. This is expected from the fact that the gravitational theory is only keeping track of a single block running between shockwaves in the scattering process. As shown in section~\ref{sec:confblock} this is also the case in the CFT if only one block contributes. Non-perturbative corrections at finite $N$ should presumably fix this glitch. We comment on this in the conclusions.

\subsection{AdS isometries and collinear transformations}
Throughout this section, we will work with AdS in Poincar\'{e} coordinates with metric
\be \label{Poincare}
ds^2=\bar{g}_{\mu\nu}dx^\mu dx^\nu = \frac{dz^2-dx^+dx^- +d\vec{x}^{\perp2}}{z^2/\ell^2} \,.
\ee
We will also set $\ell=1$ for convenience. As explained in sec.~\ref{sec:algebra}, the collinear transformations that map a light-ray onto itself are generated by a scale transformation, translations in the $x^-$ direction, and a special conformal transformation. While the scale and translational isometries of the metric \rref{Poincare} are manifest, the special conformal transformations are less obvious. They can be obtained by the usual inversion/translation/inversion procedure. In the bulk inversion is simply $z^\mu\to z^\mu/(z^\nu z_\nu)$ \cite{Freedman:1998tz}. Therefore, the most general collinear transformation in the bulk is given by
\begin{align} \label{isometriescol}
x^-&\to\quad  \frac{ax^-+b}{c\xm+d}\, ,& x^+&\to\quad  x^+ -c\frac{z^2+|\vec{x}^\perp|^{2}}{c\xm+d}\, ,  \\
z&\to\quad  \frac{z}{c\xm+d} \, , & x^i&\to\quad  \frac{x^i}{c\xm+d} \, ,\notag 
\end{align}
with $ad-bc=1$. In what follows, the $S$ transformation will be particularly relevant so we give it here explicitly
\begin{align} \label{Stransform}
x^-&\to\quad -\frac{1}{\xm}\, , & x^+&\to\quad  x^+ -\frac{z^2+|\vec{x}^\perp|^{2}}{\xm} \, ,\\
z&\to\quad  \frac{z}{\xm}\, , & x^i&\to \quad \frac{x^i}{\xm}\, .\notag
\end{align}
As we will see, the isometries \rref{isometriescol} can be used to obtain new linearized shockwave solutions corresponding to generalized ANEC operators from known exact solutions (e.g.~the usual ANEC shockwave). Moreover, the  $S$ transformation \rref{Stransform} can in some cases be further exploited to obtain new exact solutions.  We explain this in detail below.

\subsection{Review of AdS shockwaves}

Shockwave geometries  \cite{Aichelburg:1970dh,Dray:1984ha,DEath:1992plq} (see \cite{Hofman:2009ug} for an AdS perspective) are solutions to the Einstein equations of the form
\be \label{metricform}
g_{\mu\nu}=\bar{g}_{\mu\nu}+\delta g_{\mu\nu}=\bar{g}_{\mu\nu}+\epsilon h_{\mu\nu} \,,
\ee
where $\bar g_{\mu\nu}$ is the AdS metric. They have remarkable properties: they are full non-linear (i.e.~to all orders in $\epsilon$) solutions to Einstein's equations and remain so even when higher-derivative terms are added to the action. The shockwave solution in AdS takes the form
\be \label{usualshockwave}
h_{++}=\mathcal{H}(x^+) \frac{z^2}{(z^2+|\vec{x}^\perp|^{2})^3} \,,
\ee
for any function $\mathcal{H}(x^+)$. In what follows we will be mostly interested in the case where the shock is localized to a null-plane, namely 
\be
\mathcal{H}(x^+)=\delta(\xp) \,.
\ee
From the form of the metric, it is manifest that for any $\vec{x}^\perp\neq0$, there is no source turned on. At $\vec{x}^\perp=0$, there is a source for the stress-tensor with delta function support in $\xp$.\footnote{Note that because the source is localized in the $\vec{x}^\perp$ directions it is contained in the $z^{-4}$ term of the metric rather than the ordinary $z^{-2}$.} The CFT operator that couples to the $++$ component of the metric is $T_{--}$ and so this metric describes the insertion of the following operator in the path integral
\be
S_{\text{CFT}}\to S_{\text{CFT}}+ \epsilon \int d^4x \  T_{--} \delta(\xp)\delta^{(2)}(\vec{x}^\perp) \,.
\ee
In other words, shockwave geometries are the gravitational duals of exponentiated ANEC operators. These shocks can be superposed non-linearly giving access to higher $n$-point functions for the ANEC operators by taking the appropriate $\epsilon$ derivatives. Alternatively we can compute correlators of ANEC operators by propagating wavefunctions for particular states in the bulk of AdS and expanding the result in a power series in $\epsilon$. 

One perspective that we will exploit further below to obtain more general solutions in Einstein gravity is to understand the action of the collinear conformal group on this operator. When the source is localized in the $x^+$ direction this can be of great help. Concretely, the operator above is invariant under $x^-$ translations and has collinear twist $\bar J_0 =1$ and collinear weight  $J_0=2$ as can be seen from the transformations in section \ref{sec:algebra}. It is easy to see that the most general ansatz for the metric given these properties is:

\be
\delta g =  \epsilon \frac{\delta(x^+)}{z^4} f\left(\zeta \right) \,dx^+ dx^+\, , \quad\quad \text{with}\quad\quad \zeta=\frac{|\vec{x}^\perp|^2}{z^2}\, .\
\ee
This scaling ansatz turns Einstein equations into an ordinary differential equation that can be solved to yield (\ref{usualshockwave}), $f=\frac{1}{(1+\zeta)^3}$. This procedure, for sources localized in $x^+$ can be generalized to obtain exact solutions for other generalized ANEC sources. We explain these techniques in detail below.

\subsection{Generalized shocks}\label{gensho}

In this section, we write down explicit metrics corresponding to the insertion of generalized ANEC operators $L_{-1}$, $L_0$, $L_1$ and $L_2$. We will see that the story for $L_0$ is more intricate and discuss this fact. We present a linearized solution in this case. The most efficient method to obtain these metrics depends on the particular operator at hand. We consider each one individually. 

\subsubsection*{$L_{-1}$ Shocks}

Here, a form of the scaling ansatz above produces an exact solution. We are searching for a metric which corresponds to the insertion of the following operator in the path integral
\be\label{graviL-1}
 \epsilon \int d^4x \, \xm \,  T_{--} \delta(\xp)\delta^{(2)}(\vec{x}^\perp) \,.
\ee
This means we want to find a metric with collinear twist $\bar J_0 =1$ and collinear weight  $J_0=1$. The most general ansatz with this scaling that remains regular in $x^-$ is 

\bea
\delta g &=& \epsilon \left(x^- \frac{\delta(x^+)}{z^4} f\left(\zeta \right)+\frac{\delta'(x^+)}{z^2} q\left(\zeta \right) +\frac{\delta(x^+)^2}{z^2} s\left(\zeta \right)  \right) dx^+ dx^+ \nonumber\\
&+& \epsilon^2 \left(\frac{\delta'(x^+)}{z^4} t\left(\zeta \right) +\frac{\delta(x^+)^2}{z^4} r\left(\zeta \right)  \right)  \,dx^+ dx^+ \nonumber\\
&+&\epsilon \frac{\delta(x^+)}{z^4}  k\left(\zeta \right)  \vec{x}^\perp \cdot d\vec{x}^\perp  dx^+ + \epsilon \frac{\delta(x^+)}{z^4}  g\left(\zeta \right)  z\, dz \,  dx^+ \, , \label{metric-1}\\
&\text{with}& \quad \quad \zeta=\frac{|\vec{x}^\perp|^2}{z^2}\, .
\eea
Notice that in this case the scaling properties allow an $\epsilon^2$ contribution to the exact shockwave solution. In principle this is the situation. It turns out that the functions $t\left(\zeta \right)$ and $s\left(\zeta \right)$ can be set to zero as they represent sources for independent ANEC operators.\footnote{An integrated ANEC operator in the case of $s$.} This solution has some gauge freedom. In particular the general form above is not in Fefferman-Graham gauge. We can remedy the situation right away by fixing $ g\left(\zeta \right)=0$. The system then becomes a fully determined system of coupled ODEs that can be solved explicitly. We can fix the integration constants by demanding that their only source is given by the boundary operator (\ref{graviL-1}). The procedure consists in demanding that away from $\vec{x}^\perp=0$ the metric contains no components which go as $\frac{1}{z^2}$. To fix the remaining constants we demand that integrating the source over the whole transverse plane yields the uniform shockwave solution (up to an overall normalization):
\be
\delta g_{uniform} \sim \epsilon \frac{x^-}{z^2} \delta(x^+) dx^+ dx^+\, .
\ee
The result of this procedure produces the exact solution:
\bea
f\left(\zeta \right) &=&\frac{1}{(1+\zeta)^3}\, ,\\
k\left(\zeta \right) &=&-\frac{1}{\zeta (1+ \zeta)^2}\, ,\\
q\left(\zeta \right) &=& -\frac{1}{2(1+ \zeta)^2} -\frac{1}{2 (1+ \zeta)} - \frac{1}{2} \log \frac{\zeta}{1+\zeta}\, ,\\
r\left(\zeta \right) &=& \frac{1}{2 (1+ \zeta)^5} +\frac{3}{4 (1+ \zeta)^4} +\frac{3}{8 (1+ \zeta)^3} \left(1+2\log \frac{\zeta}{1+\zeta}\right)\, .
\eea
As advertised, this solution is quadratic in $\epsilon$. This term is proportional to the square of a delta function in $x^+$. This feature might seem unpleasant. Notice, however, that this property is actually gauge dependent. We used our gauge freedom to go to Fefferman-Graham gauge but we could have just as easily used it to go to a gauge where $r\left(\zeta \right)=0$, yielding an exact linear in $\epsilon$ solution.\footnote{Actually, it is the inverse metric that determines the propagation of waves in this AdS geometry and the actual source for the energy-momentum tensor, non-linearly. One could try instead to go to coordinates where the $\delta(x^+)^2$ term disappears from $g^{-1}$.} We will discuss this momentarily. In any case, calculations that depend only on the linear properties in $\epsilon$ will not be sensitive to the $\delta(x^+)^2$ term.

%


More worrisome seems to be the case that this solution has a coordinate singularity at $\vec{x}^\perp=0$ visible in $k\left(\zeta \right)$. This singularity is solely due to the choice of coordinates, which can be verified by computing the Kretschmann scalar $R_{abcd}R^{abcd}=40$, just like empty AdS. In any case this coordinate singularity is mild and disappears if one integrates the source over a small area. Carrying out this procedure is useful if one wants to explore the properties of this metric near $\vec{x}^\perp=0$. As expected, if we do so in a rotationally invariant manner, we obtain that the smeared value of the $dx^i \, dx^+$ component of the metric is indeed zero at $\vec{x}^\perp=0$. The upshot is that this singularity can also be gauged away by a change of coordinates as we now show. 

Lastly, one might be interested in going to coordinates where the term proportional to $\delta'(x^+)$ disappears. This is particularly useful when computing scattering past this shock. If we are only interested in the insertion of the dual operator in a correlation function we only care about terms that are linear in $\epsilon$. Looking at the metric (\ref{metric-1}), this is the only offensive term that might complicate the calculation, see section \ref{sec:superimposeshocks}.

We can deal with all these issues simultaneously by considering the following change of coordinates which is consistent with our scaling Ansatz:
\be
x^- \rightarrow x^- - \epsilon \, \delta(x^+) \, a(\zeta) \, , \quad \quad \text{with} \quad \zeta=\frac{|\vec{x}^\perp|^2}{z^2}\, .
\ee
Under this change of coordinates:
\bea
f\left(\zeta \right) &\rightarrow& f\left(\zeta \right)\, ,  \\
k\left(\zeta \right) &\rightarrow& k\left(\zeta \right) + 2 a'\left(\zeta\right)\, , \\
g\left(\zeta \right) &\rightarrow& g\left(\zeta \right) - 2 \zeta a'\left(\zeta\right)\, , \\
q\left(\zeta \right) &\rightarrow& q\left(\zeta \right) + a\left(\zeta\right)\, , \\
r\left(\zeta \right) &\rightarrow& r\left(\zeta \right) - f\left(\zeta \right) \, a\left(\zeta\right) \, .
\eea
Having figured out the sources in the more physical Fefferman-Graham gauge, the transformations above allow us to go other useful gauges. As promised above, it is trivial to see that the choice
\be
a=  \frac{1}{2 (1+ \zeta)^2} +\frac{3}{4 (1+ \zeta)} +\frac{3}{8} \left(1+2\log \frac{\zeta}{1+\zeta}\right)\, ,
\ee
takes us to an exact linear in $\epsilon$ solution. As we will see momentarily this procedure extends to the $L_1$ and $L_2$ shockwaves while it fails for $L_0$.

More useful for our purposes will be the choice
\be
a= \frac{1}{2(1+ \zeta)^2} +\frac{1}{2 (1+ \zeta)} + \frac{1}{2} \log \frac{\zeta}{1+\zeta} \, .
\ee

In these coordinates we simultaneously remove the $\delta'(x^+)$ term in the metric and the artificial singularity at $\vec{x}^\perp=0$. The price to pay was to depart our beloved Fefferman-Graham gauge. As we will use this metric in our scattering experiments we quote the result below for the new metric components.
\bea
f\left(\zeta \right) &=&\frac{1}{(1+\zeta)^3}\, ,\\
k\left(\zeta \right) &=&-\frac{2}{ (1+ \zeta)^3}\, ,\\
g\left(\zeta \right) &=& \frac{\zeta-1}{(1+\zeta)^3} \, ,\\
q\left(\zeta \right) &=&0\, ,\\
r\left(\zeta \right) &=& \frac{1}{4 (1+ \zeta)^4} +\frac{3}{8 (1+ \zeta)^3} \left(1+\frac{2}{3}\log \frac{\zeta}{1+\zeta}\right)\, .
\eea

\subsubsection*{$L_{0}$ Shocks}

The $L_0$ shocks turn out to be the most complicated. We have not found an explicit solution in this case. The reason is that if one imposes $J_0=0$ and $\bar J_0=1$ scaling, the resulting dimensions for $\epsilon$ allow the appearance of factors of the form $\epsilon^n \delta^{(n)}(x^+)$ or  $\epsilon^n \left(\delta(x^+)\right)^n$  for any $n>0$. Our scaling ansatz is therefore not guaranteed to produce a solution at a finite order in $\epsilon$. 

While the most general solution will certainly not be linear in $\epsilon$ one can hope that there is a gauge where that is possible, as was the case for $L_{-1}$. This can be checked explicitly. We have done so and found only complex solutions to this order.  One can hope that introducing terms quadratic in $\epsilon$ the equation can be solved for a real metric.

There is also a good argument to explain why we would not expect exact solutions that are linear in $\epsilon$ for $L_0$. Remember that $L_{\{-2,-1,0,1,2\}}$ form a multiplet under the action of the $SL(2)$ generated by $J_{\{-1,0,1\}}$. The light-ray operators transform in a five-dimensional representation of the collinear group. One can compute the invariant $SL(2)$ norm for this (non-unitary) representation. In our conventions the transformation properties of the $L$'s under the action of finite $SL(2)$ group elements is given by \eqref{eq:GeneralAnecTrafo}. The invariant norm of a generic vector

\be
v = \sum_{i=-2}^2 a_{i} L_{i} 
\ee
is given by

\be \label{norm}
 |v|^2 = \left(a_0^2 -\frac{4}{3}a_1 a_{-1} + \frac{1}{3}a_2 a_{-2}\right) \, .
\ee

Now, the fact that an exact solution can truncate to linear order is directly related to the associated vector in the algebra squaring to zero, suppressing higher order corrections. We see in the above expression that while $L_{-2}$, $L_{-1}$, $L_1$ and $L_2$ are indeed lightlike, $L_0$ has a non-zero norm. This explains the lack of exact linear solutions in this case, but gives us hope that this will be possible for $L_1$ and $L_2$. We will confirm this expectation shortly.

An interesting comment relates to CFTs in odd dimensions. Here we expect the multiplet of $L$'s to lie in an even-dimensional representation of $SL(2)$. In this case, we expect there always exists a complete basis of null-operators. Therefore, we expect, for example, that  all  associated shocks in AdS$_4$ can be made linear in $\epsilon$.

While we leave for future work the task of finding an exact solution we now present a linearized (i.e.~non-exact) solution representing the $L_0$ shock. While this could be obtained by brute force, we present here a method based on the $SL(2)$ algebra that will be crucial to find exact solutions for $L_1$ and $L_2$.

The $SL(2)$ algebra acts on a vector space and its action is therefore linear. This is very clear in the CFT as can be seen in Fig.~\ref{fig:LightRayTower}. In the gravitational setting this translates to the fact that the algebra cannot act directly on the space of bulk solutions which can be non-linear and, hence, do not manifestly exhibit the properties of a vector space. Of course, this would be the case for linearized solutions but we will learn something by thinking about the action of the group on exact solutions.

Given an exact solution, we can generate a new one by the action of a finite symmetry transformation parameterized by an $SL(2)$ group element. Notice that this technique only allows us to access solutions within the same conjugacy class. As we can see by the form of the norm \rref{norm}, solutions sourced by the $L_0$ operator have to necessarily be in a different conjugacy class than those sourced by $L_{-2}$ and $L_{-1}$. This problem does not affect $L_1$ and $L_2$. In the next subsection we will use this method and the S transformation \rref{Stransform} to obtain those solutions. Here we will have to content ourselves with a linear solution.

We act with a one-parameter family of $SL(2)$ transformations connected to the identity on an exact solution that is linear in $\epsilon$, for example the one sourced by $L_{-2}$. Notice that this produces a family of exact solutions, still linear in $\epsilon$. We can expand this solution in powers of the parameter. If the transformation is generated by, say  $e^{\lambda J_1}$, then we will have a solution of the form
\be
g= \bar g + \epsilon h + \epsilon \lambda h_1 + \epsilon \lambda^2 h_2 + \cdots
\ee
This must be an exact solution, so it must also be a linear solution. Furthermore, it must be a solution for all $\lambda$ which means that each $h_j$ for $j=1,2,\cdots$ must be a linear perturbation that solves Einstein equations at this order. Matching to the expansion of $e^{\lambda J_1}$ we learn that:
\be
h_j = \left(J_1\right)^j(h)\, .
\ee
One can check explicitly that each one of these solutions has the following source at the boundary (if $h$ corresponds to the $L_{-2}$ source \rref{usualshockwave}):
\be
\text{source}(h_j) = [ J_1,\text{ source}(h_{j-1})] \, .
\ee
Because the $L_n$'s fall in a five-dimensional representation, after acting with $J_{1}$ five times we obtain solutions that have no sources at the boundary and are pure gauge. This gives us an efficient method to obtain linearized solutions for all $L_n$'s. Here we quote the linearized solution for an $L_0$ source:

\bea
h_{zz} &=&  \frac{4 \delta(x^+)}{z^2(1+\zeta)^3}\, ,\\
h_{z+} &=& - \frac{6 x^- \delta(x^+)}{z^3 (1+\zeta)^3} +  \frac{2  \delta'(x^+)}{z (1+\zeta)^2}\, ,\\
h_{z i} &=&  \frac{4 x^i \delta(x^+)}{z^3 (1+\zeta)^3}\, ,\\
h_{++} &=&  \frac{6 \left(x^-\right)^2 \delta(x^+)}{z^4 (1+\zeta)^3} - \frac{3 x^- \delta'(x^+)}{z^2 (1+\zeta)^2}+ \frac{ \delta''(x^+)}{2 (1+\zeta)}\, ,\\
h_{+-} &=&  \frac{ \delta(x^+)}{z^2 (1+\zeta)^2}\, ,\\
h_{+i} &=&  -\frac{6 x^i  x^-\delta(x^+)}{z^4 (1+\zeta)^3} +\frac{2 x^i \delta'(x^+)}{z^2 (1+\zeta)^2}\, ,\\
h_{ij} &=&  \frac{4 x^i  x^j \delta(x^+)}{z^4 (1+\zeta)^3}\, .
\eea
In the following section we will obtain exact solutions by acting with the $S$ transformation \rref{Stransform} on the $L_{-2}$ and $L_{-1}$ shocks. Notice that the exact solution for $L_0$ must be self-dual under $S$. As expected, we have failed in finding an exact solution linear in $\epsilon$ with this property but hopefully this fact can be used to find a solution containing higher orders in $\epsilon$. We leave this for future work.

\subsubsection*{$L_{1}$ and $L_2$ Shocks}

Having learned our lessons in the previous cases we are now ready to obtain exact linear in $\epsilon$ solutions for $L_{1}$ and $L_2$ sources. As for $L_0$, the scaling Ansatz fails to produce a finite order in $\epsilon$ guess\footnote{Although it gives us a systematic way to organize the solution order by order in $\epsilon$.} for the solution. We will instead use the transformation properties of the generalized ANEC operators under the collinear transformations \eqref{eq:GeneralAnecTrafo} as was done in the previous case. Here, however, we can use the finite group element corresponding to the $S$ transformation \rref{Stransform} to obtain exact solutions. This is possible as $L_1$ and $L_2$ are, correspondingly, in the same orbit as $L_{-1}$ and $L_{-2}$. The procedure is straightforward and metrics for both $L_1$ and $L_2$ share similar properties. Here we quote the $L_2$ metric only for succinctness, starting from the usual $L_{-2}$ shock \rref{usualshockwave}

\bea \label{L2metric}
h_{zz}&=&4 \delta\left(x^+- (1+\zeta)\frac{z^2}{\xm}\right) \frac{(x^-)^2 }{z^2 (1+\zeta)^3}\, , \notag \\
h_{z+}&=&-2 \delta\left(x^+- (1+\zeta)\frac{z^2}{\xm}\right)\frac{(x^-)^3}{z^3 (1+\zeta)^3} \, ,  \notag \\
h_{z-}&=&-2 \delta\left(x^+- (1+\zeta)\frac{z^2}{\xm}\right)\frac{x^- }{ z (1+\zeta)^2} \, ,\notag \\
h_{zi}&=&4 \delta\left(x^+- (1+\zeta)\frac{z^2}{\xm}\right) \frac{x^i(x^-)^2 }{ z^3(1+\zeta)^3} \, ,  \notag \\
h_{++}&=&\delta\left(x^+- (1+\zeta)\frac{z^2}{\xm}\right) \frac{(x^-)^4 }{ z^4(1+\zeta)^3}  \, , \notag \\
h_{+-}&=&\delta\left(x^+- (1+\zeta)\frac{z^2}{\xm}\right) \frac{(x^-)^2 }{ z^2(1+\zeta)^2} \, ,   \\
h_{+i}&=&-2 \delta\left(x^+- (1+\zeta)\frac{z^2}{\xm}\right) \frac{x^i(x^-)^3 }{ z^4(1+\zeta)^3}\, ,   \notag \\
h_{--}&=&\delta\left(x^+- (1+\zeta)\frac{z^2}{\xm}\right) \frac{1}{ (1+\zeta)}  \, , \notag \\
h_{-i}&=&-2\delta\left(x^+- (1+\zeta)\frac{z^2}{\xm}\right) \frac{x^i \xm }{ z^2 (1+\zeta)^2}\, ,   \notag \\
h_{ij}&=& 4\delta\left(x^+- (1+\zeta)\frac{z^2}{\xm}\right) \frac{x^i x^j (x^-)^2  }{z^4 (1+\zeta)^3} \, . \notag
\eea
The first surprise is perhaps that the shockwave is no longer localized directly on a null-plane everywhere in the bulk. This is a natural consequence of the way that we obtained the metric through a change of coordinates. Note however that at the boundary, the source is still localized on the light-ray $x^+=0$, $\vec{x}^\perp=0$ . The functional form of the metric restricts support to $\vec{x}^\perp=0$. In the bulk, the situation is more complicated. Notice that if we expanded the delta function in its derivatives we would obtain all powers of $\left(\frac{z^2}{x^{-}}\right)^{n} \delta^{(n)}(x^+)$ which is dimensionless by our scalings. In our previous scaling ansatz this was forbidden by demanding only positive powers of $x^-$ appear in the solution. Once a term like this shows up, the expansion cannot truncate to preserve regularity. There might be other gauges where only regular terms appear and truncate to finite order in $x^-$. The price to pay, however, will probably be the inclusion of arbitrarily high powers of $\epsilon$. It would be nice to see if a simpler solution exists. 

\begin{figure}[t!]
\centering
\includegraphics[width=0.45\textwidth]{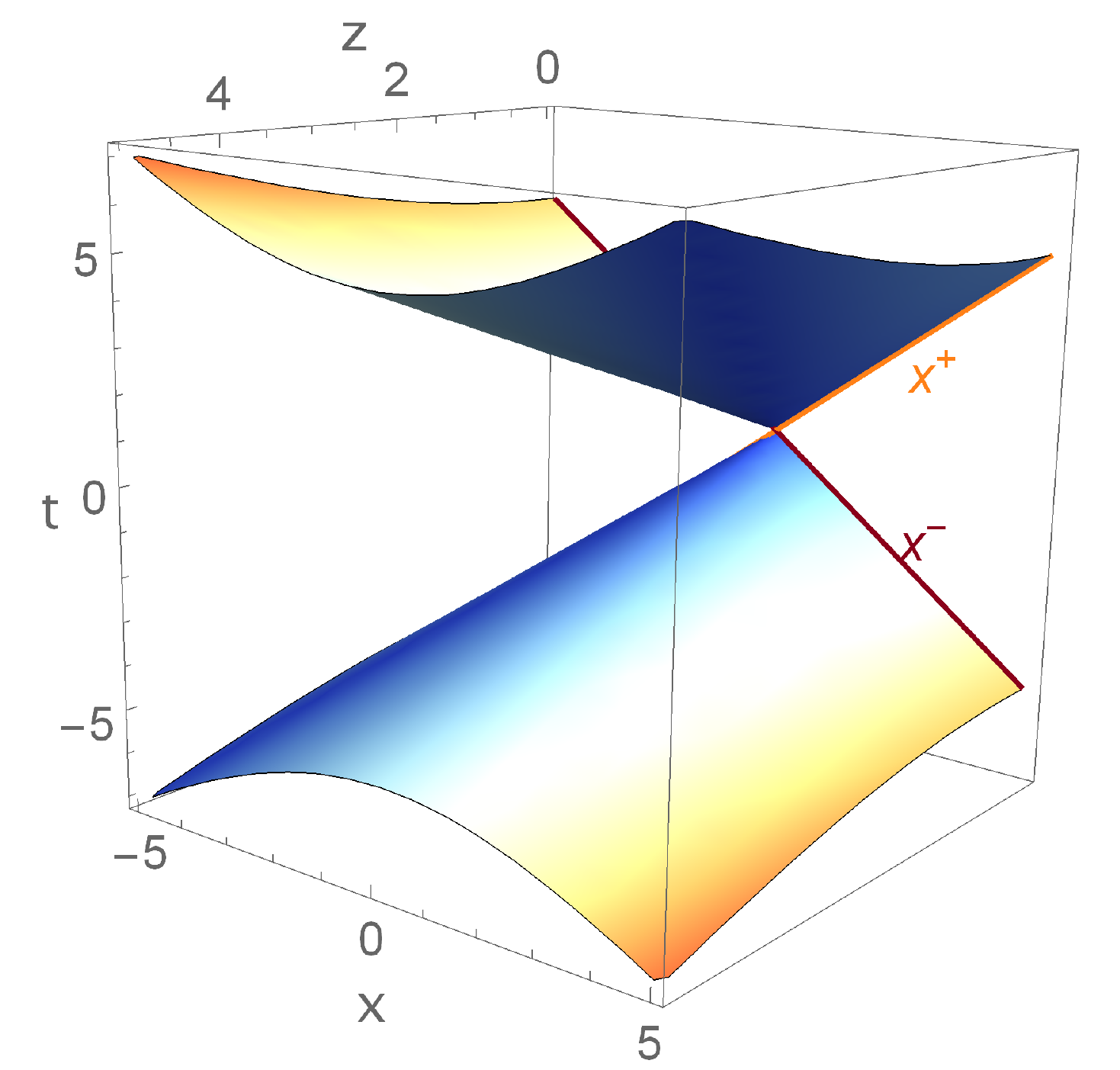}
\caption{The support of the shock inside the bulk of AdS, plotted at $\vec{x}^\perp=0$. The magnitude of the delta function is $\xm$-dependent such that it vanishes (this is displayed in blue) on the lightlike directions $x^+$ at the boundary.}
\label{hyperbolas}
\end{figure}

To see the way the shockwave propagates in the bulk, it is slighlty more convenient to rewrite
\be
\delta\left(x^+-\frac{z^2+|\vec{x}^\perp|^2}{\xm}\right) = |\xm | \delta\left(-x^+\xm+z^2+|\vec{x}^\perp|^2\right) \,,
\ee
and we see that the shock lies on two hyperbolae, i.e.~at all points null-separated from the origin, but that the delta function has a magnitude that is $\xm$ dependent. This is represented in Fig.~\ref{hyperbolas}.

Note that the fact that the magnitude of the delta function is $\xm$ dependent is what allows the limit towards the boundary to give the correct result yielding support on a single light-ray of the light-cone at the boundary.

\subsection{Superposing shocks \label{sec:superimposeshocks}}

Having obtained shockwave solutions, we are now ready to discuss the superposition of them and the propagation of particles in these backgrounds. This procedure will allow us to compute correlators in the holographic setup. In some cases (i.e.~when the support of the shockwaves does not overlap in the bulk), it is trivial to obtain exact solutions by linear superposition of the shockwaves discussed in the previous section. For example two usual $L_{-2}$ shockwaves can be trivially superposed by placing them at different $x^+$ and $\vec{x}^\perp$ positions as:

\be\label{doublemetric}
\delta g=   \epsilon_1 \frac{\delta(x^+)}{z^4} f\left(\frac{|\vec{x}^\perp|^2}{z^2} \right) \,dx^+ dx^+ +  \epsilon_2 \frac{\delta(x^+-y^+)}{z^4} f\left(\frac{|\vec{x}^\perp-\vec{y}^\perp|^2}{z^2} \right) \,dx^+ dx^+\, .
\ee

Furthermore, since the solution above is completely smooth one can easily take the limit $y^+ \rightarrow 0$ to obtain a shockwave localized on a single null plane. The limit is clearly independent of the sign of $y^+$, which determines the time ordering of these perturbations for an incoming particle. This might make us think that the commutator of the sources immediately vanishes as a consequence. While this is true for the case above, we will see explicitly that this reasoning is incorrect for more general sources. In particular we will use the results above to compute the commutator $[ L_{-2}, L_{-1}]$.\footnote{More precisely, we will compute this commutator in states created by scalar operators.}

Let us now go over these computations for three different examples of shockwave superpositions: $L_{-2} \oplus L_{-2}$, $L_{-2} \oplus L_{-1}$ and $L_{2} \oplus L_{2}$.

\subsubsection*{ $L_{-2} \oplus L_{-2}$ superposition}

It is known that ANEC operators commute \cite{Kologlu:2019bco}. This was already observed in the gravitational setting in \cite{Hofman:2008ar}. This fact has important consequences for the space of allowed gravitational theories  \cite{Belin:2019mnx}. Let us review this computation following mostly~\cite{Hofman:2008ar}.

We consider scalar perturbations in a shockwave background created by $L_{-2}$ insertions. All we need is the form of the Laplace operator in this curved space\footnote{We disregard mass terms in this discussion as they play no role.}. Consider first the metric \rref{doublemetric} with $\epsilon_2$ turned off
\be
4 \partial_- \partial_+ \phi + \frac{3}{z} \partial_z \phi - \partial_z^2 \phi - \nabla^2 \phi + 4 \epsilon_1  \frac{\delta(x^+)}{z^2} f\left(\frac{|\vec{x}^\perp|^2}{z^2} \right) \partial_-^2 \phi =0\, ,
\ee
where $\nabla^2$ is the Laplace operator in flat transverse space $\vec{x}^\perp$. This equation can be solved exactly. Away from the shock, the equation is trivially solved by AdS evolution. At the shock we just need to integrate across the delta function. There, the only coordinate that varies rapidly is $x^+$, so we can disregard regular terms that do no involve $x^+$ derivatives. Integrating the resulting equation we obtain:

\be
\partial_-\phi(x^+=0^+) = e^{ - \frac{\epsilon_1}{z^2} f\left(\frac{|\vec{x}^\perp|^2}{z^2} \right) \partial_-} \partial_-\phi(x^+=0^-) \, .
\ee
It turns out this is all we need to compute correlators of $L_{-2}$ insertions in scalar states. Assume we know the wave function corresponding to scalar states on the null surface $x^+=0$. Then the expectation value of the exponentiated ANEC operator is computed as \cite{Hofman:2008ar}:

\be
\left\langle \phi_{out} | e^{\epsilon L_{-2}} | \phi_{in} \right\rangle \sim  \int dx^- \frac{dz \, dx^1 dx^2}{z^3} i \phi^*_{out}e^{\epsilon L_{-2}} \partial_- \phi_{in}+ c.c. \, .
\ee
Notice that this expression amounts to the integral over the light-ray parametrized by $x^-$ and the three dimensional hyperboloid given by $(z,x^1,x^2)$. We write the symbol $\sim$ as we are disregarding overall normalizations that can be obtained easily by knowing the charges of the states involved. We will be mostly interested in the transverse space dependence of the observables above.

If one is interested in the expectation value of $L_{-2}$ all one needs to do is to expand the expression above and keep only the linear term in $\epsilon$. Concretely:
\be
\left\langle \phi | L_{-2}(\vec{y}^\perp) | \phi \right\rangle \sim  \int dx^- dz \, dx^1 dx^2 i \phi^*  \frac{z}{\left(z^2 + |\vec{x}^\perp-\vec{y}^\perp|^2\right)^3}\partial_-^2 \phi+ c.c.\, ,
\ee
where above we have inserted the operator $L_{-2}$ at an arbitrary position $\vec{y}^\perp$ in transverse space. If we are interested in computing this for conformal collider experiments where we imagine the shockwave is sourced at the conformal boundary of Minkowski space, this calculation amounts to the computation of the energy flux at infinity. The wave functions for scalar states with definite timelike momentum $q^0$ are delta-function localized in the hyperboloid at $z=1$ and $\vec{x}^\perp=0$ and are plane wave-like in $x^-$ going as $e^{i q^0 x^-}$, see \cite{Hofman:2008ar}. In this case we obtain:
\be\label{L-2flat}
\left\langle \phi | L_{-2}(\vec{y}^\perp) | \phi \right\rangle \sim \frac{1}{(1+|\vec{y}^\perp|^2)^3}\, .
\ee
where we have stripped above overall coefficients not depending on $\vec{y}^\perp$. The map between the transverse coordinates and the $S^2$ at infinity in the collider experiment is:
\be
y^1 = \frac{n^1}{1+ n^3}\, , \quad\quad y^2= \frac{n^2}{1+n^3}\, , \quad\quad \text{with} \quad  \left(n^1\right)^2+\left(n^2\right)^2+\left(n^3\right)^2=1\, ,
\ee
and the surface elements are related by:
\be
d^2y^\perp = \frac{d^2\Omega}{(1 + n^3)^2} \, .
\ee
This implies that the operators on the plane and the sphere are related as\cite{Hofman:2008ar}:

\be
 L_{-2}(\vec{y}^\perp)=  (1 + n^3)^3 \mathcal{E}(n^i) \, .
\ee
The power of 3 above can be understood as coming from the fact that the ANEC operator has collinear twist 1 adding to the two powers coming from the transformation of the measure.

Plugging these results in \rref{L-2flat} above we find
\be
\left\langle \phi | \mathcal{E}(n^i) | \phi \right\rangle \sim 1\, .
\ee
The result is independent of the angle in the celestial sphere, as it should be for a scalar operator evaluated on a scalar state. The actual normalization is $\frac{q^0}{4\pi}$ to reproduce the total energy of the state  upon integration.

We can now tackle the insertion of two shocks as in \rref{doublemetric}. We will be interested in the computation of the commutator $[L_{-2}(\vec{y}^\perp), L_{-2}(0)]$ so we will be considering shockwaves inserted at an infinitesimal distance from each other in light-cone time $y^+$. Notice that while the metric is smooth under $y^+ \rightarrow 0$, the solution for the propagation of perturbations on top of it depends generically on the ordering of the shocks. This is because the formal solution to the Laplace equation across the shock is 
\be
\partial_-\phi(x^+=0^+) = e^{ - \frac{\epsilon_2}{z^2} f\left(\frac{|\vec{x}^\perp-\vec{y}^\perp|^2}{z^2} \right) \partial_-} e^{ - \frac{\epsilon_1}{z^2} f\left(\frac{|\vec{x}^\perp|^2}{z^2} \right) \partial_-} \partial_-\phi(x^+=0^-) \, .
\ee
This is completely analogous to  solutions in gauge theory given by path ordered exponentials. In this simple case, however, we see right away that the action of both exponential operators commute and the ordering is not important. Concretely,
\be
\ba
&\left\langle \phi | [ L_{-2}(\vec{y}^\perp), L_{-2}(0)] | \phi \right\rangle \\
& \qquad \sim \int dx^- \frac{dz \, dx^1 dx^2}{z^3} i \phi^*  \left[\frac{z^4}{\left(z^2 + |\vec{x}^\perp-\vec{y}^\perp|^2\right)^3}\partial_-,\frac{z^4}{\left(z^2 + |\vec{x}^\perp|^2\right)^3} \partial_- \right]\partial_- \phi+ c.c. \\
& \qquad \sim 0, 
\ea
\ee
as expected.

\subsubsection*{ $L_{-2} \oplus L_{-1}$ superposition}

Let us now perform the equivalent computation for this more interesting case. Here we consider the shockwave metric:
\begin{align}
\delta g&=\epsilon_1 \delta(x^+) \frac{|\vec{x}^\perp|^2-z^2}{(z^2 +|\vec{x}^\perp|^2)^3}   z\, dz \,  dx^+ - \epsilon_1 \delta(x^+)  \frac{2 z^2}{(z^2 +|\vec{x}^\perp|^2)^3}    \vec{x}^\perp \cdot d\vec{x}^\perp dx^+ \label{doublemetric2} \\
 &\phantom{=}+ \epsilon_1  \delta(x^+) x^- \frac{z^2}{\left(z^2 + |\vec{x}^\perp|^2\right)^3} \,dx^+ dx^+ +  \epsilon_2 \delta(x^+-y^+) \frac{z^2}{\left(z^2 + |\vec{x}^\perp-\vec{y}^\perp|^2\right)^3}\,dx^+ dx^+\,\nn\, .
\end{align}
We have chosen to represent the $L_{-1}$ shock in the coordinates where the $\delta'(x^+)$ term is absent. Furthermore we have only kept track of terms linear in $\epsilon_1$ as they will be the only ones of importance for the calculation at hand. 

For now, let us set $\epsilon_2 \rightarrow 0$ and consider the resulting Laplace equation:
\begin{align}
4 \partial_- \partial_+ \phi &+ \frac{3}{z} \partial_z \phi - \partial_z^2 \phi - \nabla^2 \phi \\
&=-4 \epsilon_1\delta(x^+)\frac{z^4}{(z^2+ |\vec{x}^\perp|^2)^3}\left( 1 + x^i \partial_i + x^- \partial_- - \frac{|\vec{x}^\perp|^2-z^2}{2 z^2} z \partial_z \right)\partial_- \phi\, .\nn
\end{align}
Going through the same steps as before (and moving the source to an arbitrary point $\vec{y}^\perp$) we can calculate:
\be
\ba
&\left\langle \phi | L_{-1}(\vec{y}^\perp) | \phi \right\rangle \sim\\
&\int dx^- dz \, d^2 x^\perp i \phi^*  \frac{z}{\left(z^2 + |\vec{x}^\perp\!-\!\vec{y}^\perp|^2\right)^3}\!\left( 1 \! + \! (x^i-y^i) \partial_i \! + \! x^- \partial_- \! - \! \frac{|\vec{x}^\perp\!-\!\vec{y}^\perp|^2\!-\!z^2}{2 z^2} z \partial_z \right)\!\partial_- \phi+ c.c.
\ea
\ee
Once again, considering a localized wave function on the hyperboloid and integrating by parts we get:
\bea
\left\langle \phi | L_{-1}(\vec{y}^\perp) | \phi \right\rangle \!\! &\sim& \!\! \left( \# - \partial_i \frac{(x^i-y^i)}{2}  + \partial_z \frac{|\vec{x}^\perp-\vec{y}^\perp|^2-z^2}{4 z}  \right)\frac{z}{\left(z^2 + |\vec{x}^\perp-\vec{y}^\perp|^2\right)^3}\Big|_{z=1,\vec{x}^\perp=0}\nonumber\\
&\sim& \!\! \frac{1}{(1+|\vec{y}^\perp|^2)^3}\, .
\eea
In the above computation, the delta-function localized momentum states need to be regularized to compute the action of $x^- \partial_-$ on the wave function. The physics is completely equivalent to that of the term including a derivative of the momentum delta function in \rref{L1holo}. For our purposes, it suffices to say that this term produces a constant, independent of $\vec{y}^\perp$. Fixing the normalization would amount to demanding that, upon integration over transverse space, the dilatation charge is reproduced. As in \rref{normdelta} the correct factor is $\frac{i \Delta}{4\pi}$.\footnote{While we have disregarded the mass of the bulk scalar in this section, as it does not affect the scattering of the shockwave, it does control the scaling of the wave function with the energy $q^0$. This is where the $\Delta$ originates in the normalization at hand.}

This result, once again, corresponds to a uniform flux in the celestial sphere. In this case, the associated charge is the Lorentzian boost symmetry in the plane $x^- \rightarrow \lambda^{-1} x^-$, $x^+ \rightarrow \lambda x^+$. Under a conformal transformation this symmetry maps to the dilatation symmetry in the conformal collider picture \cite{Hofman:2008ar}. This is also a scalar operator, so we do not expect any angular dependence.

It is now straightforward to compute the commutator of $L_{-1}$ and $L_{-2}$ shocks. 
\be
\ba
&\left\langle \phi | [ L_{-2}(\vec{y}^\perp), L_{-1}(0)] | \phi \right\rangle \\
&\sim\int \frac{dx^- d^2x^\perp dz}{z^3} \phi \! \left[\frac{z^4}{\left(z^2 \!+\! |\vec{x}^\perp\!-\!\vec{y}^\perp|^2\right)^3}\partial_-,\frac{z^4}{\left(z^2\! +\! |\vec{x}^\perp|^2\right)^3}\!\left(\! 1\! +\! x^i \partial_i\! +\! x^- \partial_-\! -\! \frac{|\vec{x}^\perp|^2\!-\!z^2}{2 z^2} z \partial_z\! \right)\! \right]\!\partial_- \phi \!+ \!c.c. \\
&\sim\int \frac{dx^- d^2x^\perp dz \,}{z^3} \phi \frac{z^6\left(2|\vec{x}^\perp|^2 |\vec{x}^\perp-\vec{y}^\perp|^2 +(|2 \vec{x}^\perp -\vec{y}^\perp|^2-2 |\vec{y}^\perp|^2 ) z^2 + 2 z^4\right) }{\left(z^2 + |\vec{x}^\perp|^2\right)^3\left(z^2 + |\vec{x}^\perp-\vec{y}^\perp|^2\right)^4}\partial_-^2 \phi+ c.c\label{comm12}\\
&\sim \frac{2 - |\vec{y}^\perp|^{ 2}}{(1 +|\vec{y}^\perp|^2)^4}\, .
\ea
\ee
This commutator is not zero in contrast to free field theory computations in previous sections of this work. It matches, however, the computation in conformal field theory when only one block propagates between the shockwaves in section \ref{sec:confblock}. A clear way to state this result is in conformal collider variables. Defining\footnote{All operators in the global five dimensional multiplet of generalized ANEC operators have the same conformal twist, so they pick up the same factor $ (1 + n^3)^3 $ when mapping to collider variables.}
\be
 L_{-1}(\vec{y}^\perp)=  (1 + n^3)^3 \mathcal{D}(n^i) \, ,
\ee
we obtain:
\be
\left\langle \phi | [  \mathcal{E}(m^i), \mathcal{D}(n^i)] | \phi \right\rangle \sim \left(1 + 3 \, \vec{m} \cdot \vec{n} \right)\, .
\ee

Once again, the normalization can be easily obtained by integrating over the $S^2$. The angular dependence however is striking and matches the CFT result \rref{resultconfblo}.

Notice that by looking at the result \rref{comm12} we see that, as far as the scalar field is concerned, one can consider the evolution across the commutator as provided by an effective metric. It is given by:
\be
\delta g =   \epsilon \delta(x^+) \frac{z^4\left(2|\vec{x}^\perp|^2 |\vec{x}^\perp-\vec{y}^\perp|^2 +(|2 \vec{x}^\perp -\vec{y}^\perp|^2-2 |\vec{y}^\perp|^2 ) z^2 + 2 z^4\right) }{\left(z^2 + |\vec{x}^\perp|^2\right)^3\left(z^2 + |\vec{x}^\perp-\vec{y}^\perp|^2\right)^4}\,dx^+ dx^+
\ee
This metric does not satisfy the Einstein equations. This implies that this commutator cannot be expressed in terms of sources for the boundary energy momentum components alone. An interesting direction here would be to compute the bulk energy momentum tensor that could support this solution. This way, one could understand if composite operators related to $\phi$ could account for this commutator by back-reacting on the metric. We will not pursue this here.

\subsubsection*{ $L_{2} \oplus L_{2}$ superposition}

\begin{figure}[t!]
\centering
\includegraphics[width=0.60\textwidth]{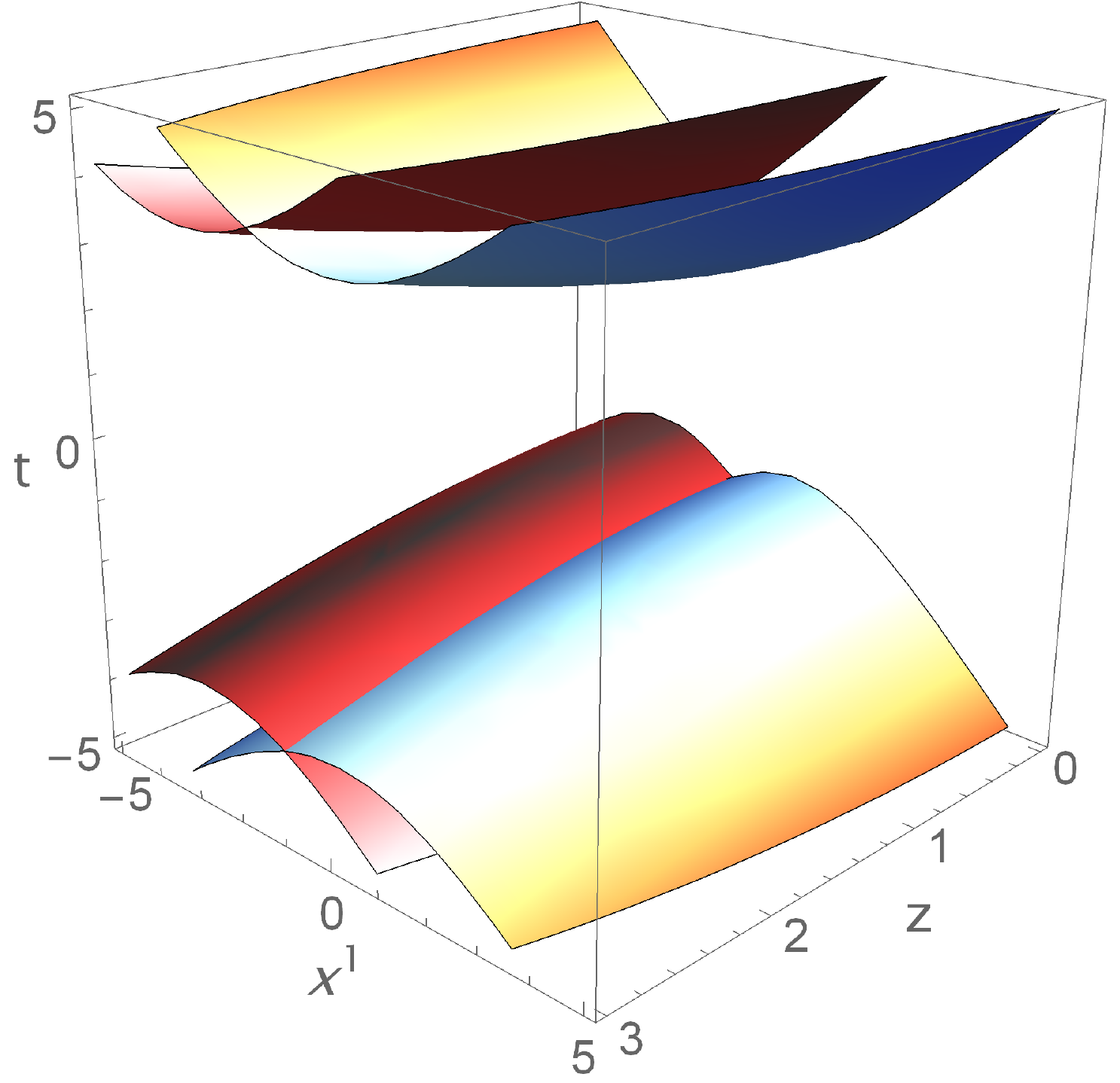}
\caption{Intersection of two $L_2$ shocks, where the sources are separated in the $x^+$ and $x^1$ directions. This is plotted at finite $x^3$ and $x^2=0$. }
\label{intesertshock}
\end{figure}

Here, we include a short discussion on the superposition properties of $L_2$ shocks. A similar discussion would apply to $L_1$ shocks as well. The novelty in this case is that two $L_2$ sources located at different $\vec{x}^\perp$ and $x^+$ do have intersecting support in the bulk as displayed in figure \ref{intesertshock}. 

\begin{figure}[t!]
\centering
\includegraphics[width=0.60\textwidth]{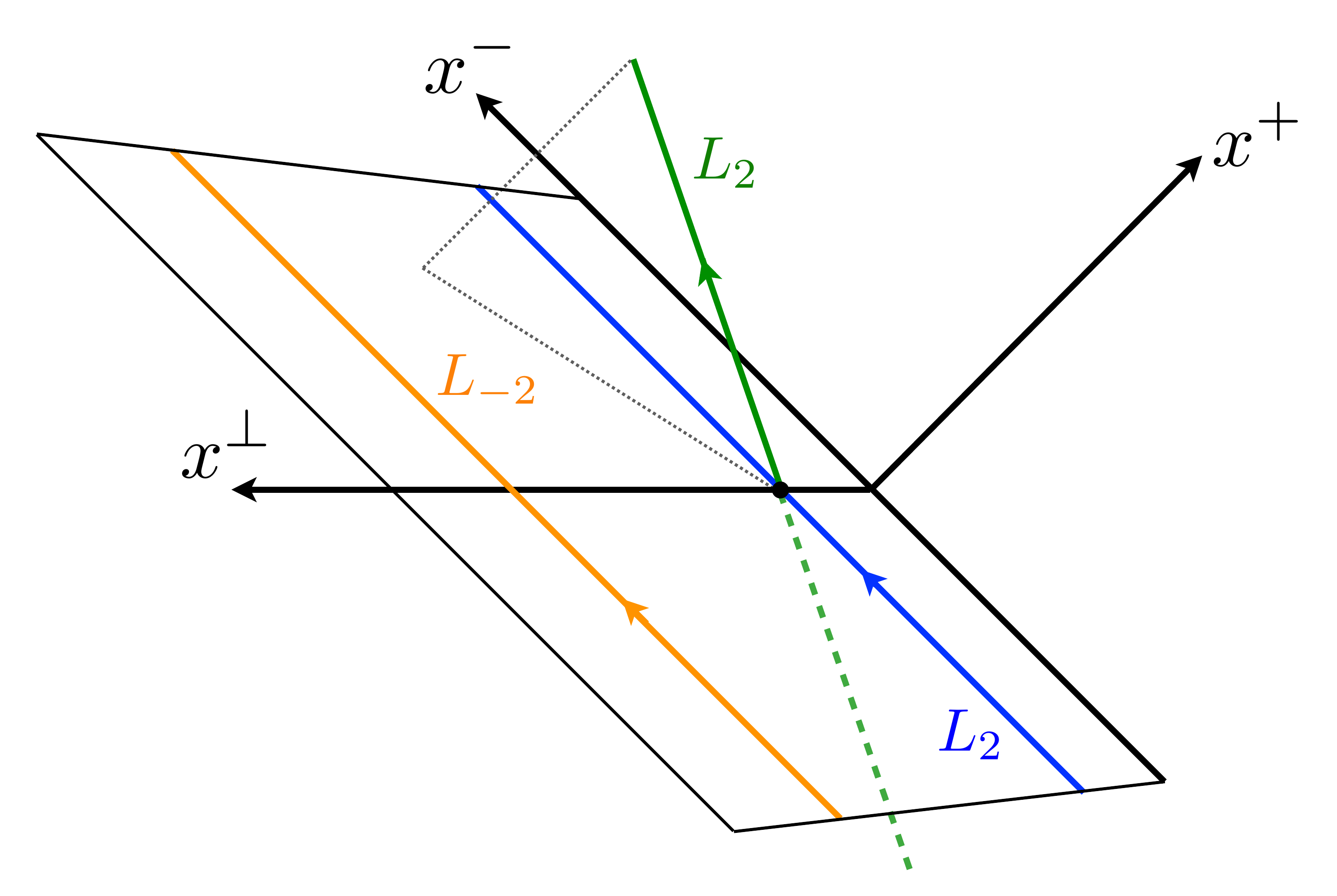}
\caption{The blue and orange rays represent two $L_{-2}$ operators located on the same null-plane. If we act with the $S$ transformation on this setup, the blue ray maps to itself, while the orange ray maps to the green one. We can see that it has left the null-plane and now intersects the blue ray at an angle. Since the original blue and orange $L_{-2}$ operators commute, the configuration after the $S$ transformation must also commute. }
\label{L2commuting}
\end{figure}

Therefore these solutions cannot be superposed. One could of course solve for the linear propagation of one shock on top of the other. This would be enough to repeat the type of calculations from the previous section. Let us focused instead in a particular type of configuration of these shocks that can be superposed. Let us start with the superposition of two usual $L_{-2}$ shockwaves
\be
\delta g = \delta(\xp) \left(\eps_1\frac{z^2}{(z^2+|\vec{x}^\perp|^2)^3}+\eps_2\frac{z^2}{(z^2+|\vec{x}^\perp-\vec{y}^\perp|^2)^3} \right) \, ,\ee
which is an exact solution of Einstein's equation. Now, we can apply the diffeomorphism \rref{Stransform}. The transformation of the shockwave inserted at $\vec{x}^\perp=0$ is the solution given in \rref{L2metric}. The part proportional to $\epsilon_2$ ends up being more complicated. Looking at the sources proportional to $\epsilon_2$ as $z\rightarrow 0$ , one can easily check that the new operator is located at
\bea
x^i&=&-y^i \xm \notag \\
\xp &=& |\vec{y}^\perp|^2 \xm \,.
\eea
The operator therefore intersects the previous light-ray at $\xm=0$ with an angle dependent on $\vec{y}^\perp$. This is represented in Fig. \ref{L2commuting}. Because the original two ANEC shocks commute, the resulting two shocks must also commute, even though they intersect.

The $\left(x^-\right)^4$ dependence of the source here seems to make the amount of energy near $x^-=0$ soft enough that it allows for the crossing with another operator. It would be interesting to understand better why such operators can cross, and whether it is interesting from a more phenomenological point of view. One could imagine applications to quark-gluon plasma physics where these operators correspond to dragging of nucleons in the boundary gauge theory, see \cite{Janik:2010we} for example.

\section{Conclusion and future directions}\label{sec:conclu}

The problem of bootstrapping non-trivial $d>2$ CFTs remains one of the most interesting open problems in high-energy physics. While the solution of this problem for generic theories might very well be out of reach, one could hope that the addition of extra simplifying assumptions, like supersymmetry, large $N$ and/or large gap might provide a lamppost where this program can be carried out to completion. In recent years, important (non-trivial) constraints coming from unitarity of UV complete QFTs have proven very helpful in reducing the landscape of allowed consistent theories. These come in the form of positivity bounds or, more generally, sum rules that all consistent QFTs must satisfy. When applied to holographic (i.e.~large $N$, large gap) CFTs these tools become quite powerful. 

A crucial role in this program has been played by light-ray operators. They appear behind constraints in central charges \cite{Hofman:2008ar, Hofman:2016awc,Belin:2019mnx}, computations of entanglement entropy \cite{Faulkner:2016mzt, Casini:2017roe}, unitarity constraints in QFT \cite{Hartman:2015lfa, Hartman:2016dxc, Hartman:2016lgu} and recent sum rules \cite{Kologlu:2019bco}. While computing all correlation functions of a CFT might not be possible even at large $N$, one could ask if the subsector spanned by light-ray operators can be solved in some form. Some hope that this might be possible was presented in \cite{Casini:2017roe} and \cite{Belin:2019mnx}. In a similar manner that all $n$-point functions of the energy momentum are fixed for $d=2$ CFTs, how much does the algebra of operators and unitarity constrain the correlation functions of light-ray operators?

We have explored this problem in this present work. In particular, we have studied the algebra of $L_n =\int dx^- \left(x^-\right)^{n+2} T_{--}(x)$ operators both in free field theories and holographic CFTs both in a QFT setup and from the point of view of AdS bulk gravity. We list our findings and comment on them, including a discussion on some future directions.

We have proposed a formalism to compute correlation functions of light-ray operators in states created by some CFT operator. Throughout this work we have focused on states created by scalar operators. The technique amounts to computing light-ray integrals by complex contour techniques taking into consideration the time ordering of operators through $i \epsilon$ prescriptions. While this technique is just an efficient calculational method when the real integrals involved are convergent, it amounts to a regularization prescription when they are not. This is an important point of contact to keep in mind when comparing the results presented here to those of \cite{Kologlu:2019bco,Kologlu:2019mfz}. A physical interpretation is to consider matrix elements of these operators on states localized enough in the $x^-$ direction. From the conformal collider experiment perspective this relates to the assumption that most of the radiation will be captured at the calorimeters after a finite time.

We then considered the action of the collinear conformal group that leaves the light-ray invariant on generalized ANEC operators. We found that there exists a five-dimensional subalgebra spanned by $L_n$ with $n \in \{-2, -1, 0, 1, 2\}$ which is closed under the action of this group.  They annihilate the conformal vacuum and therefore have vanishing 2-point functions.

Operators outside this finite set can have non-vanishing two-point correlators in the vacuum giving rise to a central term for the infinite dimensional algebra. We find that this central term is infinite at vanishing $x^+$ separation, in agreement with suggestions in \cite{Casini:2017roe}. It is important to remark that this term is not the naive central term expected by the form of the Virasoro algebra suggested in \cite{Casini:2017roe} (see however \cite{Huang:2020ycs} for a previous appearance of this term). Our finite complex contour integrals cannot produce that term as it is forbidden by the collinear group. Said differently, in our computations there is no IR divergence. The lack of this extra scale severely constrains the form of a central term to the one presented in \rref{eq:centralcharge}.

One might attempt to change the normalization of the generalized ANEC operators to absorb this divergence as $L_n \rightarrow \epsilon^{n+1} L_n$. This regularizes the central term while preserving the form of \rref{eq:Virasoro4dintro}. This amounts to the insertion of an explicit UV cutoff scale $\epsilon^{-1}$. Furthermore, the expectation value of the $L_n$'s themselves might become trivial or divergent under this prescription. It would be interesting to pursue this in future work.

We further computed correlators involving one and two insertions of operators in the five-dimensional global subalgebra in scalar states. We first considered this in free field theory. We found that commutators involving $L_1$ and $L_2$ failed to commute at finite spacelike separations. This non-commutativity behaved as $|\vec{x}^\perp|^{-2}$ at short distances for non identical operators and is therefore non-integrable. This implies that it is not possible to have a well defined algebra of light-ray operators for free field theories. As was initially argued in~\cite{Kologlu:2019bco,Kologlu:2019mfz}, this non-commutativity arises when the light-ray integrals of the Wightman function are not absolutely convergent, though the integrals of the double commutator still converge.


One future direction to consider is the inclusion of other components of the energy-momentum tensor in the definition of the light-ray algebra. Notice, from \rref{eq:CollTransLn}, that the action of the collinear algebra, away from $\vec{x}^\perp=0$, mixes the other components of $T_{\mu \nu}$ with $T_{--}$. If one included those terms in the definition of the new light-ray operators one might be able to soften (or even cancel) the finite separation contribution to commutators. From the point of view of conformal colliders this amounts to considering the flux related to other charges beside the ones associated to  translations and dilatations. This is worth exploring further.

It would be interesting to understand how our results change by including interactions. The obvious arena to push this agenda is to consider generalized ANEC operators in weakly-coupled $\mathcal{N}=4$ Super Yang-Mills theory, building on~\cite{Belitsky:2013bja,Belitsky:2013ofa,Henn:2019gkr,Moult:2019vou}.

For holographic CFTs, there is only one conformal block that can propagate between the insertions of generalized ANEC operators. This has the effect of further enhancing non-commutativity all the way down to $[L_{-2}, L_{-1}]$ which we compute in \rref{resultconfblo} and is in conflict with expectations for any finite $N$ CFT \cite{Cordova:2018ygx}. This is explained in terms of a discontinuity  in the infinite $N$ limit of Regge trajectories. In this case the non-commutativity is integrable. This is the familiar behavior at strong coupling, where the short distance singularities get softened as a consequence of operators acquiring large anomalous dimensions \cite{Hofman:2008ar}. It would be interesting to see if this feature persists generically and allows the construction of a light-ray algebra in this case. We will return to this briefly when we discuss our holographic results.

Of course, the commutativity of $[L_{-2}, L_{-1}]$ should be restored by non perturbative effects. This is quite interesting as we see that an IR sensitive observable is affected at order 1 in the large $N$ limit. It seems irresistible to suggest an analogy with the black hole information paradox. In that case it is the fact that we care about late time observables that complicates the situation. Is the present discussion a bootstrap version of this type of phenomenon? Sum rules of the form \rref{sumrule1} and the general discussion of \cite{Kologlu:2019bco} can provide a hint on how to control this problem explicitly. This is yet another interesting direction to pursue in the future.  

We now turn to our computations in AdS gravity.

We have found new exact shockwave solutions that are dual to the insertions of exponentiated generalized global ANEC operators. The operator $L_0$ has resisted producing an exact dual shockwave. A potential way forward would be to consider an $S$ self-dual scaling ansatz at finite order in $\epsilon$. This problem seems tractable and we leave it for future work.

Using these shockwaves we computed the propagation of perturbations in their background. We have used this to compute the holographic commutators of generalized ANEC operators obtaining full agreement with the results in section \ref{sec:confblock}.

We also computed the effective metric created by the commutator of shockwaves and stated that it does not satisfy the Einstein equations. It would be nice to understand what type of (multi-trace) operators are responsible for the bulk energy-momentum tensor producing these solutions. Further understanding here could shed light on the operators involved in the holographic light-ray algebra and could help making progress in understanding the sum-rules that give the non-perturbative completion to these calculations.

We left for the future the computation  of commutators involving $L_{1}$ and $L_2$ in the gravitational setup. The fact that these shocks intersect in the bulk changes qualitatively the nature of this experiment. Finally, it would be nice to explore phenomenological applications of the shockwaves presented here. For example, in the understanding of nucleon scattering in Quark-Gluon plasma physics \cite{Janik:2010we}

In conclusion, the study of light-ray operators has already provided important results in constraining the space of consistent UV complete QFTs. While their properties are strikingly simpler than those of local operators, and particularly so for holographic CFTs, the understanding of their algebra and bootstrapping of their correlation functions remain yet  out of reach. Still, the simple geometric action of the conformal group on them and their inherent Lorentzian nature make these objects the ideal avenue to further the understanding of the landscape of allowed theories. We hope to see important progress in this area in the coming years.

\section*{Acknowledgements}

It is a pleasure to thank Tarek Anous, Mert Besken, Horacio Casini, Shira Chapman, Jan de Boer, Liam Fitzpatrick, Austin Joyce, Denis Karateev, Gabor Sarosi, John Stout, Gonzalo Torroba, and Sasha Zhiboedov for discussions. A.B.\ and D.H.\ would like to acknowledge the workshop ``Bootstrapping String Theory" where interesting discussions on this topic took place. D.H.\ and G.M.\ are supported in part  by the ERC starting grant {\scriptsize{GENGEOHOL}} (grant agreement No 715656). M.W.\ is partly supported by the Simons Collaboration on the Nonperturbative Bootstrap and the National Centre of Competence in Research SwissMAP funded by the Swiss National Science Foundation.

\appendix
\addtocontents{toc}{\protect\setcounter{tocdepth}{1}}

\section{Useful contour integrals for three-point functions \label{ap:UsefulContourIntegrals3pt}}

When we compute three-point correlators involving light-ray operators in section \ref{sec:3ptF}, we encounter various integrals over $x^-$, which can be evaluated by closing the contour in the upper or lower half-plane. In general, the pole structure of the correlation function implies that one direction is easier to evaluate than the other. For global operators, the way we close the contour does not matter and every possible contour yields the correct answer. 

\subsubsection*{Lower Half-Plane (Operator to Left)}

First, we can close the contour in the lower half-plane, in which case we pick up the OPE singularity with the operator to the left in the correlator (i.e.~the singularity where $x_2$ hits $x_1$):
\be
x_2^- = \x{1}{2} - i\epsilon\, .
\ee
For correlators involving the stress tensor, the pole will be at most third order, which means we only need to evaluate three integrals:

\begin{align}
\int_{-\infty}^\infty dx_2^- f(x_2^-) \fr{1}{x_{12}^2 x_{23}^{2b}} &= \frac{2\pi i (-1)^{b+1}}{x_{12}^+ (x_{23}^+)^b} \bigg[\fr{f(\x{1}{2})}{(\x{1}{2} - \x{3}{2})^b} \bigg],\nonumber \\
\int_{-\infty}^\infty dx_2^- f(x_2^-) \fr{1}{x_{12}^4 x_{23}^{2b}} &= \frac{2\pi i (-1)^{b+1} }{(x_{12}^+)^2 (x_{23}^+)^b}\bigg[\fr{f'(\x{1}{2})}{(\x{1}{2} - \x{3}{2})^b} - b \fr{f(\x{1}{2})}{(\x{1}{2} - \x{3}{2})^{b+1}}\bigg], \label{eq:FintegralsLeft}\\
\int_{-\infty}^\infty dx_2^- f(x_2^-) \fr{1}{x_{12}^6 x_{23}^{2b}} &= \frac{2\pi i (-1)^{b+1}}{(x_{12}^+)^3 (x_{23}^+)^b} \bigg[\half \fr{f''(\x{1}{2})}{ (\x{1}{2} - \x{3}{2})^b} -  \fr{bf'(\x{1}{2})}{(\x{1}{2} - \x{3}{2})^{b+1}} + \half  \fr{b(b+1)f(\x{1}{2})}{(\x{1}{2} - \x{3}{2})^{b+2}}\bigg].\nonumber
\end{align}

\subsubsection*{Upper Half-Plane (Operator to Right)}

If we instead close the contour in the upper half-plane, we pick up the OPE singularity with the operator to the right in the correlator (where $x_2$ hits $x_3$):
\be
x_2^- = \x{3}{2} + i\epsilon.
\ee
We then need to evaluate three integrals that are similar to the previous case:
\begin{align}
\int_{-\infty}^\infty dx_2^- f(x_2^-) \fr{1}{x_{12}^{2a} x_{23}^2} &=\frac{ 2\pi i (-1)^{a+1}}{(x_{12}^+)^a x_{23}^+} \bigg[  \fr{f(\x{3}{2})}{ (\x{1}{2} - \x{3}{2})^a} \bigg], \nonumber\\
\int_{-\infty}^\infty dx_2^- f(x_2^-) \fr{1}{x_{12}^{2a} x_{23}^4} &= \frac{2\pi i (-1)^a }{(x_{12}^+)^a (x_{23}^+)^2}\bigg[ \fr{f'(\x{3}{2})}{ (\x{1}{2} - \x{3}{2})^a} + \fr{af(\x{3}{2})}{(\x{1}{2} - \x{3}{2})^{a+1}} \bigg], \label{eq:FintegralsRight}\\
\int_{-\infty}^\infty dx_2^- f(x_2^-) \fr{1}{x_{12}^{2a} x_{23}^6} &= \frac{2\pi i (-1)^{a+1}}{(x_{12}^+)^a (x_{23}^+)^3} \bigg[ \half \fr{f''(\x{3}{2})}{ (\x{1}{2} - \x{3}{2})^a}+\fr{af'(\x{3}{2})}{(\x{1}{2} - \x{3}{2})^{a+1}}+ \half \fr{a(a+1) f(\x{3}{2})}{(\x{1}{2} - \x{3}{2})^{a+2}} \bigg].\nonumber
\end{align}

\section{Useful contour integrals for four-point functions in free field theory \label{ap:UsefulContourIntegrals4pt}}

To compute the four-point functions involving two light-ray operators, the starting point is the four-point function of two local stress-energy tensors $\braket{\phi(x_1)T_{--}(x_2)T_{--}(x_3)\phi(x_4)}$. In free field theory, this is just a Wick contraction exercise, starting from  
\be
T_{--}(x)= \fr{1}{6\pi^2}\left[ (\p_-\phi(x))^2 - \fr{1}{2} \phi(x) \p_-^2 \phi(x)\right]\, ,\label{eq:Tmmfreefieldapp}
\ee  
that yields  
\begin{align}
&\braket{\phi(x_1)T_{--}(x_2)T_{--}(x_3)\phi(x_4)} \nonumber\\
&=\frac{1}{3(2\pi^2)^2}\frac{(x_{23}^+)^4}{x_{14}^2 x_{23}^{12}}+\frac{1}{(2\pi^2)^2}\left[\left(\frac{\partial}{\partial x_2^-}\frac{1}{x_{12}^2}\right)\left(\frac{\partial^2}{\partial x_2^- \partial x_3^-}\frac{1}{x_{23}^2}\right)\left(\frac{\partial}{\partial x_3^-}\frac{1}{x_{34}^2}\right)+ (2\leftrightarrow 3)\right]\nonumber\\
&+ \frac{1}{(12\pi^2)^2}\left[\left(\frac{\partial^2}{\partial(x_2^-)^2}\frac{\partial^2}{\partial(x_3^-)^2}\frac{1}{x_{12}^2x_{23}^2x_{34}^2}\right)+ (2 \leftrightarrow 3)\right]\label{eq:phiTmmTmmphicompapp}\\
&-\frac{1}{6(2\pi^2)^2}\left\lbrace\left[\frac{\partial^2}{\partial(x_2^-)^2}\left(\frac{1}{x_{12}^2}\left(\frac{\partial}{\partial x_3^-} \frac{1}{x_{23}^2}\right)\left(\frac{\partial}{\partial x_3^-}\frac{1}{x_{34}^2}\right)\right)+(2\leftrightarrow 3)\right]+ (1\leftrightarrow 4) \right\rbrace\nonumber\, .
\end{align}
This result is identical, up to an overall normalization, to the one in \cite{Cordova:2018ygx}. 

Because we are interested in the four-point functions that involve global light-ray operators, we already explained how the final answer is insensitive to the way we close the contours to evaluate the integrals. We thus evaluate these integrals by closing both contours outwards, meaning we integrate $x_2^-$ by picking up the singularity when $x_2\rightarrow x_1$, and $x_3^-$ by picking up the singularity when $x_3 \rightarrow x_4$. It implies that only  the terms with a denominator of the form $x_{12}^a x_{34}^b$ for arbitrary $a$ and $b$ will survive integration.  These are the terms that have the first topology of figure \ref{figFeynman}.  This implies that to perform the integrals, we can concentrate solely on the following terms in the four-point function
\be
\ba
&\<\phi(x_1) T_{--}(x_2) T_{--}(x_3) \phi(x_4)\> \\
& \qquad \supset \fr{1}{x_{12}^2 x_{23}^{10} x_{34}^2} \bigg( \fr{1}{36\pi^4} (x_{23}^+)^4 + \fr{1}{9\pi^4} x_{23}^4 (x_{23}^+)^2  \left( \fr{x_{12}^+}{x_{12}^2} + \fr{x_{23}^+}{x_{23}^2} \right) \left( \fr{x_{23}^+}{x_{23}^2} + \fr{x_{34}^+}{x_{34}^2} \right) \\
& \qquad \qquad \qquad \qquad \quad + \, \fr{1}{36\pi^4} x_{23}^8 \left( \fr{x_{12}^+}{x_{12}^2} + \fr{x_{23}^+}{x_{23}^2} \right)^2 \left( \fr{x_{23}^+}{x_{23}^2} + \fr{x_{34}^+}{x_{34}^2} \right)^2 \bigg).
\ea
\label{eq:PhiTTPhiappendix}
\ee
To compute the four-point function involving two global light-ray operators 
\be
\<\phi(x_1) \Ecal_f(x_2) \Ecal_g(x_3) \phi(x_4)\> \equiv \int_{-\infty}^\infty dx_2^- f(x_2^-) \int_{-\infty}^\infty dx_3^- g(x_3^-) \<\phi(x_1) T_{--}(x_2) T_{--}(x_3) \phi(x_4)\>,
\ee
we just need to integrate \eqref{eq:PhiTTPhiappendix}. Let us step through the various terms in eq.~\eqref{eq:PhiTTPhiappendix} separately, using the general expressions in~\eqref{eq:FintegralsLeft} and \eqref{eq:FintegralsRight}. The different terms have poles up to third order in both $x_2^-$ and $x_3^-$. The integrals are 
\be\label{eq:4ptint1}
\int dx_2^- f(x_2^-) \int dx_3^- g(x_3^-) \fr{(x_{23}^+)^4}{x_{12}^2 x_{23}^{10} x_{34}^2} = (2\pi i)^2  \fr{(x_{23}^+)^4f(\x{1}{2}) g(\x{4}{3})}{x_{12}^+ x_{34}^+ [-x_{23}^+(\x{1}{2} - \x{4}{3}) + |\vec{x}_{23}^\perp|^2]^5}\, .
\ee
and
\be
\ba
&\int dx_2^- f(x_2^-) \int dx_3^- g(x_3^-) \fr{(x_{23}^+)^2}{x_{12}^2 x_{23}^6 x_{34}^2} \left( \fr{x_{12}^+}{x_{12}^2} + \fr{x_{23}^+}{x_{23}^2} \right) \left( \fr{x_{23}^+}{x_{23}^2} + \fr{x_{34}^+}{x_{34}^2} \right) \\
& \qquad = (2\pi i)^2 \Bigg( 21 f(\x{1}{2}) g(\x{4}{3}) \fr{(x_{23}^+)^4}{x_{12}^+ x_{34}^+ [-x_{23}^+(\x{1}{2} - \x{4}{3}) + |\vec{x}_{23}^\perp|^2]^5} \\
& \qquad \qquad \qquad + \, 4 \left( f'(\x{1}{2}) g(\x{4}{3}) - f(\x{1}{2}) g'(\x{4}{3}) \right) \fr{(x_{23}^+)^3}{x_{12}^+ x_{34}^+ [-x_{23}^+(\x{1}{2} - \x{4}{3}) + |\vec{x}_{23}^\perp|^2]^4} \\
& \qquad \qquad \qquad - \, f'(\x{1}{2}) g'(\x{4}{3}) \fr{(x_{23}^+)^2}{x_{12}^+ x_{34}^+ [-x_{23}^+(\x{1}{2} - \x{4}{3}) + |\vec{x}_{23}^\perp|^2]^3} \Bigg).\label{eq:4ptint2}
\ea
\ee
Finally, we have the third term, which has poles up to third order in both $x_2^-$ and $x_3^-$, leading to the resulting expression
\be
\ba
&\int dx_2^- f(x_2^-) \int dx_3^- g(x_3^-) \fr{1}{x_{12}^2 x_{23}^2 x_{34}^2} \left( \fr{x_{12}^+}{x_{12}^2} + \fr{x_{23}^+}{x_{23}^2} \right)^2 \left( \fr{x_{23}^+}{x_{23}^2} + \fr{x_{34}^+}{x_{34}^2} \right)^2 \\
& \qquad = (2\pi i)^2 \Bigg( 131 f(\x{1}{2}) g(\x{4}{3}) \fr{(x_{23}^+)^4}{x_{12}^+ x_{34}^+ [-x_{23}^+(\x{1}{2} - \x{4}{3}) + |\vec{x}_{23}^\perp|^2]^5} \\
& \qquad \qquad \qquad + \, 38 \Big( f'(\x{1}{2}) g(\x{4}{3}) - f(\x{1}{2}) g'(\x{4}{3}) \Big) \fr{(x_{23}^+)^3}{x_{12}^+ x_{34}^+ [-x_{23}^+(\x{1}{2} - \x{4}{3}) + |\vec{x}_{23}^\perp|^2]^4} \\
& \qquad \qquad \qquad - \, 14 f'(\x{1}{2}) g'(\x{4}{3}) \fr{(x_{23}^+)^2}{x_{12}^+ x_{34}^+ [-x_{23}^+(\x{1}{2} - \x{4}{3}) + |\vec{x}_{23}^\perp|^2]^3} \\
& \qquad \qquad \qquad + \, 3 \Big( f''(\x{1}{2}) g(\x{4}{3}) + f(\x{1}{2}) g''(\x{4}{3}) \Big) \fr{(x_{23}^+)^2}{x_{12}^+ x_{34}^+ [-x_{23}^+(\x{1}{2} - \x{4}{3}) + |\vec{x}_{23}^\perp|^2]^3} \\
& \qquad \qquad \qquad - \, \fr{3}{2} \Big( f''(\x{1}{2}) g'(\x{4}{3}) - f'(\x{1}{2}) g''(\x{4}{3}) \Big) \fr{x_{23}^+}{x_{12}^+ x_{34}^+ [-x_{23}^+(\x{1}{2} - \x{4}{3}) + |\vec{x}_{23}^\perp|^2]^2} \\
& \qquad \qquad \qquad + \, \fr{1}{4} f''(\x{1}{2}) g''(\x{4}{3}) \fr{1}{x_{12}^+ x_{34}^+ [-x_{23}^+(\x{1}{2} - \x{4}{3}) + |\vec{x}_{23}^\perp|^2]} \Bigg).\label{eq:4ptint3}
\ea
\ee
Combining these results yields the four-point function of light-ray operators.

\section{Aside on delta functions\label{ap:deltafunction}}

In this section, we want to review how to extract a perpendicular delta function $\de^{(2)}(\vec{x}_{23}^\perp)$ from the expression with $i\epsilon$ that we encounter as commutators. We need to evaluate the $\epsilon\rightarrow 0$ limit of expressions of the schematic form:
\benn
\fr{(i\epsilon)^{a-1}}{(|\vec{x}|^2 + i\epsilon q)^a}.
\eenn
Let us now try to systematically evaluate such expressions, in order to extract the delta function contribution. 

To start, let us quickly review how to evaluate the familiar expression
\be
y(x) = \fr{1}{x + i\epsilon}.
\ee
We can think of $y(x)$ as a distribution satisfying the relation
\be
x \cdot y(x) = 1.
\ee
The general solution to this constraint is clearly
\be
y(x) = \Pcal \fr{1}{x} + c_0 \de(x),
\ee
where $\Pcal$ indicates the principal value. We can then fix the coefficient $c_0$ by integrating over the region $-b \leq x \leq b$, take the limit $\epsilon \ra 0$, then take $b \ra 0$ so that we only pick up the delta function. Evaluating this integral and taking $\epsilon \ra 0$, we find
\be
\int_{-b}^b dx\, \fr{1}{x + i\epsilon} = \log(b + i\epsilon) - \log(-b + i\epsilon) = - i\pi.
\ee
We therefore find $c_0 = - i\pi$, giving us the familiar identity
\be
\fr{1}{x + i\epsilon} = \Pcal \fr{1}{x} - i\pi \de(x).
\ee
Integrating this expression against a test function shows that this relation is correct in the distribution sense. 

Let us now try to generalize this analysis to evaluate the expression
\benn
\lim_{\epsilon\rightarrow 0}\,\, \fr{(i\epsilon)^{a-1}}{(|\vec{x}|^2 + i\epsilon q)^a},
\eenn
where $\vec{x} = (r\cos\theta,r\sin\theta)$ is a vector in $\mathbb{R}^2$. We expect that this expression contains a delta function $\de^{(2)}(\vec{x})$. To determine whether this is the case, and to determine the coefficient, we can follow the procedure above, though now we'll integrate over a disk of radius $b$ around the origin,
\be
\ba
\int d^2x \fr{(i\epsilon)^{a-1}}{(|\vec{x}|^2 + i\epsilon q)^a} &= \int_0^b dr \int_0^{2\pi} d\theta \,  \fr{r(i\epsilon)^{a-1}}{(r^2 + i\epsilon q)^a} 
= \fr{\pi}{(n-1)q^{a-1}}.
\ea
\ee
We thus obtain the identity
\be
\lim_{\epsilon\rightarrow 0}\,\,\fr{(i\epsilon)^{a-1}}{(|\vec{x}|^2 + i\epsilon q)^a} = \fr{\pi}{(a-1)q^{a-1}} \de^{(2)}(\vec{x})\, \qquad  \qquad (a > 1).
\label{eq:DeltaIdentity}
\ee
Integrating \eqref{eq:DeltaIdentity} against a test function proves that this is the correct relation and that it stands as a distribution. 

What happens when $a=1$? In this case, we want to understand the following quantity
\be 
\lim_{\epsilon\rightarrow 0}\,\,\frac{1}{|\vec{x}|^2+i\epsilon}  = \lim_{\epsilon\rightarrow 0}\,\,\left(\frac{|\vec{x}|^2}{|\vec{x}|^4+\epsilon^2} -i\frac{\epsilon}{|\vec{x}|^4+\epsilon^2}\right)\, . \label{eq:n1delta}
\ee
If we consider the real part of \eqref{eq:n1delta} and integrate it on a disk of radius $b$, we get 
\begin{align}
\int_{-\infty}^{\infty}d^2x\frac{|\vec{x}|^2}{|\vec{x}|^4 + \epsilon^2} 
= 2\pi \int_0^b dr \frac{r^3}{r^4 + \epsilon^2}= \pi\log\left(1 + \frac{b^4}{\epsilon^2}\right)\, .
\end{align}
In the limit $\epsilon\rightarrow 0$, it diverges and we conclude that the real part is not integrable. For the imaginary part, we get 
\begin{align}
\int_{-\infty}^{\infty}d^2x\frac{\epsilon}{|\vec{x}|^4 + \epsilon^2}  =\int_0^b dr \int_0^{2\pi}d\theta\, \frac{r\epsilon }{r^4 + \epsilon^2}= \pi \arctan\left(\frac{b^2}{\epsilon}\right)\, ,
\end{align}
and the limit $\epsilon\rightarrow 0$ gives $\pi^2/2$. We thus see that the imaginary part is integrable and has a well-defined limit when $\epsilon\rightarrow 0$. This is the part responsible for our finite separation contribution. It is thus impossible to extract a meaningful transverse delta function in this case for $a=1$.

\section{Commutators of global light-ray operators\label{ap:commutators}}

In this appendix, we want to list the results of the commutator of two global light-ray operators for the cases where there is no finite transverse separation contribution and where $x_2^+ = x_3^+$. We will present two examples in some details and list the results of the other computations. 

\subsection{The commutator $\comm{L_{-1}}{L_{-2}}$}

In terms of transverse delta functions, the two orderings are 
\begin{align}
\<\phi(x_1) L_{-1}(x_2) L_{-2}(x_3) \phi(x_4)\>& = \frac{\de^{(2)}(\vec{x}_{23}^\perp)}{\pi x_{12}^+ x_{24}^+}\left[- \fr{6\x{1}{2}}{(\x{1}{2} - \x{4}{2})^4} +  \fr{2}{ (\x{1}{2} - \x{4}{2})^3} \right]\, ,\\
\<\phi(x_1) L_{-2}(x_3) L_{-1}(x_2) \phi(x_4)\> &= \frac{\de^{(2)}(\vec{x}_{23}^\perp)}{\pi x_{12}^+x_{24}^+}\left[- \fr{6\x{4}{2}}{(\x{1}{2} - \x{4}{2})^4}  -  \fr{2}{ (\x{1}{2} - \x{4}{2})^3} \right]\, .
\end{align}
Combining them results in the commutator  
\be
\<\phi(x_1) \comm{L_{-1}(x_2)}{L_{-2}(x_3)} \phi(x_4)\> = -\fr{2}{\pi} \fr{1}{x_{12}^+ x_{24}^+ (\x{1}{2} - \x{4}{2})^3} \de^{(2)}(\vec{x}_{23}^\perp),
\ee
which is also what \eqref{eq:PhiComm} yields if you use $f(x^-) = x^-$ and $g(x^-) = 1$. This exactly matches the 3-pt function involving $L_{-2}$ that we computed in \eqref{eq:ThreePtFct} (multiplied by $-i\delta^{(2)}(\vec{x}_{23}^\perp))$:
\be
\<\phi(x_1) \comm{L_{-1}(x_2)}{L_{-2}(x_3)} \phi(x_4)\> = -i \de^{(2)}(\vec{x}_{23}^\perp) \<\phi(x_1) L_{-2}(x_2) \phi(x_4)\>.
\label{eq:Lm1Lm2Comm2}
\ee

\subsection{The commutator $\comm{L_0}{L_{-2}}$}

In terms of transverse delta functions, the two orderings are 
\begin{align}
&\<\phi(x_1) L_0(x_2) L_{-2}(x_3) \phi(x_4)\> &= \frac{\de^{(2)}(\vec{x}_{23}^\perp)}{\pi x_{12}^+x_{24}^+}\left[- \fr{6(\x{1}{2})^2}{(\x{1}{2} - \x{4}{2})^4} +  \fr{4\x{1}{2}}{(\x{1}{2} - \x{4}{2})^3} - \fr{1}{3(\x{1}{2} - \x{4}{2})^2} \right], \nonumber\\
&\<\phi(x_1) L_{-2}(x_3) L_0(x_2) \phi(x_4)\> & = \frac{\de^{(2)}(\vec{x}_{23}^\perp)}{\pi x_{12}^+x_{24}^+}\left[ - \fr{6(\x{4}{2})^2}{(\x{1}{2} - \x{4}{2})^4}  -  \fr{4\x{4}{2}}{(\x{1}{2} - \x{4}{2})^3}  -  \fr{1}{3 (\x{1}{2} - \x{4}{2})^2} \right].\nonumber
\end{align}
Combining them results in the commutator  
\be
\<\phi(x_1) \comm{L_0(x_2)}{L_{-2}(x_3)} \phi(x_4)\> = -\fr{2}{\pi} \fr{\x{1}{2} + \x{4}{2}}{x_{12}^+ x_{24}^+ (\x{1}{2} - \x{1}{4})^3} \de^{(2)}(\vec{x}_{23}^\perp),
\ee
which matches the three-point function with $L_{-1}$:
\be
\<\phi(x_1) \comm{L_0(x_2)}{L_{-2}(x_3)} \phi(x_4)\> = -2i \de^{(2)}(\vec{x}_{23}^\perp) \<\phi(x_1) L_{-1}(x_2) \phi(x_4)\>.
\ee

\subsection{Remaining commutators}
The different cases are given by 
\begin{align}
\<\phi(x_1) \comm{L_{1}(x_2)}{L_{-2}(x_3)} \phi(x_4)\> &= -\fr{1}{\pi} \fr{(\x{1}{2})^2+4\x{1}{2}\x{4}{2}  + (\x{4}{2})^2}{x_{12}^+ x_{24}^+ (\x{1}{2} - \x{1}{4})^3} \de^{(2)}(\vec{x}_{23}^\perp)\, ,\\
&=-3i \de^{(2)}(\vec{x}_{23}^\perp) \<\phi(x_1) L_{1}(x_2)\phi(x_4)\>\, .
\end{align} 
and 
\begin{align}
\<\phi(x_1) \comm{L_{2}(x_2)}{L_{-2}(x_3)} \phi(x_4)\> &= -\fr{4}{\pi} \fr{\x{1}{2}\x{4}{2}(\x{1}{2}+ \x{4}{2})}{x_{12}^+ x_{24}^+ (\x{1}{2} - \x{1}{4})^3} \de^{(2)}(\vec{x}_{23}^\perp)\, ,\\
&=-4i \de^{(2)}(\vec{x}_{23}^\perp) \<\phi(x_1) L_{0}(x_2)\phi(x_4)\>\, .
\end{align}
and 
\begin{align}
\<\phi(x_1) \comm{L_{0}(x_2)}{L_{-1}(x_3)} \phi(x_4)\> &=-\fr{1}{3\pi} \fr{(\x{1}{2})^2+4\x{1}{2}\x{4}{2}  + (\x{4}{2})^2}{x_{12}^+ x_{24}^+ (\x{1}{2} - \x{1}{4})^3} \de^{(2)}(\vec{x}_{23}^\perp)\, ,\\
&=-i \de^{(2)}(\vec{x}_{23}^\perp) \<\phi(x_1) L_{0}(x_2)\phi(x_4)\>\, .
\end{align}
and 
\begin{align}
\<\phi(x_1) \comm{L_{1}(x_2)}{L_{-1}(x_3)} \phi(x_4)\> &= -\fr{2}{\pi} \fr{\x{1}{2}\x{4}{2}(\x{1}{2}+ \x{4}{2})}{x_{12}^+ x_{24}^+ (\x{1}{2} - \x{1}{4})^3} \de^{(2)}(\vec{x}_{23}^\perp)\, ,\\
&=-2i \de^{(2)}(\vec{x}_{23}^\perp) \<\phi(x_1) L_{0}(x_2)\phi(x_4)\>\, .
\end{align}
and finally
\begin{align}
\<\phi(x_1) \comm{L_{2}(x_2)}{L_{-1}(x_3)} \phi(x_4)\> &= -\fr{6}{\pi} \fr{(\x{1}{2}\x{4}{2})^2}{x_{12}^+ x_{24}^+ (\x{1}{2} - \x{1}{4})^3} \de^{(2)}(\vec{x}_{23}^\perp)\, ,\\
&=-3i \de^{(2)}(\vec{x}_{23}^\perp) \<\phi(x_1) L_{0}(x_2)\phi(x_4)\>\, .
\end{align}
This concludes the discussion of all the cases where the algebra \eqref{eq:Virasoro4d} is satisfied in free field theory, and where the $f''(x^-)g''(x^-)$ contribution to the commutator is vanishing. 

\subsection{Integrating commutators in the perpendicular direction  } \label{app:deltaperp}
Instead of extracting a transverse delta function for our commutators, we can also evaluate integrals in the perpendicular direction instead. Evaluating the $\vec{x}_3^\perp$ integral allows us to read of the coefficient multiplying $\delta^{(2)}(\vec{x}_{23}^\perp)$, thus providing a useful consistency check. We will do this in some details for the case $[L_{-2},L_{-2}]$. 
%
%
We need to compute both orderings, obtaining
\begin{align}
\<\phi(x_1) L_{-2}(x_2) L_{-2}(x_3) \phi(x_4)\> &= -\fr{24}{\pi^2} \fr{(x_{23}^+-i\epsilon)^4}{(x_{12}^+-i\epsilon) (x_{34}^+-i\epsilon) [-(x_{23}^+-i\epsilon)(\x{1}{2}-\x{4}{3}-i\epsilon) + |\vec{x}_{23}^\perp|^2]^5},\nonumber\\
\<\phi(x_1) L_{-2}(x_3) L_{-2}(x_2) \phi(x_4)\>&=  -\fr{24}{\pi^2} \fr{(x_{23}^++i\epsilon)^4}{(x_{13}^+-i\epsilon) (x_{24}^+-i\epsilon) [(x_{23}^++i\epsilon)(\x{1}{3} - \x{4}{2}-i\epsilon) + |\vec{x}_{23}^\perp|^2]^5}.\nonumber
\end{align}
Prior to evaluating the commutator, we will first integrate over $\vec{x}_3^\perp$. To make the resulting integral simpler, we'll set the transverse components of the remaining three operators to zero:
\be
\vec{x}_1^\perp = \vec{x}_2^\perp = \vec{x}_4^\perp = 0.\label{eq:simpllimitapp}
\ee
We can then write $\vec{x}_3^\perp = (r\cos\theta,r\sin\theta)$, and the coordinates $\x{i}{j}$ become
\be
\x{1}{2} = x_1^-, \qquad \x{4}{3} = x_4^- + \fr{r^2}{x_{34}^+}, \qquad \x{1}{3} = x_1^- - \fr{r^2}{x_{13}^+}, \qquad \x{4}{2} = x_4^-.\label{eq:simplifyicationcoordinates}
\ee
When inserting \eqref{eq:simplifyicationcoordinates} into both orderings, we get two expressions that have no dependence on $\theta$, and very simple dependence on $r$, so we only need to evaluate the general integral
\be
\int d^2x^\perp \fr{1}{(A+B|\vec{x}^\perp|^2)^n} = \int_0^\infty dr^2 \fr{\pi}{(A+Br^2)^n} = \fr{\pi}{(n-1)A^{n-1} B}.
\ee
Using this general integral, the two orderings after integrating over $\vec{x}_3^\perp$ are given by
\begin{align}
\int d^2x_3^\perp \<\phi(x_1) L_{-2}(x_2)L_{-2}(x_3) \phi(x_4)\> &= -\fr{6}{\pi} \fr{1}{(x_{12}^+-i\epsilon) (x_{24}^+-i\epsilon) (x_{14}^--i\epsilon)^4},\nonumber\\
\int d^2x_3^\perp \<\phi(x_1)L_{-2}(x_3) L_{-2}(x_2) \phi(x_4)\> &= -\fr{6}{\pi} \fr{1}{(x_{12}^+-i\epsilon) (x_{24}^+-i\epsilon) (x_{14}^--i\epsilon)^4}.\nonumber
\end{align}
As we can see, these two orderings give the exact same expression, such that the commutator vanishes 
\be
\int d^2x_3^\perp \<\phi(x_1) \comm{L_{-2}(x_2)}{L_{-2}(x_3)} \phi(x_4)\> = 0.\nn
\ee
Note that this integrated commutator vanishes even when $x_{23}^+ \neq 0$ (i.e.~when the light-ray operators are not on the same null surface). It is clear that the same procedure can be repeated for any of the correlators we already presented. It gives the correct answer for all these cases, and we present 
\begin{align}
\int d^2x_3^\perp \<\phi(x_1) \comm{L_{-1}(x_2)}{L_{-2}(x_3)} \phi(x_4)\> &= \fr{2}{\pi} \fr{1}{x_{12}^+ x_{24}^+ (x_{14}^-)^3}=-i \<\phi(x_1) L_{-2}(x_2) \phi(x_4)\>\, ,\nn\\
\int d^2x_3^\perp \<\phi(x_1) \comm{L_0(x_2)}{L_{-2}(x_3)} \phi(x_4)\> &= -\fr{2}{\pi} \fr{x_1^- + x_4^-}{x_{12}^+ x_{24}^+ (x_{14}^-)^3}=  -2i \<\phi(x_1) L_{-1}(x_2) \phi(x_4)\>\, , \nn\\
\int d^2x_3^\perp \<\phi(x_1) \comm{L_1(x_2)}{L_{-2}(x_3)} \phi(x_4)\> &= - \fr{1}{\pi} \fr{(x_1^-)^2 + 4 x_1^- x_4^- + (x_4^-)^2}{x_{12}^+ x_{24}^+(x_{14}^-)^3}=-3i \<\phi(x_1) L_0(x_2) \phi(x_4)\>\, .\nn
\end{align}

\section{Non-local operator from the $T\times T$ OPE \label{ap:L1L1OPEnonlocal}}
\subsection{Derivation of the non-local operator}
In this appendix, we want to explain how we can get the non-local operator that reproduces the commutator when inserted into correlation functions with scalar external states. This allows us to write the leading singularity of the $[L_1(x_2),L_1(x_3)]$ commutator as a non-local operator. This is the resummed version of the infinite sum we presented in section \ref{sec:L1L1OPE}. The computations are done in free field theory, where the stress energy tensor is given as in equation \eqref{eq:Tmmfreefieldapp}. The TT OPE follows from considering
\begin{align}
T_{--}(x)T_{--}(y) \sim \alpha^2\left[(\partial_-\phi(x))^2 -\frac{1}{2}\phi\partial_-^2\phi(x)\right]\left[(\partial_-\phi(y))^2 -\frac{1}{2}\phi\partial_-^2\phi(y)\right]\, ,
\end{align}
where $\alpha = \frac{1}{6\pi^2}$. We want to do one Wick contraction in each product of four fields above while leaving two fields uncontracted and normal ordered. Doing this, we obtain
\begin{align}
T_{--}(x)T_{--}(y) &=  4\alpha^2 \braket{\partial_-\phi(x)\partial_-\phi(y)}:\partial_-\phi(x)\partial_-\phi(y):\label{eq:TTWICKCONAPP}\\
&\phantom{=}-\alpha^2\left[\braket{\partial_-\phi(x)\phi(y)}:\partial_-\phi(x)\partial_-^2\phi(y):+\braket{\partial_-\phi(x)\partial_-^2\phi(y)}:\partial_-\phi(x)\phi(y):\right]\nonumber\\
&\phantom{=}-\alpha^2\left[\braket{\phi(x)\partial_-\phi(y)}:\partial_-^2\phi(x)\partial_-\phi(y): + \braket{\partial_-^2\phi(x)\partial_-\phi(y)}:\phi(x)\partial_-\phi(y):\right]\nonumber\\
&\phantom{=}+ \frac{\alpha^2}{4}\left[\braket{\phi(x)\phi(y)}:\partial_-^2\phi(x)\partial_-^2\phi(y): + \braket{\phi(x)\partial_-^2\phi(y)}:\partial_-^2\phi(x)\phi(y):\right.\nonumber\\
&\phantom{=}+\left.\braket{\partial_-^2\phi(x)\phi(y)}:\phi(x)\partial_-^2\phi(y):+\braket{\partial_-^2\phi(x)\partial_-^2\phi(y)}:\phi(x)\phi(y):\right]\nonumber
\end{align}
To derive the commutator (once inserted into three-point functions with appropriate external states), we need to integrate this as $x^-$ goes to $y^-$.
Using \eqref{eq:TTWICKCONAPP}, and integrating $x_2^-$ around $x_3^-$, we get 
\begin{align}
[L_m&(x_2),L_n(x_3)] = -\int dx_3^- \frac{(2\pi i)(x_3^{-})^{n+2}}{\pi^4(x_{23}^+-i\epsilon)} \left[\frac{(m+2)(m+1)m(m-1)}{144}(\x{3}{2})^{m-2}\phi(\x{3}{2})\phi(x_3)\right.\nn\\
&+(m+2)(m+1)m(\x{3}{2})^{m-1}\left(\frac{\partial_- \phi(\x{3}{2})\phi(x_3)}{18}-\frac{\phi(\x{3}{2})\partial_-\phi(x_3)}{36}\right)\nn\\
&+(m+2)(m+1)(\x{3}{2})^{m}\left(\frac{19\partial_-^2\phi(\x{3}{2})\phi(x_3)}{144}-\frac{7\partial_-\phi(\x{3}{2})\partial_-\phi(x_3)}{36}+ \frac{\phi(\x{3}{2})\partial_-^2\phi(x_3)}{144}\right)\nonumber\\
&+(m+2)(\x{3}{2})^{m+1}\left(\frac{\partial_-^3\phi(\x{3}{2})\phi(x_3)}{8}-\frac{\partial_-^2\phi(\x{3}{2})\partial_-\phi(x_3)}{3}+\frac{\partial_-\phi(\x{3}{2})\partial_-^2\phi(x_3)}{24}\right)\nonumber\\
&+\left.(\x{3}{2})^{m+2}\left(\frac{\partial_-^4\phi(\x{3}{2})\phi(x_3^-)}{24}-\frac{\partial_-^3\phi(\x{3}{2})\partial_-\phi(x_3)}{6}+ \frac{\partial_-^2\phi(\x{3}{2})\partial_-^2\phi(x_3^-)}{24}\right)\right] \, ,
\end{align}
where all products of fields are normal ordered. Let us explain the notation in the last equation. $\phi(\x{3}{2})$ is the field $\phi$ evaluated at the position $\phi(\x{3}{2}) \equiv \phi(x_2^+, \x{3}{2},\vec{x}_2^\perp)$. This happens because we evaluated the $dx_2^-$ integral at the location of the pole where $x_{23}^2=0$, which is $\x{3}{2}$. In addition, the minus derivatives are $x_3^-$ minus derivatives.

For $[L_1(x_2),L_1(x_3)]$, which is the easiest case with a contribution at finite separation, we get 
\begin{align}
&[L_1(x_2),L_1(x_3)]= -\frac{(2\pi i)}{\pi^4(x_{23}^+-i\epsilon)}\int dx_3^- (x_3^-)^3\label{eq:L1L1comintapp}\\
&\qquad  \phantom{=} \times \left\lbrace \left[\frac{1}{3}\partial_-\phi(\x{3}{2}) +\frac{19}{24}\x{3}{2}\partial_-^2\phi(\x{3}{2}) + \frac{3}{8}(\x{3}{2})^2\partial_-^3\phi(\x{3}{2})+ \frac{1}{24}(\x{2}{3})^3\partial_-^4\phi(\x{2}{3})\right]\right.\phi(x_3)\nn\\
&\qquad \phantom{=}-\left[\frac{1}{6}\phi(\x{3}{2})+\frac{7}{6}(\x{3}{2})\partial_-\phi(\x{3}{2})+(\x{3}{2})^2\partial_-^2\phi(\x{3}{2})+\frac{1}{6}(\x{3}{2})^3\partial_-^3\phi(\x{3}{2})\right]\partial_-\phi(x_3)\nonumber\\
&\qquad \phantom{=}+\left.\left[\frac{1}{24}(\x{3}{2})\phi(\x{3}{2})+\frac{1}{8}(\x{3}{2})^2\partial_-\phi(\x{3}{2})+\frac{1}{24}(\x{3}{2})^3\partial_-^2\phi(\x{3}{2})\right]\partial_-^2\phi(x_3)\right\rbrace\, .\nn
\end{align}
Once inserted into three-point functions with scalar external states, the non-local operator \eqref{eq:L1L1comintapp} reproduces the result \eqref{eq:L1L1comm} for $\braket{\phi(x_1)[L_1(x_2),L_1(x_3)]\phi(x_4)}$.  Let us see how this works. 

Once we compute the three-point function of \eqref{eq:L1L1comintapp} with $\phi$ as external states, every term in the three-point function is of the form  $\braket{\phi(x_1):\partial_-^{n_1}\phi(\x{3}{2})\partial_-^{n_2}:\phi(x_4)}$, which has two different Wick contractions  
\begin{align}
\braket{\phi(x_1)\partial_-^{n_1}\phi(\x{3}{2})\partial_-^{n_2}\phi(x_3)\phi(x_4)} &= \braket{\phi(x_1)\partial_-^{n_1}\phi(x_{3,2}^-)}\braket{\partial_-^{n_2}\phi(x_3)\phi(x_4)} \label{eq:Wick1app}\\
&\phantom{=}+\braket{\phi(x_1)\partial_-^{n_2}\phi(x_3)}\braket{\partial_-^{n_1}\phi(\x{3}{2})\phi(x_4)}\nn\, .
\end{align}
When performing the $dx_3^-$ integral of \eqref{eq:L1L1comintapp}, we want to close the contour in the upper-half plane. It means that the two poles are at positions
\begin{align} 
x_3^-& = \x{4}{3}+ i \epsilon= x_4^- + \frac{|x_{34}^\perp|^2}{x_{34}^+} + i\epsilon\label{eq:x34pole1app}\, ,\\
 x_3^- &= \x{4}{2} -\frac{|x_{23}^\perp|^2}{x_{23}^+-i\epsilon} + i\epsilon = x_4^- - \frac{|x_{23}^\perp|^2}{x_{23}^+-i\epsilon} + \frac{|x_{24}^\perp|^2}{x_{24}^+} + i\epsilon\label{eq:x34pole2app}\, .
\end{align}
If we insert \eqref{eq:L1L1comintapp} in a three-point function and use only the first Wick contraction \eqref{eq:Wick1app}, performing the integral by computing the residue at the location of the pole \eqref{eq:x34pole1app}, we arrive to
\begin{align}
\left.\braket{\phi(x_1)[L_1(x_2),L_1(x_3)]\phi(x_4)}\right|_1
= - \ \fr{1}{\pi^2} \fr{\x{1}{2} \x{4}{3}}{x_{12}^+ x_{24}^+ |\vec{x}_{23}^\perp|^2}\, .
\end{align}
On the other hand, if we use the second line of \eqref{eq:Wick1app} and perform the integral picking up the pole as \eqref{eq:x34pole2app}, we get 
\begin{align}
\left.\braket{\phi(x_1)[L_1(x_2),L_1(x_3)]\phi(x_4)}\right|_2
= \frac{1}{\pi^2}\frac{\x{4}{2}\x{1}{3}}{x_{12}^+x_{24}^+|x_{23}^\perp|^2}
\end{align} 
Adding both results yields the commutator we already derived using $\braket{\phi(x_1)T_{--}(x_2)T_{--}(x_3)\phi(x_4)}$ that is \eqref{eq:L1L1comm}. 

If the goal is to reproduce only the leading singularity in an expansion as $\vec{x}_2^\perp \rightarrow \vec{x}_3^\perp$, which is given by 
\begin{align}
\left.\braket{\phi(x_1)[L_1(x_2),L_1(x_3)]\phi(x_4)} \right|_{\vec{x}_2^\perp\rightarrow \vec{x}_3^\perp}
&= \frac{2(x_{23}^\perp)^I\left[-(x_{13}^\perp)^I x_{34}^+ \x{4}{3} + (x_{34}^\perp)^I x_{13}^+ \x{1}{3}\right]}{\pi^2 |x_{23}^\perp|^2(x_{13}^+x_{34}^+)^2}+\dots\, , \label{eq:L1L1commexp1app}
\end{align}
then only the last line of equation \eqref{eq:L1L1comintapp} is needed. This implies that you can get the whole finite separation contribution to the commutator by just considering 
\begin{align}
&\left.\braket{\phi(x_1)[L_1(x_2),L_1(x_3)]\phi(x_4)}\right|_{\text{l.s.}} = -\int dx_3^- \frac{(2\pi i)(x_3^-)^3}{\pi^4(x_{23}^+-i\epsilon)}\left[\frac{1}{24}(\x{3}{2})\braket{\phi(x_1)\phi(\x{3}{2})\partial_-^2\phi(x_3)\phi(x_4)}\right.\nn\\
&\qquad \qquad +\left.\frac{1}{8}(\x{3}{2})^2\braket{\phi(x_1)\partial_-\phi(\x{3}{2})\partial_-^2\phi(x_3)\phi(x_4)}+\frac{1}{24}(\x{3}{2})^3\braket{\phi(x_1)\partial_-^2\phi(\x{3}{2})\phi(x_3)\phi(x_4)}\right]\, ,\label{eq:L1L1toberesummedapp}
\end{align}
where the two central operators of each term are normal ordered. 
%


Finally, we want to comment on the following. We can expand \eqref{eq:L1L1toberesummedapp} as $x_2$ goes to $x_3$. This produces an infinite sum that is similar to the one we proposed in \eqref{eq:L1L1comopf}. This suggests that the non-local operator \eqref{eq:L1L1toberesummedapp} is the resummed version of the infinite sum \eqref{eq:L1L1comopf}. Let us explain this in more details. If we consider \eqref{eq:L1L1toberesummedapp} and expand $\phi(\x{3}{2})$ around $x_3$,  this will produce a bilocal operator built out of $\phi(x_3)$ and its derivatives. Because throughout this work, we have evaluated correlators at $x_2^+=x_3^+$, the Taylor expansion of $\phi(\x{3}{2})$ cannot have $\partial_+$ derivatives, and is given by
\be 
\phi(x_2) = \sum_{p=0}^\infty \frac{1}{p!}\sum_{k=0}^{p}\begin{pmatrix} p\\ k\end{pmatrix}(x_{23}^-\partial_-)^k((x_{23}^\perp)^I \partial_I)^{p-k}\phi(x_3)\, .\label{eq:phi2expand}
\ee
The leading singularity in the commutator $\left.[L_1(x_2),L_1(x_3)]\right|_{\text{l.s.}}$ is of the form $(x_{23}^\perp)^I/|x_{23}^\perp|^2$, and we thus want to focus on terms in the expansion \eqref{eq:phi2expand} that have exactly one $\partial_I$ derivative. They are given by $k=p-1$ in  \eqref{eq:phi2expand}. Moreover, $x_2^- = \x{3}{2}$ such that $x_{23}^- = -\frac{i|x_{23}^\perp|^2}{i\epsilon}$, and the expansion becomes
\begin{align}
:\phi(x_{2,3}^-)\phi(x_3): = \sum_{p=0}^{\infty}\frac{1}{p!}\left(-\frac{|x_{23}^\perp|^2}{i\epsilon}\right)^{p}(x_{23}^\perp)^I :\partial_-^p\partial_I\phi(x_3)\phi(x_3): \, .\label{eq:phi2expand2}
\end{align}
Inserting this expansion in the last line of \eqref{eq:L1L1toberesummedapp}, we obtain an infinite sum that resembles \eqref{eq:L1L1comopf}. In principle, one can obtain the coefficients $a_m$ by explicitly comparing these two sums. 

Finally, note that it is clear that one can also just resum \eqref{eq:L1L1toberesummedapp} with \eqref{eq:phi2expand2}. Unsurprisingly, this reproduces the full contribution at finite separation of the $[L_1(x_2),L_1(x_3)]$ commutator.

\section{Details of the $[L_1,L_1]$ expansion \label{ap:L1L1OPE}}
In this appendix, we want to give more details on the computation of the $L_1 L_1$ OPE to investigate the leading singularity in the commutator of $[L_1(x_2),L_1(x_3)]$ at finite transverse separation. We are going to compute the leading operators in the infinite sum of equation \eqref{eq:L1L1comopf}, that we remind here for convenience 
\be 
\left.[L_1(x_2),L_1(x_3)]\right|_{\text{l.s.}} = \sum_{m=0}^\infty a_m\mathcal{G}^m\, .
\ee  
with 
\be
\mathcal{G}^m(x_i) \equiv \frac{(x_{23}^\perp)^I}{|x_{23}^\perp|^2} \int dx_3^- (x_3^-)^{m+3} \<\phi(x_1) \, \phi\overset{\leftrightarrow}{\partial}_I\overset{\leftrightarrow\,\,\,}{\partial^m_-} \phi(x_3) \, \phi(x_4)\>.\label{eq:L1L1comopaap2}
\ee

\subsection{$m=0$ }
When $m=0$ in \eqref{eq:L1L1comopaap2}, the leading operator is $\partial_I(\phi^2(x_3))$ and the three-point function is 
\begin{align}
\mathcal{G}^{0}(x_i)
&=\int dx_3^- \frac{4(x_{23}^\perp)^I}{|x_{23}^\perp|^2}(x_3^-)^3 \left[(x_{13}^\perp)^I\frac{1}{x_{13}^4x_{34}^2}-(x_{34}^\perp)^I\frac{1}{x_{13}^2 x_{34}^4}\right]\, .
\end{align}
These two terms are never going to mix so we can compute them independently, and we denote the term we are considering by a subscript indicating which perpendicular vector will be summed over with $(x_{23}^\perp)^I$. Let us consider
\begin{align}
\left.\mathcal{G}^0(x_i)\right|_{13}&=\int dx_3^- \frac{4(x_{23}^\perp)^I}{|x_{23}^\perp|^2}\left[(x_{13}^\perp)^I\frac{(x_3^-)^3}{x_{13}^4x_{34}^2}\right]\, .\label{eq:Gosum}
\end{align}
This expression has three poles in $x_3^-$. The first one is when $x_{13}^2=0$, the second when $x_{34}^2=0$ and the last one when $x_3^-\rightarrow \infty$. We will come back to the contributions from the pole at infinity shortly. Let us compute all of these and indicate which pole we are considering by a superscript 
\begin{align}
\left.\mathcal{G}^0(x_i)\right|_{13}^{x_3\rightarrow x_1}&=\frac{4(2\pi i)}{|x_{23}^\perp|^2}(x_{23}^\perp)^I(x_{13}^\perp)^I \left[\frac{(\x{1}{3})^2(-2\x{1}{3} + 3\x{4}{3})}{x_{34}^+(x_{13}^+)^2(\x{1}{3}-\x{4}{3})^2}\right] \, ,\nn\\
\left.\mathcal{G}^0(x_i)\right|_{13}^{x_3\rightarrow x_4}&=\frac{4(2\pi i)}{|x_{23}^\perp|^2}(x_{23}^\perp)^I(x_{13}^\perp)^I\left[\frac{-(\x{4}{3})^3}{x_{34}^+(x_{13}^+)^2(\x{1}{3}-\x{4}{3})^2}\right]\, ,\nn\\
\left.\mathcal{G}^0(x_i)\right|_{13}^{x_3\rightarrow \infty}&=\frac{-4(2\pi i)}{|x_{23}^\perp|^2}(x_{23}^\perp)^I(x_{13}^\perp)^I \left[\frac{(2\x{1}{3}+ \x{4}{3})}{x_{34}^+(x_{13}^+)^2}\right]\nn\, .
\end{align}
Summing the three contributions gives zero as expected, but this indicates that we need to take care of the pole at infinity for this operator. We can do the same computation with the second term of \eqref{eq:Gosum}. It gives
\begin{align}
\left.\mathcal{G}^0(x_i)\right|_{34}^{x_3\rightarrow x_1}&= -\frac{4(2\pi i)}{|x_{23}^\perp|^2}(x_{23}^\perp)^I(x_{34}^\perp)^I\left[\frac{(\x{1}{3})^3}{x_{13}^+(x_{34}^+)^2(\x{1}{3}-\x{4}{3})^2}\right]\, ,\nn\\
\left.\mathcal{G}^0(x_i)\right|_{34}^{x_3\rightarrow x_4}&= -\frac{4(2\pi i)}{|x_{23}^\perp|^2}(x_{23}^\perp)^I(x_{34}^\perp)^I\left[\frac{(\x{4}{3})^2(2\x{4}{3}-3\x{1}{3})}{x_{13}^+(x_{34}^+)^2(\x{1}{3}-\x{4}{3})^2}\right]\, ,\nn\\
\left.\mathcal{G}^0(x_i)\right|_{34}^{x_3\rightarrow \infty}&= -\frac{4(2\pi i)}{|x_{23}^\perp|^2}(x_{23}^\perp)^I(x_{34}^\perp)^I\left[\frac{-(\x{1}{3}+2\x{4}{3})}{x_{13}^+(x_{34}^+)^2}\right]\nn\, .
\end{align}
The limit we described in the main text, and that we remind here, has been designed such that the contribution of the pole at infinity vanishes (at least at leading order), such that the subtlety at infinity disappears. The limit is the following :
\be 
|x_{13}^\perp| = |x_{34}^\perp| = |x^\perp|\rightarrow \infty,\qquad\qquad  x_{34}^+=x_{13}^+ = x^+\, .
\ee
Let us add the two terms for the pole at infinity, and take the limit we just described. This yields 
\begin{align}
\left.\mathcal{G}^0(x_i)\right|_{13+34}^{x_3\rightarrow \infty }
&=(2\pi i) \frac{(x_{23}^\perp)^I}{|x_{23}^\perp|^2}\frac{|x^\perp|^2}{(x^+)^4}\left[-(x_{13}^\perp)^I+ (x_{34}^\perp)^I \right]+\dots \nonumber
\end{align}
Because $(x_{13}^\perp)^I  =(x_{34}^\perp)^I = (x^\perp)^I$, this vanishes. The other poles give
\begin{align}
\left.\mathcal{G}^0(x_i)\right|_{13+34}^{x_3\rightarrow x_4}&=-\left.\mathcal{G}^0(x_i)\right|_{13+34}^{x_3\rightarrow x_1}=-6(2\pi i)\frac{(x_{23}^\perp)^I(x^\perp)^I}{|x_{23}^\perp|^2}\frac{|x^\perp|^2}{(x^+)^4}+\dots \nn
\end{align}
They have exactly the same functional form as the leading singularity of the commutator in this limit $\left.[L_1(x_2),L_1(x_3)]\right|_{\text{l.s}}$ (cf \eqref{eq:L1L1x2closex3limit}). We will now present the general terms in this sum, splitting $m$ into even and odd numbers and explaining the general features. 

\subsection{Even $m$}

The leading operators that appear in the sum are given by \eqref{eq:Ji1is}. We remind them here for convenience
\be 
J_{i_1\dots i_s}(x)= \sum_{k=0}^{s}h(k,s)\partial_{i_1}\dots \partial_{i_k}\phi(x)\,\partial_{i_{k+1}}\dots \partial_{i_s}\phi(x)-\text{traces}\, , \label{eq:Opappearapp}
\ee
with 
\be 
h(k,s)=\frac{(-1)^k}{\Gamma(k)^2\Gamma(s-k)^2}\, .
\ee
For even $m=2p$, we want to compute 
\be 
\mathcal{G}^{2p}(x_i)=\int dx_3^- \frac{(x_{23}^\perp)^I}{|x_{23}^\perp|^2}(x_3^-)^{3+2p}  \partial_I\braket{\phi(x_1)J_{(2p)}(x_3)\phi(x_4)}\, .
\ee
The operator $J_{(2p)}$, where the subscript denote the number of $-$ indices, can be constructed as 
\begin{align}
J_{(2p)} = \sum_{\ell=0}^{p-1}\, 2\,h(\ell ,2p):\partial_-^{2p-\ell}\phi\partial_-^{\ell}\phi : + h(p,2p):\partial_-^{p}\phi \partial_-^{p}\phi\:, .
\end{align}
The location of the poles is the same as for the case $m=0$, but the pole at infinity is present in the sum only for $p=\lbrace 0,1\rbrace$. We will come back to this shortly. As for the $m=0$ case, their contribution vanishes in the limit we are considering. For generic $p$ and in the limit we are interested in, the outcome is 
\begin{align}
\left.\mathcal{G}^{2p}(x_i)\right|_{13+34}^{x_3\rightarrow x_4} &= (2\pi i)\frac{(-1)^p2^{1-2p}(1 +2p)(3+2p)}{(2p-1)\Gamma(1 + p)^2}\frac{(x_{23}^\perp)^I(x^\perp)^I}{|x_{23}^\perp|^2} \frac{|x^\perp|^2}{(x^+)^4}\, . 
\end{align}
We then see that all the even terms contribute to $\left.[L_1(x_2),L_1(x_3)]\right|_{\text{l.s}}$ in the limit we consider.  

\subsection{Odd $m$}
We can now consider the case where $m=2q+1$ is odd. We want to compute 
\be 
\mathcal{G}^{2q+1}(x_i)=\int dx_3^- \frac{(x_{23}^\perp)^I}{|x_{23}^\perp|^2}(x_3^-)^{4+2q}\braket{\phi(x_1)J_{(2p+1)I}(x_3)\phi(x_4)}\, .
\ee
The operator $J_{(2p)I}$ can be constructed as 
\begin{align}
J_{(2q+1)I} &= \sum_{\ell=0}^{q+1}(2q+2-\ell)h(\ell,2q+2)\left(:\partial_-^{2q+1-\ell}\partial_I\phi \partial_-^\ell\phi : + :\partial_-^\ell \phi \partial_-^{2q+1-\ell}\partial_I\phi:\right)\nn\\
& \phantom{=}+\sum_{\ell=1}^{q}\ell\,  h(\ell,2q+2)\left(:\partial_-^{2q+2-\ell}\phi \partial_-^{\ell-1}\partial_I\phi:+:\partial_-^{\ell-1}\partial_I\phi \partial_-^{2q+2-\ell}\phi:\right) \, .
\end{align}
Reproducing the computation we did for arbitrary $q$ in the limit we are interested in yields 
\begin{align}
\left.\mathcal{G}^0(x_i)\right|_{13+34}^{x_3\rightarrow x_4} &= (2\pi i)\frac{(-1)^q 2^{-2q}(3 + 2q)}{(1+2q)\Gamma(q+1)\Gamma(q+2)}\frac{(x_{23}^\perp)^I(x^\perp)^I}{|x_{23}^\perp|^2} \frac{|x^\perp|^2}{(x^+)^4}\, . 
\end{align}
We then see that all the odd terms all contribute at leading order to $\left.[L_1(x_2),L_1(x_3)]\right|_{\text{l.s}}$ in the limit we consider. In addition, only $q=0$ has a contribution coming from a pole at infinity.  We conclude that all terms in the expansion contribute at the same order and thus the infinite sum does not truncate. We are just resumming an infinite number of coefficients. 

\subsection{Poles at infinity}
Let us now explain why only the first three terms with $m=0,\, 1,\, 2$ have a contribution coming from evaluating a residue at infinity. For this discussion, we use our intuition from two-dimensional CFT. 

In this setup, the relevant scaling dimension is the collinear weight $h$. When performing a light-ray integral on an arbitrary operator $\mathcal{O}(x)$ with collinear weight $h$, the resulting light-ray operator has collinear weight $h-1$. In two dimensions, the first operator that acts non-trivially on the vacuum on the right is schematically 
\be 
\int dx^- \frac{1}{x^-}\mathcal{O}(x)\, ,
\ee
which has collinear weight $h$. Using inversion, the first operator that acts non-trivially on the vacuum on the left is then
\be 
\int dx^- (x^-)^{2h-1}\mathcal{O}(x)\, .
\ee
Because acting non-trivially on the vacuum is directly related with having a pole at infinity, this implies that for any operator $\mathcal{O}(x)$ dressed with a power of $(x^-)$ smaller than $2h-1$, there will be no poles at infinity.

Let us now do some collinear weight counting. The rules are the following: $\phi(x)$ has $h=1/2$, $\partial_-$ has $h=1$ and $\partial_I$ has $h=1/2$. In addition, when considering operators with even $m$ that are of the form $\partial_I J_{(2p)}$ we do not count the $\partial_I$ because it can be stripped off from the integral without changing the behaviour at infinity. 

For a given $m$, the collinear weight of the leading operator is 
\be 
h(m) = 1 + m + \frac{1}{2}( m\;\mathrm{mod}\; 2)\, .
\ee
We have a pole at infinity if $2h-1 \leq m-3$, which is the power of $(x_3^-)$ that appears in our infinite sum. This implies 
\be 
2 + 2m + (m\;\mathrm{mod}\; 2) \leq m-3 \qquad \rightarrow \qquad m\leq 2-(m\;\mathrm{mod}\; 2)\, .
\ee
This inequality is satisfied provided $m\leq 2$, which indicates that only the three first terms in our sum have a contribution coming from a pole at infinity. This is exactly the behaviour we witnessed when performing these integrals explicitly.  

\section{Conformal blocks at finite position}
\label{app:NullPlaneBlocks}

In section~\ref{sec:confblock}, we analyzed the contribution of the $\Ocal$ conformal block to the commutators of light-ray operators. However, we specifically considered the case where the light-ray operators were all inserted at future null infinity, which simplified the calculation significantly. In this appendix, we consider the case where the two light-ray operators are instead inserted on the same null slice at some finite $x^+$, to confirm the non-vanishing commutator at finite transverse separation. Because this setup is more complicated, we will focus on the specific case where $\Ocal$ has dimension $\De=2$, which corresponds to $\phi^2$ for the case of free field theory. Our computation of the $\Ocal$ conformal block will use the Mathematica package {\tt CFTs4D} presented in~\cite{Cuomo:2017wme} and will largely follow the same methodology and notation introduced there.

\subsection{Lightning review of spinning correlators}

A general four-point function of operators in traceless symmetric representations of the Lorentz group (each labeled by their spin $j_i$) can always be written in the form
\be
\<\Ocal_{j_1}(x_1) \Ocal_{j_2}(x_2) \Ocal_{j_3}(x_3) \Ocal_{j_4}(x_4)\> = \Kcal_4(x_i) \sum_I g^I(u,v) \, \mathbb{T}_4^I(x_i),
\ee
where the RHS is a sum over all possible four-point function tensor structures $\mathbb{T}_4^I$, which are completely fixed by conformal symmetry, based on the spins $j_i$. Each tensor structure is multiplied by a corresponding scalar function $g^I(u,v)$, which is a function of the standard conformally-invariant cross-ratios
\be
u \equiv \fr{x_{12}^2 x_{34}^2}{x_{13}^2 x_{24}^2} = z\bar{z}, \qquad v \equiv \fr{x_{14}^2 x_{23}^2}{x_{13}^2 x_{24}^2} = (1-z)(1-\bar{z}).
\label{eq:UVdef}
\ee
Finally, the overall kinematic factor $\Kcal_4$ is fixed by the scaling dimensions and spins of the external operators,
\be
\mathcal{K}_4(x_i) = \left(\fr{x_{24}}{x_{14}}\right)^{\kappa_1 - \kappa_2} \left(\fr{x_{14}}{x_{13}}\right)^{\kappa_3 - \kappa_4} \fr{1}{x_{12}^{\kappa_1 + \kappa_2} x_{34}^{\kappa_3 + \kappa_4}},
\ee
where $\kappa_i \equiv \De_i + j_i$.

The set of tensor structures $\mathbb{T}_4^I$ depends on the spins of the four operators, but they can all be constructed from the two building blocks
\be
H_{ij}^{\mu\nu} = x_{ij}^2 \eta^{\mu\nu} - 2 x_{ij}^\mu x_{ij}^\nu, \qquad V_{k,ij}^\mu = \fr{x_{ki}^2 x_{kj}^2}{x_{ij}^2} \left( \fr{x_{ki}^\mu}{x_{ki}^2} - \fr{x_{kj}^\mu}{x_{kj}^2} \right).
\label{eq:HandV}
\ee

We can compute the four-point function by inserting a complete set of intermediate states. These can be arranged into irreducible representations of the conformal group, each associated with a primary operator $\Ocal'$, resulting in the conformal partial wave decomposition
\be
\<\Ocal_{j_1}(x_1) \Ocal_{j_2}(x_2) \Ocal_{j_3}(x_3) \Ocal_{j_4}(x_4)\> = \sum_{\Ocal'} \sum_{a,b} \lambda^a_{\Ocal_1 \Ocal_2 \Ocal'} \lambda^b_{\Ocal' \Ocal_3 \Ocal_4} W^{ab}_{\Ocal'}(x_i).
\label{eq:CPWexpansion}
\ee
The indices $a,b$ label the set of allowed tensor structures for three-point functions involving $\Ocal'$ with the external operators, and $\lambda^a$ are the associated OPE coefficients,
\be
\<\Ocal_{j_1}(x_1) \Ocal_{j_2}(x_2) \Ocal_{j_3}(x_3)\> = \Kcal_3(x_i) \sum_a \lambda^a_{\Ocal_1\Ocal_2\Ocal_3} \, \mathbb{T}_3^a(x_i),
\ee
with the familiar three-point function kinematic factor
\be
\mathcal{K}_3(x_i) = \fr{1}{x_{12}^{\kappa_1 + \kappa_2 - \kappa_3} x_{23}^{\kappa_2 + \kappa_3 - \kappa_1} x_{13}^{\kappa_1 + \kappa_3 - \kappa_2}}.
\ee

The functions $W^{ab}_{\Ocal'}$ in eq.~\eqref{eq:CPWexpansion} are known as \emph{conformal partial waves}, and encode the contribution to a four-point function from a given pair of three-point function tensor structures associated with $\Ocal'$. These individual conformal partial waves can each be decomposed into four-point function tensor structures,
\be
W^{ab}_{\Ocal'}(x_i) = \Kcal_4(x_i) \sum_I G^{I,ab}_{\Ocal'}(u,v) \, \mathbb{T}_4^I(x_i).
\ee
The functions $G^{I,ab}_{\Ocal'}$ are referred to as \emph{conformal blocks}, and encode the contribution of a pair of three-point function tensor structures for $\Ocal'$ to a particular four-point function tensor structure. Their structure is completely fixed by conformal symmetry, with many efficient techniques for computing their exact expressions.

In this work, we are specifically interested in the case where two of the external operators are scalars ($j_1 = j_4 = 0$), and the other two are the stress tensor ($j_2 = j_3 = 2$).

\subsection{Computing the conformal partial wave}

Concretely, we would like to compute the contribution of $\Ocal$ to the four-point function
\benn
\<\Ocal(x_1) T_{--}(x_2) T_{--}(x_3) \Ocal(x_4)\>.
\eenn
As discussed in the previous section, this contribution can be decomposed into a sum over four-point function tensor structures $\mathbb{T}_4^I$, which can all be built from the five building blocks:
\be
H_{23}^{++}, \quad V_{2,13}^+, \quad V_{2,14}^+, \quad V_{3,14}^+, \quad V_{3,24}^+.
\ee
Schematically, we have three types of combinations,
\be
(H_{23}^{++})^2, \quad H_{23}^{++} V_2^+ V_3^+, \quad (V_2^+)^2 (V_3^+)^2,
\ee
with two options each for $V_2$ and $V_3$. There are therefore $1 + 4 + 9 = 14$ different four-point tensor structures.

In general, there is a distinct conformal partial wave $W^{ab}_\Ocal$ for every incoming and outgoing three-point function tensor structure. Fortunately, because the exchanged operator $\Ocal$ is a scalar, for this case there is only one three-point tensor structure, and therefore only a single conformal partial wave.

The typical strategy for constructing conformal partial waves for external states with spin is to act with particular differential operators, known as weight-shifting operators, on the known conformal blocks for scalar external states~\cite{Costa:2011dw,Karateev:2017jgd}. For our particular case, where the spinning external operator is the stress tensor, with $\De_T=4$ and $j_T=2$, the seed conformal block is
\be
G^{\textrm{seed}}_\Ocal(u,v) = \fr{z \zb}{z-\zb} \Big( k^{(\De-2,2-\De)}_\De(z) k^{(\De-2,2-\De)}_{\De-2}(\zb) - k^{(\De-2,2-\De)}_{\De-2}(z) k^{(\De-2,2-\De)}_\De(\zb) \Big),
\ee
where $k^{(\alpha,\beta)}_\Delta(x)$ is defined as
\be
k^{(\alpha,\beta)}_\Delta(x) = x^\fr{\Delta}{2} {}_2F_1\Big(\tfr{1}{2}(\Delta - \alpha),\tfr{1}{2}(\Delta+\beta);\Delta;x\Big).
\ee
Because the exchanged operator is a scalar, we can rewrite this seed block in the more useful form~\cite{Dolan:2011dv}
\be
G^{\textrm{seed}}_\Ocal(u,v) = \sum_{n=0}^\infty \sum_{m=0}^\infty \fr{(\De-1)_n}{n!} \fr{\G^2(n+m+1)}{m! (\De)_{2n+m}} u^{\fr{\De}{2}+n} (1-v)^m,
\ee
which can be partially resummed to obtain
\be
G^{\textrm{seed}}_\Ocal(u,v) = \sum_{n=0}^\infty \fr{n! (\De-1)_n}{(\De)_{2n}} u^{\fr{\De}{2}+n} {}_2F_1(n+1,n+1;\De+2n;1-v).
\label{eq:SeedBlock}
\ee

We then need to construct the appropriate weight-shifting operator and act on the seed conformal block to obtain the conformal partial wave. The result is remarkably complicated, but fortunately we are only interested in a subset of the full expression. In particular, we are interested in terms which are nonzero when we integrate over both $x_2^-$ and $x_3^-$ to obtain a correlation function involving light-ray operators. This means we only need to focus on terms with poles when $x_2 \ra x_1$ and $x_3 \ra x_4$. Such poles come from the overall kinematic factor
\be
\mathcal{K}_4(x_i) = \fr{1}{x_{12}^{\De + 6} x_{34}^{\De+6}} \left(\fr{x_{24}}{x_{14}}\right)^{\De - 6} \left(\fr{x_{14}}{x_{13}}\right)^{6 - \De},
\ee
multiplied by powers of $u^{\fr{\De}{2}+n}$ coming from derivatives of the seed block. Because higher-order terms in $u$ are less singular, in practice we therefore only need the first three terms in the block expansion~\eqref{eq:SeedBlock}.

In addition, we are interested only in contributions to the finite separation commutator of operators on the same null slice. We can therefore set $x_{23}^+ = 0$, in which case all tensor structures containing $H_{23}$ vanish.

For general $\Ocal$, the most singular term has a third-order pole in both $x_2^-$ and $x_3^-$, and takes the simple form
\be
\<\Ocal(x_1) T_{--}(x_2) T_{--}(x_3) \Ocal(x_4)\>\Big|_\Ocal \supset \fr{4 (x_{12}^+)^2 (x_{34}^+)^2 x_{14}^{4-2\De}}{x_{12}^6 x_{34}^6 x_{13}^2 x_{24}^2} {}_2F_1(1,1;\De;1-v),
\ee
which corresponds to the first term in the expansion of the seed block~\eqref{eq:SeedBlock}. The remaining less singular terms have the same basic structure, but include sums of multiple hypergeometric functions with various arguments.

For the rest of this appendix, we will focus on the specific case $\De=2$, which corresponds to the operator $\phi^2$ in free field theory, though these results hold for any scalar operator with the same scaling dimension in any CFT. In this case, we obtain the full set of singular terms:
\be
\ba
&\<\phi^2(x_1) T_{--}(x_2) T_{--}(x_3) \phi^2(x_4)\>\Big|_{\phi^2} \\
& \qquad \supset -\frac{4 (x_{12}^+)^2 (x_{24}^+)^2 \log v}{(1-v) x_{12}^6 x_{13}^2 x_{24}^2 x_{34}^6} -\frac{8 (x_{12}^+) (x_{24}^+) (1-v+\log v)}{(1-v)^2 x_{12}^4 x_{13}^2 x_{24}^2 x_{34}^4} \left( \fr{(x_{12}^+)^2}{x_{12}^2 x_{13}^2} + \fr{(x_{24}^+)^2}{x_{24}^2 x_{34}^2} \right) \\
& \qquad \qquad - \, \frac{2\left(3-4v+v^2+2 \log v\right)}{(1-v)^3 x_{12}^2 x_{13}^2 x_{24}^2 x_{34}^2} \left( \fr{(x_{12}^+)^4}{x_{12}^4 x_{13}^4} + \fr{(x_{24}^+)^4}{x_{24}^4 x_{34}^4} \right) \\
& \qquad \qquad - \, \frac{52 (x_{12}^+)^2 (x_{24}^+)^2 (2(1-v)+(1+v) \log v)}{(1-v)^3 x_{12}^4 x_{13}^4 x_{24}^4 x_{34}^4} \\
& \qquad \qquad - \, \frac{40 (x_{12}^+) (x_{24}^+) ((1-v) (5+v)+2 (1+2v) \log v)}{(1-v)^4 x_{12}^2 x_{13}^4 x_{24}^4 x_{34}^2} \left( \fr{(x_{12}^+)^2}{x_{12}^2 x_{13}^2} + \fr{(x_{24}^+)^2}{x_{24}^2 x_{34}^2} \right) \\
& \qquad \qquad - \, \frac{292 (x_{12}^+)^2 (x_{24}^+)^2 \left(3(1-v^2)+(1+4v+v^2)\log v\right)}{(1-v)^5 x_{12}^2 x_{13}^6 x_{24}^6 x_{34}^2},
\ea
\label{eq:Phi2Block}
\ee
where we have suppressed any overall OPE coefficient.

\subsection{Light-ray operator commutators}

Now that we have the singular terms from the $\phi^2$ partial wave, we can integrate to obtain the contribution to correlators of light-ray operators. As a simple example, let's first consider the case where both operators are the ANEC operator $L_{-2}$. In this case, we simply need to evaluate the integral
\be
\<\phi^2(x_1) L_{-2}(x_2) L_{-2}(x_3) \phi^2(x_4)\>\Big|_{\phi^2} = \int dx_2^- \, dx_3^- \<\phi^2(x_1) T_{--}(x_2) T_{--}(x_3) \phi^2(x_4)\>\Big|_{\phi^2}.
\ee
In practice, this integration is rather straightforward, as we simply pick up the poles
\be
x_2^- = x_{1,2}^-, \qquad x_3^- = x_{4,3}^-.
\ee

As we can see from eq.~\eqref{eq:Phi2Block}, the singular terms are largely functions of $v$, so the resulting expression is mostly dependent on $v$ evaluated at the singular points, which we indicate by
\be
\tilde{v} \equiv v\Big|_{\substack{x_2^- = x_{1,2}^- \\ x_3^- = x_{4,3}^-}} = \fr{x_{14}^2 |\vec{x}_{23}^\perp|^2}{x_{12}^+ x_{24}^+(x_{1,2}^- - x_{4,2}^-)(x_{1,3}^- - x_{4,3}^-)}.
\ee
Note that $\tilde{v}$ is simplified somewhat by the fact that $x_{23}^+=0$.

Evaluating this integral, we then obtain the resulting partial wave contribution to a light-ray operator correlator (up to an overall numerical coefficient),
\be
\<\phi^2(x_1) L_{-2}(x_2) L_{-2}(x_3) \phi^2(x_4)\>\Big|_{\phi^2} = - \fr{864 \Big(3(1-\tilde{v}^2) + (1 + 4\tilde{v} + \tilde{v}^2) \log \tilde{v} \Big)}{(x_{12}^+)^2 (x_{24}^+)^2 (x_{1,2}^- - x_{4,2}^-)^3(x_{1,3}^- - x_{4,3}^-)^3(1-\tilde{v})^5}.
\ee
One important feature of this expression, apart from its notable simplicity relative to the full partial wave, is that it is clearly symmetric under the exchange $x_2 \lra x_3$, due largely to the symmetric nature of $\tilde{v}$. Because of this symmetry, the resulting commutator is clearly zero, as expected,
\be
\<\phi^2(x_1) \comm{L_{-2}(x_2)}{L_{-2}(x_3)} \phi^2(x_4)\>\Big|_{\phi^2} = 0,
\ee
In fact, the integral of each individual singular term in eq.~\eqref{eq:Phi2Block} has this same structure, such that no cancellation between distinct terms is needed to ensure that ANEC operators commute for $\phi^2$ exchange.

Finally, let's repeat this procedure for $\comm{L_{-1}}{L_{-2}}$. Using~\eqref{eq:Phi2Block}, we can compute the two orderings, then take the difference to obtain:
\be \label{finitecompos}
\<\phi^2(x_1) \comm{L_{-1}(x_2)}{L_{-2}(x_3)} \phi^2(x_4)\>\Big|_{\phi^2} = -\fr{216 \Big((1-\tilde{v})(7+16\tilde{v}+\tilde{v}^2) + 2(1 + 7\tilde{v} + 4\tilde{v}^2) \log \tilde{v} \Big)}{(x_{12}^+)^2 (x_{24}^+)^2 (x_{1,2}^- - x_{4,2}^-)^2(x_{1,3}^- - x_{4,3}^-)^3(1-\tilde{v})^5},
\ee
again, up to an overall coefficient. We therefore find that $\Ocal$ exchange leads to a \emph{nonzero} commutator at finite transverse separation, as seen in section~\ref{sec:confblock} from correlators on the celestial sphere. In free field theory, this nonzero contribution must therefore cancel with the infinite tower of two-particle operators in the $\phi^2 \times T$ OPE to ensure that the commutator vanishes in the full correlator, as we would have seen in section \ref{TphiOPEsec} had we considered the state $\phi^2$.

\bibliographystyle{ytphys}
\bibliography{ref}

\end{document}